\documentclass[pra,showpacs,twocolumn,superscriptaddress,aps]{revtex4-1}
\usepackage[utf8]{inputenc}
\usepackage{float}
\usepackage{epsfig,epstopdf}
\usepackage{amsmath,amssymb,amsfonts}
\usepackage{bm,graphicx,subfigure,float}
\usepackage{geometry}
\usepackage{booktabs}
\usepackage{longtable,multirow,supertabular,xspace}
\usepackage{color}
\usepackage[table]{xcolor} 
\usepackage{pgfplots} 
\usepackage{tikz}
\usetikzlibrary{spy}

\usepackage[colorlinks=true, linkcolor=blue, citecolor=blue, urlcolor=black]{hyperref}
\begin{document}
\title{Multi-vortex Bose-Einstein condensate: examining the role of interaction range using Gaussian potential }
\author{Md Hamid }\email{hamid160520@st.jmi.ac.in}
\affiliation{Department of Physics, Jamia Millia Islamia (A Central University), New Delhi 110025, India.} 
\author{M. A. H. Ahsan}\email{mahsan@jmi.ac.in}
\affiliation{Department of Physics, Jamia Millia Islamia (A Central University), New Delhi 110025, India.}
\date{\today}
\begin{abstract}
\indent 
We present exact diagonalization study
on a system of 
$10 \leq N \leq 24$ spinless bosons interacting via repulsive Gaussian potential, harmonically confined in $xy$-plane with an externally impressed rotation about the $z$-axis. The two-body interaction strength
in the Gaussian potential is taken in the strongly interacting regime with values of interaction range in the regime $0\leq \sigma \leq 1$ where $\sigma$ is the width of the Gaussian in the interaction potential, in units of the oscillator length $a_{\perp}$ of the confining potential. The diagonalization of the $N$-body Hamiltonian matrix, in subspaces of total angular momentum 
in the regime $0\le L_{z} \le 4N$ corresponding to the filling fraction $\nu\lesssim 3.2$, is carried out 
to obtain the variationally exact ground-state wavefunction and the corresponding eigenvalue. 
It is found that an increase in interaction range $\sigma$ leads to (a) a systematic decrease in energy, (b) an increase in the critical angular velocity $\Omega_{c_{i}}$ of the $i${th} vortex state and (c) an increase in the largest eigenvalue $\lambda_{1}$, corresponding to the condensate fraction, of the one-particle reduced density matrix (OPRDM).
The von Neumann entropy $S_{1}\left(L_{z},\{\sigma\}\right)$, 
quantifying the quantum entanglement between the particles in the many-body ground state, is largely
found to decrease with increase in interaction range $\sigma$, in given subspace of total angular momentum $L_{z}$. Crossings in von Neumann entropy  
$S_{1}\left(L_{z},\{\sigma\}\right)$ for several of the angular momentum states are observed with variation in interaction range $\sigma$. The response of the Bose-Einstein condensate to rotation is examined through $L_{z}\left(\sigma\right)-\Omega\left(\sigma\right)$ stability graph for several values of interaction range $\sigma$. A vortex state with longer plateau (longer rotational angular velocity span)
on the $L_{z}-\Omega$ stability graph is considered to be more stable. Consequently, the two-vortex state is found to be less stable compared to the single-vortex, the three-vortex and the four-vortex states. The three-vortex and the four-vortex states have comparable span on the $L_{z}-\Omega$ stability graph. The internal structure of the condensate, as depicted by the conditional probability distribution (CPD) in the body-fixed frame, exhibits characteristic features with variation in interaction range $\sigma$.
One such feature observed in the present work is the merging of the cores of the two-vortex state as $\sigma$ is increased.
\end{abstract}
\maketitle
\section{\label{introd} INTRODUCTION}
\indent The advancements in cooling, trapping and manipulation techniques~\cite{Metcalf99,Juliette09,Pitaevski03,Feshbach62,CourseCXL} of cold atoms have led to creation of ultracold dilute atomic gas systems in laboratories with tunable parameters like interaction, size, density, effective dimensionality. This has opened up the possibility of investigating quantum many-body effects such as the effects of quantum statistics, quantum fluctuations due to uncertainty relation between the conjugate observables, interaction between the particles {\it etc}, in a controlled manner. Dilute vapours of weakly interacting alkali atoms such as $^{23}$Na,\ $^{39}$K,\ $^{87}$Rb, with repulsive~\cite{Davis95,Modugno01,Anderson_95}, and $^{7}$Li with attractive interaction~\cite{Bradley95,jmf_gunn97}, confined in magnetic trap and cooled to nanokelvin temperatures have been created in laboratories. One of the landmark achievements in the evolving research has been the realization of Bose-Einstein condensation (BEC) in trapped atomic Bose  gases~\cite{Anderson_95,Bradley95,Davis95,Bradley95}. Dilute fermionic gas of $^{40}$K atoms at very low temperatures have also been created in the laboratories~\cite{Regal04,Salasnich16}. An interesting phenomena in atomic gases is the crossover between Bose-Einstein condensation (BEC) and Bardeen-Cooper-Schrieffer (BCS) superconductivity~\cite{Regal03,Alavi21}. The BEC-BCS crossover involves gradual transition between a gas of weakly interacting Bose atoms (BEC regime) and a gas of weakly interacting pairs of fermionic atoms (BCS regime) with interatomic interaction varied from repulsive to attractive using techniques like Feshbach resonance~\cite{Feshbach62}.
\\
\indent A fundamental characteristic of Bose-Einstein condensate is its response to rotation leading to nucleation of quantized vortices~\cite{Leggett91}.
The quantized vortices have been observed in superfluid $^{4}$He as stable topological objects~\cite{Donney_91}.
In type-$II$ superconductors, quantized vortices arrange themselves in regular triangular lattice pattern which leads to the formation of Abrikosove lattice~\cite{Abrikosov_04}.
\\
\indent The short-range interatomic interaction in dilute atomic gases has theoretically been described in terms of singular $\delta$-function potential~\cite{Two_Cold_Busch97,lewenstein06,hengfan09,Xia01} with single parameter $a_{s}$, the $s$-wave scattering length. The sign of $a_{s}$ is positive for repulsive interaction~\cite{Inou98,Lang09,Luca_Salasnich09,pethick_camb} and negative for attractive interaction~\cite{Bradley95}. 
In recent years, there has been studies~\cite{Pete19,Mujal17,Rost13,Imran_15,Imran_16,Imran_the,Imran_17,Hamid_two, Imran_20,Imran_EJPD23} using
Gaussian type of interaction potential which is smooth and provides control over the interatomic interaction through two parameters, namely, the strength of the interaction as measured by the s-wave scattering length $a_{s}$ and the range of interaction $\sigma$ as measured by the width of the Gaussian potential. However, there has been no systematic study of the effect of interaction range on multi-vortex sates in BEC. This work presents such a study.
\\ 
\indent The interatomic interaction gives rise to phase rigidity to the ground-state many-body wavefunction which manifests itself in successive appearance of vortices with quantized circulations~\cite{Ahsan_Kumar} at a series of critical rotational velocities $\Omega_{c_{i}},\ i=1,2, \cdots$, in response to the externally impressed rotation. Madison {\it et.al.}~\cite{Madison_00} stirred cloud of $^{87}$Rb atoms using focussed laser beam which led to the experimental realization of quantized vortices in Bose-Einstein condensed atomic gases~\cite{Chevy00,Compton09,Cornell99,Cooper01,UziLandman07,Williams10,Fried98}.
\\
\indent When the number of vortices become large, it becomes convenient to introduce the filling fraction $\nu=N/N_{v}$, defined as the ratio of the number of bosons $N$ to the average number of vortices $N_{v}$. The filling fraction ${\nu}$ is a controlling parameter for vortex-lattice to vortex-liquid quantum phase transition at zero temperature~\cite{Cooper01,MacDonald02}. On the basis of exact diagonalization studies with periodic boundary conditions on a quasi-2D system, it has been suggested that the vortex-lattice melts~\cite{GIMenon96,Cooper01} due to quantum fluctuations when the filling fraction is of order $\nu \geq 6$~\cite{Cooper01} and a meanfield description becomes possible. For $\nu <6$, the ground-state is expected to be a strongly correlated entangled state~\cite{Cooper01,hengfan09} and the mean-field description breaks down. Our present study is limited to the angular momentum regime $0\leq L_{z} \leq 4N$ which corresponds to $\nu \lesssim 3.2$ necessitating a quantum-many-body description such as Exact diagonalization as reported in the present work.
\\
\indent This paper is organised as follows: In Sec.~\ref{TheHamiltonian_ch2}, we describe the Hamiltonian for harmonically confined spinless bosons interacting via Gaussian type of potential. Sec.~\ref{sec:res} presents results on several quantities of interest like the energy eigenvalues, the critical rotational angular velocity $\Omega_{c}$, the one particle reduced density matrix(OPRDM), the von Neumann entropy and the conditional probability distribution(CPD).
In Sec.~\ref{sec:summary}, we present summary and conclusion of our study. In Appendix, we present expressions in beyond lowest Landau level approximation for (a) one-particle density (b) conditional probability distribution(CPD) and (c) the stirring term in the Hamiltonian which goes over smoothly to Eq.(9), Eq.(14) and Eq.(17), respectively, in reference~\cite{lewenstein06} in the lowest Landau level approximation. In the Table~\ref{Tab:n16_sp3}, we summarize our numerical results on the many-body ground state energy in the laboratory frame, the three largest eigenvalues of the one-particle reduced density matrix (OPRDM) and the corresponding one-particle quantum numbers in the total angular momentum regime $0\leq L_{z} \leq 4N$ for $N=16$. 
\section{THE SYSTEM AND THE HAMILTONIAN}
\label{TheHamiltonian_ch2}
\indent We consider a system of $N$ spinless bosons each of mass $M$,
interacting via a Gaussian type of potential, harmonically confined in symmetric $x$-$y$ plane with an externally impressed rotation $\mathbf{\Omega}=\Omega{\hat{z}}$ about the z-axis.
The Hamiltonian for the system in the co-rotating frame is given by
\begin{eqnarray}
{{\textbf{H}}}^{rot} &=& {{\textbf{H}}^{lab}}-{{\Omega}}\,\mbox{L}_{z}^{lab},
\label{eq:Hamil_gen}
\end{eqnarray}
where
$\mbox{L}_{z}^{lab}=\sum_{i}^{N}\mbox{l}_{z_i}$ is the total angular momentum of the system about $z$-axis in the laboratory frame.
The Hamiltonian for the system in the laboratory frame is given by
\begin{eqnarray}
{\mathbf{H}}^{lab}&=& \sum^{N}_{i=1} \left[\frac{{\mathbf{p}}_{i}^{2}}{2M} + 
V({\bf r}_{i})\right]
+\sum_{{i}<{j}}U({{\bf r}_{i},{\bf r}_{j}}).
~~~~\label{eq:Hamil_Ch2}
\end{eqnarray}
The first and the second terms in Eq.~(\ref{eq:Hamil_Ch2}) are the kinetic energy and the harmonic trap potential, respectively. The harmonic trap potential is given by
\begin{eqnarray}
V(\mathbf{r})&=&\frac{1}{2}M(\omega_{\perp}^{2}r_{\perp}^{2} + \omega_{z}^{2}z^{2}) 
=\frac{1}{2}M\omega_{\perp}^{2}\left(r_{\perp}^{2} +\lambda_{z}^{2}z^{2}\right),\nonumber\\
\label{eq:trap_pot}
\end{eqnarray}
where $r_{\perp}=\sqrt{x^{2}+y^{2}}$ is the radial distance from the axis of rotation and $\omega_{\perp}$, $\omega_{z}$ are the radial and axial trap frequencies, respectively, with the anisotropy parameter(aspect ratio) $\lambda_{z}\equiv{\omega_{z}}/{\omega_{\perp}}$. In terms of these trap frequencies and the mass of the particle, we define the harmonic oscillator lengths {\it i.e.} the radial length $a_{\perp}=\sqrt{\frac{\hbar}{M\omega_{\perp}}}$ and the axial length $a_{z}=\sqrt{\frac{\hbar}{M\omega_{z}}}$. All energies are to be measured in units of $\hbar\omega$ and
all lengths in the units of $a_{\perp}$. We assume our system to be highly oblate spheroidal with $\lambda_{z}\gg 1$ \cite{DalfovoStringari18,baym_pethick_96}.
\\
\indent The third term in Eq.~(\ref{eq:Hamil_Ch2}) is the particle-particle Gaussian type of interaction potential~\cite{Pete19,Ahsan_Kumar,Imran_15,Imran_20,Mujal17} given by
\begin{eqnarray}
U(\mathbf{r,r^{\prime}})&=&
g_{2}\left(\frac{1}{\sqrt{2\pi}\sigma_{\perp}}\right)^{2} \exp\left[ -\frac{(r_{\perp}-r_{\perp}^{\prime})^{2}}{2\sigma^{2}_{\perp}} \right] \nonumber\\
&& \times \left(\frac{1}{\sqrt{2\pi}\sigma_{z}}\right) \exp \left[-\frac{\left(z-z^{\prime}\right)^{2}}{2\sigma^{2}_{z}} \right]
\end{eqnarray}
where $g_{2}=\frac{4\pi\hbar^{2} a_{s}}{M}$ is the interaction strength and $a_{s}$ is the s-wave scattering length.
The parameters $\sigma_{\perp}$($\sigma_{z}$) is the particle-particle interaction range in the $x$-$y$ plane (along the $z-$axis). In view of $\lambda_{z}\gg 1$, 
we trace out $z$-degree of freedom from the interaction potential $U(\mathbf{r,r^{\prime}})$ and scale it with unit of energy $\hbar\omega_{\perp}$ to obtain the dimensionless Gaussian type of interaction potential in quasi-2D plane
\begin{eqnarray}
U(\mathbf{r,r^{\prime}})
&=&4\pi \frac{a_{s}}{a_{z}}\frac{1}{\sqrt{2\pi}}\frac{1}{\sqrt{{1+(\sigma_{z}/a_{z})^{2}}}}\left(\frac{1}{\sqrt{2\pi}\sigma_{\perp}/a_{\perp}}\right)^{2}
\nonumber \\
&\times&
\exp\left({-\frac{1}{2(\sigma_{\perp}/a_{\perp})^2}\left[\frac{({\bf r}_{\perp}-{\bf r}_{\perp}^{\prime})}{a_{\perp}} \right]^{2}}\right). \nonumber
\end{eqnarray}
Since particles are confined to quasi-2D plane, there is no excitation along the $z$-axis. Taking $\sigma_{z}/a_{z} \ll 1$, the effective interaction potential in quasi-2D plane with $\sigma_{\perp} \equiv \sigma$ becomes
\begin{eqnarray}
U({\bf r},{\bf r}^{\prime})
&=&\underbrace{ 4\pi \frac{a_{s}}{a_{\perp}}\frac{1}{\sqrt{2\pi\lambda_{z}}}}_{U_{0}}
\left(\frac{1}{\sqrt{2\pi}\sigma/a_{\perp}}\right)^{2}
\nonumber \\
&\times&\exp\left({-\frac{1}{2(\sigma/a_{\perp})^2}\left[\frac{({\bf r}_{\perp}-{\bf r}^{\prime}_{\perp})}{a_{\perp}} \right]^{2}}\right) ~~~~~~~~
\label{eq:int_pot}
\end{eqnarray}
where, $U_{0}=4\pi \frac{a_{s}}{a_{\perp}}\frac{1}{\sqrt{2\pi\lambda_{z}}}$ is the effective 2D interaction strength. Hence forward, instead of the term effective 2D interaction strength, we will use the term interaction strength $U_{0}$. 
\\
\indent The system described by the Hamiltonian ${\bf H}^{\mbox{lab}}$ in Eq.~(\ref{eq:Hamil_Ch2}) is subjected to an externally impressed rotation. In experiments, this is accomplished through a focused laser beam, used to stir the Bose-Einstein condensate to a desired angular momentum state, leading to creation of vortices beyond a critical velocity~\cite{Madison_00,Chevy00}. Theoretically, this is achieved by introducing a small perturbative potential~\cite{lewenstein06} which after scaling by the unit of energy $\hbar\omega_{\perp}$ with $A$ as the dimensionless small parameter becomes:
\begin{eqnarray}
V_{p}&=&\frac{1}{\hbar\omega_{\perp}}\frac{A}{2}M\omega_{\perp}^{2}
\left(x^{2}-y^{2}\right)
\nonumber\\&=&
\frac{A}{2}
\left(\frac{r_{\perp }}{a_{\perp}}\right)^{2}\cos(2\phi)
\label{eq:Azp}
\end{eqnarray}
The perturbation potential $V_{p}$ breaks the azimuthal symmetry in the $x$-$y$ plane due to the apperence of azimuthal angle $\phi$ on the right hand side of Eq.~(\ref{eq:Azp}).  
\\
\indent The full Hamiltonian now becomes ${\bf H}={\bf H}^{\mbox{lab}}+Vp$. 
Assuming $A\ll 1$, we 
ignore the perturbative potential once the condensate has been set into rotation with a given total angular momentum $L_{z}$.
The exact diagonalization of the Hamiltonian is thus performed in subspaces of total angular momentum $L_{z}$. Discussion on the perturbation potential $V_{p}$ in beyond lowest Landau level approximation is presented in Appendix~\ref{App:anisot}. 
\section{\label{sec:res} RESULTS AND DISCUSSIONS}
\indent We present our results for the ground-state of $N=10,12,16$ spinless bosons in the beyond lowest Landau level approximation and for $N=20,24$
in the lowest Landau level approximation. The values of anisotropy parameter(aspect ratio) and the trap frequency are taken $\lambda_{z}={\omega_{z}}/{\omega_{\perp}}
=\sqrt{8}$ and 
$\omega_{\perp}=\frac{2\pi \times 220}{\sqrt{8}}=
2\pi\times 77.78$ Hz respectively, with axial frequency $\omega_{z}=2\pi\times220$ Hz~\cite{DalfovoStringari18,baym_pethick_96}. Due to large  anisotropy of the trap, there will practically be no excitation along the z-axis. The parameters have been chosen corresponding to $^{87}$Rb atom for which the trap length turns out to be $a_{\perp}=\sqrt{\hbar /(M\omega_{\perp})}=1.222 \mu m$~\cite{DalfovoStringari18}.
\\
\indent The dimensionless parameter for particle-particle interaction strength for a system of $N$ bosons, in the mean-field, is proportional to $N{a_{s}}/{a_{\perp}}$\cite{baym_pethick_96}.
As the number of particles increases, the dimensionality of the many-particle Hilbert space increases exponentially. 
Therefore, instead of increasing $N$, 
we increase the scattering length to $a_{s}=1000a_{0}~$\cite{Ahsan_Kumar}, where $a_{0}$ is the Bohr radius, so as to get the value of the parameter $N{a_{s}}/{a_{\perp}}$ closer to experimental situation~\cite{DalfovoStringari18} with $U_{0}=0.2171$ and $N=10$ to $24$. The rotational velocity $\Omega$ is taken in slowly to moderately rotating regime, with angular momenta $0 \le L_{z} \le 4N$ corresponding to the filling fraction $\nu \lesssim 3.2$ where $\nu=N/N_{v}$, $N_{v}$ being the number of vortices.
In the present work we investigate the effect of the interaction range $\sigma$, as define in Eq.~(\ref{eq:int_pot}), on the properties of the condensate such as the critical angular velocity, the condensate fraction, the nucleation of vortices and the merging of the cores of vortices {\it etc}. In the quasi-2D $x$-$y$ plane, the inter-particle spacing is proportional to $ \frac{a_{\perp}}{\sqrt{N}}$ which for $N=16$ is $0.25a_{\perp}$.
To examine the role of $\sigma$ on quantum many-body effects in the system, for a given interaction strength of $U_{0}$, the interaction range
$\sigma$ is varied from $0.1 \ \mbox{to}\ 1.0$ (in units of $a_{\perp}$).
\\
\indent In our present study 
for moderately to strongly interacting system and slowly to moderately rotating regime~\cite{Ahsan_Kumar}, the beyond-lowest Landau level(LLL) approximation is considered a better approximation over LLL approximation~\cite{Imran_15,Imran_16,Imran_20}. For computational feasibility, we select~\cite{Ahsan_Kumar} a finite number of one-particle basis states to construct the many-particle basis states to determine the variationally exact ground state wavefunction $\Psi_{0}\left({\bf r}_{1},{\bf r}_{2},\cdots {\bf r}_{N}\right)$ given by the eigenvalue equation:
\begin{eqnarray}
{\bf H}^{lab}\Psi_{0}\left({\bf r}_{1},{\bf r}_{2},\cdots, {\bf r}_{N}\right)=&&E^{lab}_{0}\left(L_{z},\{U_{0},\sigma\}\right)\nonumber\\
&&\Psi_{0}\left({\bf r}_{1},{\bf r}_{2},\cdots, {\bf r}_{N}\right)~~~~~
\label{eq:eig_eq_E0}
\end{eqnarray}
where ${\bf H}^{lab}$ is the many-body Hamiltonian in Eq.~(\ref{eq:Hamil_gen}) and $E^{lab}_{0}\left(L_{z},\{U_{0},\sigma\}\right)$ the corresponding interacting ground state energy in the laboratory frame for a given number of bosons $N$ in a given subspace of the total angular momentum $L_{z}$. The ground state energy $E^{lab}_{0}$ depends parametrically on $U_{0}$, $\sigma$ and $L_{z}$ is the constant of motion.\\
\indent We present our results 
in Table~\ref{Tab:n16_sp3} on the ground-state energy eigenvalues for three values of interaction range $\sigma$ for $N=16$ and angular momentum regime $0\leq L_{z}\leq 4N$. We observe that the ground-state energy decreases systematically with increase in interaction range $\sigma$. We discuss this further in Section.\ref{sec:von_S1} with reference to Fig.\ref{fig:envss}.
This decrease in ground-state energy may be attributed to increase in quantum fluctuation with increase in interaction range $\sigma$. 
For analogy, consider, for example, a system of two spin-$\frac{1}{2}$ particles described by the Hamiltonian $J S_{1}\cdot S_{2}$ with $J>0$. For quantum spins, one has for the components of spin satisfying
$[S_{\alpha},S_{\beta}]=i\hbar\epsilon_{\alpha\beta\gamma}S_{\gamma}$ whereas for classical spins the commutation relation is ignored.
When the spins are taken as classical, the ground-state of the two-spin system corresponds to anti-parallel configurations $\uparrow\downarrow$ or $\downarrow\uparrow$ with eigenvalue $-J\frac{1}{4}$. 
However, when the spins are taken quantum mechanically, the ground-state corresponds to the superposed (arising from the  indistinguishablity of spins) configuration $\frac{\uparrow\downarrow-\downarrow\uparrow}{\sqrt{2}}$ with eigenvalue $-J\frac{3}{4}$.
Hence, with increase in the coupling strength $J$, the ground-state energy of the two-spin system $-J\frac{3}{4}$ decreases. Analogically, in our system, for a given interaction strength $U_{0}$, the interaction range $\sigma$ plays a role similar to the coupling strength $J$ in a two-spin system, leading to the reduction in the ground-state energy due to quantum fluctuation. 
\subsection{Critical angular velocity}
\indent When the condensate in the laboratory frame is subjected to an externally impressed rotation with angular velocity $\Omega$ (measured in units of $\omega_{\perp}$), the non-zero angular momentum states $\{L^{i}_{z} \neq 0,\left(i=1,2,3,\cdots\right)\}$ successively become the ground state in the corotating frame at the corresponding critical angular velocities $\{\Omega_{c_{i}},\left(i=1,2,3,\cdots\right)\}$ given by \cite{Ahsan_Kumar,Imran_20,Imran_15} 
\begin{table*}[t]
\caption{\label{tab:Oc_table1_sp3} The many-body ground state energy $E^{lab}_{0}\left(L_{z},\{U_{0},\sigma\}\right)$ in the laboratory frame for $N=16$ bosons with interaction strength $\mbox{U}_{0}=0.2171$ and interaction range $\sigma=0.30,0.50,0.75$ in the total angular momentum regime $0\leq L_{z}\leq 4N$. For stable angular momentum states, the one-particle quantum number $\left(n,m^{1}\right)$ corresponding to the largest eigenvalue $\lambda_{1}$ of OPRDM and the critical angular velocity $\{\Omega_{{c}_{i}}\left(\sigma\right), i=1,2,3,\cdots\}$ are given.}
\begin{tabular}{cccc ccccc}
	\hline \hline
\multirow{2}{*}{$L_{z}^{i}$}&\multirow{2}{*}{$(n,m^{1})$}&\multirow{2}{*}{$E^{lab}_{0}\left(\sigma=0.30\right)$}&\multirow{2}{*}{$E^{lab}_{0}\left(\sigma=0.50\right)$}&\multirow{2}{*}{$E^{lab}_{0}\left(\sigma=0.75\right)$}&\multirow{2}{*}{$\Omega_{c_{i}}\left(\sigma=0.30\right)$}&\multirow{2}{*}{$\Omega_{ c_{i}}\left(\sigma=0.50\right)$}&\multirow{2}{*}{$\Omega_{c_{i}}\left(\sigma=0.75\right)$}\\
& & & & & & &
\\ \hline
\multirow{2}{*}{0}&{(0,0)}&\multirow{2}{*}{46.86428}&\multirow{2}{*}{46.35902}&\multirow{2}{*}{45.36785}&\multirow{2}{*}{0.0}&\multirow{2}{*}{0.0}&\multirow{2}{*}{0.0}&\multirow{2}{*}{}\\
&(2,0)& & & & & &\\ \hline
\multirow{2}{*}{16}&{(1,1)} & \multirow{2}{*}{59.25225} & \multirow{2}{*}{58.90377}& \multirow{2}{*}{58.46294}& \multirow{2}{*}{0.77425}  & \multirow{2}{*}{0.78405}  &\multirow{2}{*}{0.81844} \\
&(3,1)& 	&  &  &  &  &\\ \hline
\multirow{2}{*}{28}&{(2,2)} & \multirow{2}{*}{70.13310}& \multirow{2}{*}{69.86716}& \multirow{2}{*}{69.55952} & \multirow{2}{*}{0.90674}  & \multirow{2}{*}{0.91362}  &\multirow{2}{*}{0.92471}  \\
& (4,4) & & & & & &  \\ \hline
\multirow{2}{*}{30}&{(2,2)} & \multirow{2}{*}{71.99933} & \multirow{2}{*}{71.73464} & \multirow{2}{*}{71.43247} & \multirow{2}{*}{} & \multirow{2}{*}{} &\multirow{2}{*}{0.93648}  \\ 
&(4,2) & & &  & & &\\ \hline
\multirow{2}{*}{32}&{(2,2)} & \multirow{2}{*}{73.83171}& \multirow{2}{*}{73.59262}& \multirow{2}{*}{73.31758} & \multirow{2}{*}{0.92465}  & \multirow{2}{*}{0.93136}  &\multirow{2}{*}{0.94256}  \\
&(4,2) & & & & & &\\ \hline
\multirow{2}{*}{36}&{(3,3)} & \multirow{2}{*}{77.55398}& \multirow{2}{*}{77.34967}& \multirow{2}{*}{77.12801} & \multirow{2}{*}{0.93255}  & \multirow{2}{*}{0.93926}  &\multirow{2}{*}{0.95262}  \\
&(5,3) & & & & & &\\ \hline
\multirow{2}{*}{39}&{(3,3)}  & \multirow{2}{*}{80.39032} & \multirow{2}{*}{80.19670} & \multirow{2}{*}{79.98856} & \multirow{2}{*}{} & \multirow{2}{*}{} &\multirow{2}{*}{0.95352} \\ 
&(5,3) & & & & & &\\ \hline
\multirow{2}{*}{48}&{(4,4)} & \multirow{2}{*}{88.89032}& \multirow{2}{*}{88.73774}& \multirow{2}{*}{88.60174} & \multirow{2}{*}{0.94469}  & \multirow{2}{*}{0.94901}  &\multirow{2}{*}{0.95702}  \\
&(6,4) & & & & & &\\ \hline
\multirow{2}{*}{52}&{(4,4)} & \multirow{2}{*}{92.79163}& \multirow{2}{*}{92.64065}& \multirow{2}{*}{93.51604}  & \multirow{2}{*}{0.97533}  & \multirow{2}{*}{0.97573} &\multirow{2}{*}{0.97464}  \\
&(6,4) & & & & & &\\ \hline
60&(5,5)&100.63525&102.50076&100.35792&0.98045&0.98079&0.98220\\ \hline\hline
\end{tabular}
\end{table*}
\begin{eqnarray}
&&\Omega_{c_{i}}\left(L_{z}^{i},{U_{0}},\sigma\right)=\nonumber\\
&&\frac{E^{lab}_{0} \left(L_{z}^{i},\{{U}_{0},\sigma\}\right)-E^{lab}_{0}\left(L^{\left(i-1\right)}_{z},\{{U}_{0},\sigma\}\right)}{L_{z}^{i}-L^{\left(i-1\right)}_{z}},~~~~
\label{eq:omega_c}
\end{eqnarray}
where $E^{lab}_{0}\left(L_{z}^{i},\{{U_{0}},\sigma\}\right)$ 
is the lowest energy state with angular momentum $L_{z}^{i}$ in the laboratory frame given by Eq.~(\ref{eq:eig_eq_E0}).
\begin{figure}[!htb]
\includegraphics[width=1.10\linewidth]{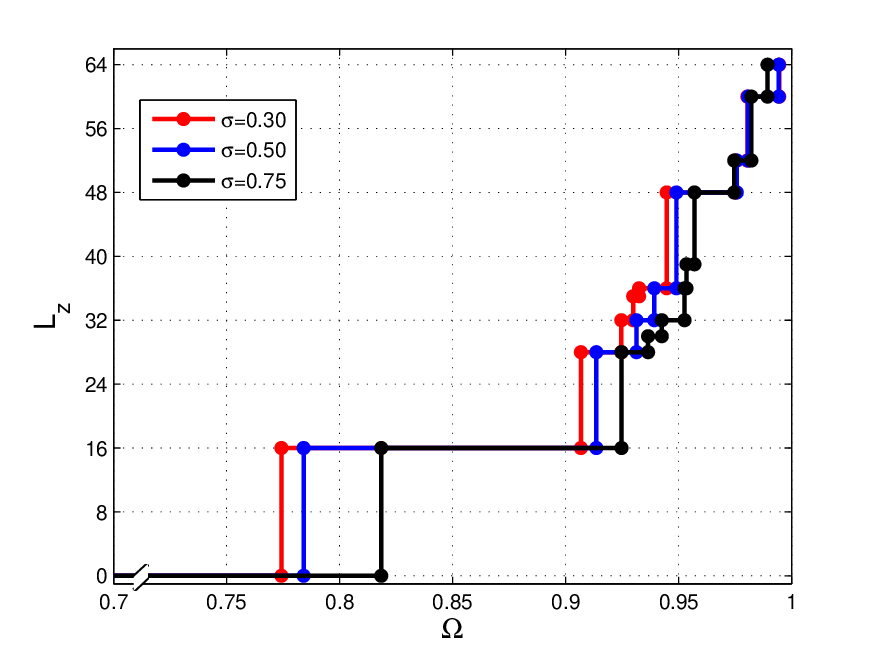}
\caption{(Color online) The $L_{z}-\Omega$ stability graph for $N=16$ with interaction strength $U_{0}=0.2171$ and three values of interaction range $\sigma =0.30,0.50,0.75$. For each value of $\sigma$, a set of critical angular velocities $\{\Omega_{ c_{i}}\left(\sigma\right), i=1,2,3,\cdots\}$ is observed where the total angular momentum $L_{z}$ takes a quantum jump.\label{fig:stbn16}}
\end{figure}
\\
\indent To analyse the response of the condensate to rotation, we present in Fig.~\ref{fig:stbn16}, the $L_{z}-\Omega$ stability graph for $U_{0}=0.2171$ and various values of interaction range $\sigma$. We observe that the angular momentum $L_{z}=0$ state remains the ground state of the system till the angular velocity attains the first critical value $\Omega_{c_{1}}$. As the system is subjected to further rotation, higher angular momentum states with $L_{z}>0$, which minimize the Helmholtz free energy at zero temperature
$F=\langle\Psi_{0}|({\bf H}^{lab}-\hbar\Omega L_{z})|\Psi_{0}\rangle$
in the co-rotating frame, become the successive ground states of the system at a series of critical angular velocities $\{\Omega_{c_{i}}, i=1,2,\cdots\}$ corresponding to certain angular momentum state $L^{i}_{z}\left(\Omega\right)$ and continue to remain the ground-state for angular velocities $\Omega_{c_i}\left(L^{i}_{z},\{U_{0},\sigma\}\right) \le \Omega <\Omega_{c_{i+1}}\left(L^{i+1}_{z},\{U_{0},\sigma \}\right)$.
Each of the ground-state corresponding to the angular momentum state $L^{i}_{z}\left(\Omega\right)$ for $\Omega_{c_i} \le \Omega <\Omega_{c_{i+1}}$ is referred to as the
stable, phase-coherent, $i^{th}$ vortical state~\cite{Fetter01,Leonardo20}.
The horizontal plateaus on the $L_{z}-\Omega$ stability graph in Fig.~\ref{fig:stbn16} represent the rotational angular-velocity-span 
\begin{eqnarray*}
\Omega_{c_{i+1}}\left(L^{i+1}_{z},\{U_{0},\sigma \}\right)-\Omega_{c_{i}}\left(\ L^{i}_{z},\{U_{0},\sigma \} \right)   
\end{eqnarray*}
of the stable angular momentum state $L^{i}_{z}$ 
and the vertical step shows the quantum jump $L^{i+1}_{z}-L^{i}_{z}$ in angular momentum at the corresponding critical angular velocity $\Omega_{c_{i}}\left(\L^{i}_{z},\{U_{0},\sigma \} \right)$. 
\\
\indent We observe from Table~\ref{tab:Oc_table1_sp3} that the value of the critical angular velocity of the $i^{th}$ vortical state increases for increase in interaction range $\sigma$ in slowly to moderately rotating regime. Thus an increase in interaction range $\sigma$ results in the onset of the next stable vortical state $L^{i+1}_{z}$ at a higher rotational angular velocity i.e. $\Omega_{c_{\left(i+1\right)}}\left( \{\sigma^{\prime}\}\right) > \Omega_{c_{\left(i+1\right)}}\left(\{\sigma\}\right)$ for $\sigma^{\prime} >\sigma$. However,
for a given vortical state $L^{i}_{z}$, the rotational angular-velocity-span of its plateau on the $L_{z}-\Omega$ stability graph decreases with increase in $\sigma$ i.e. $\left( \Omega_{c_{i+1}}\left(\{\sigma^{\prime}\}\right)-\Omega_{c_{i}}\left(\{\sigma^{\prime}\}\right)\right)<\left( \Omega_{c_{i+1}}\left(\{\sigma\}\right)-\Omega_{c_{i}}\left(\{\sigma\}\right) \right)$ for $\sigma^{\prime}>\sigma$. For example, for the stable angular momentum state $L_{z}=N=16$, the angular-velocity-spans on the $L_{z}-\Omega$ stability graph for various values of $\sigma$ are:  $\Omega_{c_{2}}(0.30)-\Omega_{c_{1}}(0.30)=0.1325$, $\Omega_{c_{2}}(0.50)-\Omega_{c_{1}}(0.50)=0.12957$ and $\Omega_{c_{2}}(0.75)-\Omega_{c_{1}}(0.75)=0.10627$.
With increase in rotational angular velocity $\Omega$, one observes a series of quantum jumps in angular momentum $\{L^{i}_{z}, i=1,2,3,\cdots\}$ on the $L_{z}-\Omega$ stability graph at the corresponding critical angular velocity $\{\Omega_{c_{i}}, i=1,2,3,\cdots\}$. The angular momentum states corresponding to plateaus on the $L_{z}-\Omega$ stability graph for different interaction range $\sigma$ are given in bold in Table~\ref{Tab:Local_min_s1}. For example, for the single-vortex state $L_{z}=N=16$, the jump to the next stable angular momentum state $L_{z}=28$ for various values of $\sigma$ are: $\Omega_{{c}_{2}}({\sigma=0.30})=0.907,\ \Omega_{{c}_{2}}({\sigma=0.50})=0.914$ and $\Omega_{{c}_{2}}({\sigma=0.75})=0.925$. Thus an increase in interaction range $\sigma$ leads to an increase in critical angular velocity $\Omega_{c_{i}}$ {\it i.e.} $\Omega_{c_{i}}\left(N,L^{i}_{z},\{U_{0},\sigma^{\prime}\}\right) > \Omega_{c_{i}}\left(N,L^{i}_{z},\{U_{0},\sigma\}\right)$ for $\sigma^{\prime}>\sigma$, as opposed to decrease in critical angular velocity $\Omega_{c_{i}}$ with increase in interaction strength $U_{0}$ and the number of particles $N$,
{\it i.e.} $\Omega_{c_{i}}\left(N,L^{i}_{z},\{U_{0}^{\prime},\sigma\}\right) < \Omega_{c_{i}}\left(N,L^{i}_{z},\{U_{0},\sigma\}\right)$ for $U_{0}^{\prime}>U_{0}$ and $\Omega_{c_{i}}\left(N^{\prime},L^{i}_{z},\{U_{0},\sigma\}\right) < \Omega_{c_{i}}\left(N,L^{i}_{z},\{U_{0},\sigma\}\right)$ for $N^{\prime}>N$~\cite{Ahsan_Kumar,Lewenstein09}. 
\\
\indent We also observe that with increase in interaction range $\sigma$, some new plateaus appear on the $L_{z}-\Omega$ stability graph in Fig.~\ref{fig:stbn16}. For example for $\sigma=0.75$, 
an additional stable plateau appears at $L_{z}=30$, which is not observed for $\sigma=0.30,0.50$. Similarly for $\sigma=0.75$, an additional stable plateau appears at $L_{z}=39$. Thus the underlying symmetry of the quantum many-body ground-state is further broken with increase in $\sigma$ beyond a critical value $\sigma_{c}$.
\begin{table*}[t]
\caption{For $N=16$ bosons interacting via Gaussian potential in Eq.~(\ref{eq:int_pot}) with interaction strength $\mbox{U}_{0}=0.2171$ and interaction range $\sigma=0.30$, stable ground-state angular momentum states and its span over $L_{z}-\Omega$ stability graph in Fig.~\ref{fig:stbn16} as observed in the co-rotating frame.}
\label{Tab:stb_span}
\begin{tabular}{|c|c|c|c|c|c|c|}
	\hline \hline
	&$L_{z}\left(\Omega=0\right)$  &$L_{z}\left(\Omega_{c_{1}}\right)$ &$L_{z}\left(\Omega_{c_{2}}\right)$ &$L_{z}\left(\Omega_{c_{3}}\right)$&$L_{z}\left(\Omega_{c_{4}}\right)$& $L_{z}\left(\Omega_{c_{5}}\right)$ \\ \hline
Stable vortical states  &0& 16 & 28 &36 &48&60  \\ \hline
	Angular momentum span& 16     &12    &8    &12   &12& \\
 \hline
Critical Angular velocity span& $\Omega_{c_{1}}-\Omega_{c_{0}}$     &$\Omega_{c_{2}}-\Omega_{c_{1}}$    &$\Omega_{c_{3}}-\Omega_{c_{2}}$    &$\Omega_{c_{4}}-\Omega_{c_{3}}$  &$\Omega_{c_{5}}-\Omega_{c_{4}}$& \\
 \hline
	&$0.77425$   &$0.13249$    &$0.02581$    &$0.01214$   & $0.00521$& \\
	\hline 
\end{tabular}
\end{table*}
\\
\indent We observe from the Table.~\ref{Tab:stb_span} that for $N=16$ with $U_{0}=0.2171$ and $\sigma=0.30$, the non-rotating angular momentum $L_{z}=0$ state remains the ground state of the system till $L_{z}\left(\Omega_{c_{1}}\right)=N$. Hence $L_{z}=0$ state remains the (stable) ground-state over a span of $L_{z}\left(\Omega_{c_{1}}\right)-L_{z}\left(\Omega=0\right)=16=N$ units of angular momentum with rotational-angular velocity span of $\Omega_{c_{1}}\left(L^{1}_{z},\{U_{0},\sigma\}\right)-\Omega_{c_{0}}\left(\L^{0}_{z},\{U_{0},\sigma \}\right)=0.77425$ on $L_{z}-\Omega$ graph.
The single-vortex state nucleates at $L_{z}=N=16$ and remains the ground state till $L_{z}=28$ with a span of $L_{z}\left(\Omega_{c_{2}}\right)-L_{z}\left(\Omega_{c_{1}}\right)=12$ units of angular momentum graph and rotational-angular velocity span of $\Omega_{c_{2}}\left(L^{2}_{z},\{U_{0},\sigma \}\right)-\Omega_{c_{1}}\left(L^{1}_{z},\{U_{0},\sigma\}\right)=0.13249$ on $L_{z}-\Omega$ graph. The two-vortex state remains the ground state of the system till $L_{z}=36$ with a span of $L_{z}\left(\Omega_{c_{3}}\right)-L_{z}\left(\Omega_{c_{2}}\right)=8$ units of angular momentum and rotational-angular velocity span of $\Omega_{c_{3}}\left(L^{3}_{z},\{U_{0},\sigma \}\right)-\Omega_{c_{2}}\left(\L^{2}_{z},\{U_{0},\sigma\}\right)=0.02581$ on $L_{z}-\Omega$ graph. 
The three-vortex state nucleates at $L_{z}=36$ and remains the ground state of the system till $L_{z}=48$ with a span of $L_{z}\left(\Omega_{c_{4}}\right)-L_{z}\left(\Omega_{c_{3}}\right)=12$ units of angular momentum and rotational-angular velocity span of $\Omega_{c_{4}}\left(L^{4}_{z},\{U_{0},\sigma \}\right)-\Omega_{c_{3}}\left(\L^{3}_{z},\{U_{0},\sigma\}\right)=0.01214$ on $L_{z}-\Omega$ graph.
The four-vortex state nucleates at $L_{z}=48$ and remains the ground state of the system till $L_{z}=60$ with a span of $L_{z}\left(\Omega_{c_{5}}\right)-L_{z}\left(\Omega_{c_{4}}\right)=12$ units of angular momentum and rotational-angular velocity span of $\Omega_{c_{5}}\left(L^{5}_{z},\{U_{0},\sigma\}\right)-\Omega_{c_{4}}\left(\L^{4}_{z},\{U_{0},\sigma\}\right)=0.00521$ on $L_{z}-\Omega$ graph. %
It is apparent from the Table.~\ref{Tab:stb_span} that the three-vortex state with triangular geometry nucleates readily\cite{Xia01_ro} from the two-vortex state 
with just $8$ units of angular momentum and remains the ground state over a span of $12$ units of angular momentum before the four-vortex state nucleates. We observe that the single-vortex state is more stable than the multi-vortex states~\cite{Madison_00}.
\subsection{The one-particle reduced density matrix in beyond lowest Landau level approximation}
\indent The one-particle reduced density matrix (OPRDM) provides a criterion for Bose-Einstein condensation~\cite{Penrose51,Landau58,penrose_Onsager,CJ_Pertik00,Yang62} in a system of bosons at zero temperature. It provides information about one-body observables for a many-body system.
\\
\indent Having obtained the variationally exact $N$-particle ground state wave-function $\Psi_{0}(\mathbf{r}_{1}, \mathbf{r}_{2},\cdots,\mathbf{r}_{N})$, the OPRDM $\rho^{\left(1\right)}(\mathbf{r},\mathbf{r}^{'})$ is found by tracing out the degrees of freedom corresponding to $\left(N-1\right)$ particles~\cite{Ahsan_Kumar}:
\begin{eqnarray*}
&&\rho^{(1)}\left(\textbf{r},\textbf{r}^{\prime}\right) \equiv  \int\int \cdots \int{d\textbf{r}_{2}{d\textbf{r}{_3}}\cdots {d\textbf{r}_{N}}} \nonumber \\ 
&\times&{\Psi_{0}}(\textbf{r},\textbf{r}_{2},\textbf{r}_{3}\cdots,\textbf{r}_{N}){\Psi^{*}_{0}}(\textbf{r}^{\prime},\textbf{r}_{2},\textbf{r}_{3}\cdots,\textbf{r}_{N}) \nonumber\\ 
&=&\sum_{n_{r}}\sum_{m}\sum_{n^{\prime}_{r}}\sum_{m^{\prime}}
\rho_{n_{r}m;n_{r}^{\prime}m^{\prime}} \nonumber\\ 
&\times&u_{n_{r},m}(\textbf{r})u^{*}_{n^{\prime}_{r},m^{\prime}}(\textbf{r}^{\prime})
\end{eqnarray*}
\\
where $u_{n_{r},m}(\textbf{r})$ is the one-particle harmonic oscillator basis-state in 2-dimension given by
\begin{eqnarray}
u_{n_{r},m}(r_{\bot}/a_{\perp},\phi)
&=&\sqrt{\frac{n_{r}!}{\pi a^{2}_{\perp}(n_{r}+|m|)!}}\ e^{im\phi}(r_{\bot}/a_{\bot})^{|m|} \nonumber\\
&\times& e^{-\frac{1}{2}(r_{\bot}/a_{\perp})^{2}}\  
L^{|m|}_{n_{r}}(r_{\bot}^{2}/a_{\perp}^{2}).\\
n_{r}=0,1,2,\cdots \ &&\mbox{ and } \ m=0,\ \pm 1,\ \pm 2, \cdots \nonumber
\label{eq:3wave_dfn}	
\end{eqnarray}
Here $L^{|m|}_{n_{r}}(r_{\bot}^{2}/a_{\bot}^{2})$ 
is the Associated Laguerre polynomial, $n_{r}$ the radial quantum number and $m$ the angular momentum quantum number.
The OPRDM $\rho^{(1)}\left(\textbf{r},\textbf{r}^{\prime}\right)$, is diagonalized to obtain 
\begin{eqnarray}
\label{eq:rdm}
\rho^{(1)}\left(\textbf{r},\textbf{r}^{\prime}\right)&=&\sum_{\mu=1}^{\cal M}\lambda_{\mu}\chi_{\mu;m^{\mu}}\left(\textbf{r}\right)\chi_{\mu;m^{\mu}}^{*}(\textbf{r}^{\prime})\\ 
\mbox{with  } \  \hat{l}_{z}\chi_{\mu,m^{\mu}}&=&m^{\mu}\chi_{\mu,m^{\mu}} \\  
\mbox{ and}\ 
\chi_{\mu,m^{\mu}}\left(\textbf{r}\right)&=&
\sum_{n_{r}^{\mu}=0,1,2,\cdots}\ 
c_{n^{\mu}_{r},m^{\mu}}
u_{n^{\mu}_{r},m^{\mu}}\left(\textbf{r}\right)
~~~~~\label{eq:chi_mu}
\end{eqnarray}
where ${\cal M}$ is the dimensionality of the one-particle Hilbert space, $\{\lambda_{\mu}\}$ are the eigenvalues corresponding to the simultaneous eigenvectors $\{\chi_{\mu;m^{\mu}}\}$ of the OPRDM and the one-particle angular momentum $\hat{l}_{z}$ with 
the eigenvalues $\{\lambda_{\mu}\}$ giving the fractions of the corresponding one-particle symmetry-broken eigenstates $\{\chi_{\mu;m^{\mu}}\}$ labelled by $\mu$ and the one-particle angular momentum quantum number $m^{\mu}$. 
The eigenvectors of OPRDM are normalized and the eigenvalues satisfy the following sum rules 
\begin{eqnarray}
\sum_{{\mu}={1}}^{{\cal M}} \lambda_{{\mu}}=1
\label{L_sum}
\mbox{  and  } \ \ \ 
\sum_{\mu=1}^{\cal M}(N\lambda_{\mu}) m^{\mu}=L_{z}.
\label{eq:eval_norm}
\end{eqnarray}
\indent For a system comprising of finite number of particles, the criterion for Bose-Einstein condensation is that one or a few of the eigenvalues $\{ \lambda_{{\mu}} \}$ of OPRDM become significantly large compared to others~\cite{penrose_Onsager,Landau58,Yang62}. 
\\
\indent As the number of particles in the condensate becomes large with $N \to \infty$, the fluctuation in the number of particles increases indefinitely with $\Delta N \to \infty$. It, therefor, follows from the commutation relation $\left[\widehat{N},\widehat{\Theta}\right]=i\hbar$
between the number operator $\widehat{N}$ and the quantum mechanical phase operator $\widehat{\Theta}$ for the condensate  that the fluctuation in the phase of the condensate vanishes $i.e.$ $\Delta \Theta \to 0$, and the system acquires a definite quantum mechanical phase leading to a situation where $\lambda_{1}\sim 1$ and $\lambda_{2} \approx \lambda_{3} \approx \cdots \ \approx 0$. However, for a system comprising of finite number of particles, $\Delta N \to$ finite and the quantum mechanical phase of the condensate fluctuates over the phases of the symmetry-broken eigenstates $\{\chi_{{\mu;m^{\mu}}}\}$ of OPRDM~\cite{Ueda06,Yang62} labelled by ${\mu}$ and the corresponding angular momentum quantum number \{$m^{\mu}$\}. Thus the OPRDM defined in Eq.~(\ref{eq:rdm}) may be viewed as the quantum mechanical average weighted by $\{\lambda_{\mu}\}$ over the corresponding symmetry-broken~\cite{Ueda06} eigenstate states $\{\chi_{{\mu;m^{\mu}}}\}$ with definite one-particle angular momentum quantum number $m^{\mu}$, where $\mu=1,2,\cdots {\cal M}$ are different fragments of OPRDM. This may result in more than one eigenvalues of OPRDM taking significantly large values as compared to others leading to the fragmentation of the rotating condensate comprising of a finite number of particles. Further, the fragmentation of the condensate, as described by OPRDM, is an interplay of parameters like the interaction strength ${U_{0}}$, the interaction range $\sigma$ for a given number of particles $N$  and angular momentum $L_{z}$.
\begin{figure}[!htb]
	\centering
\includegraphics[width=1.10\linewidth]{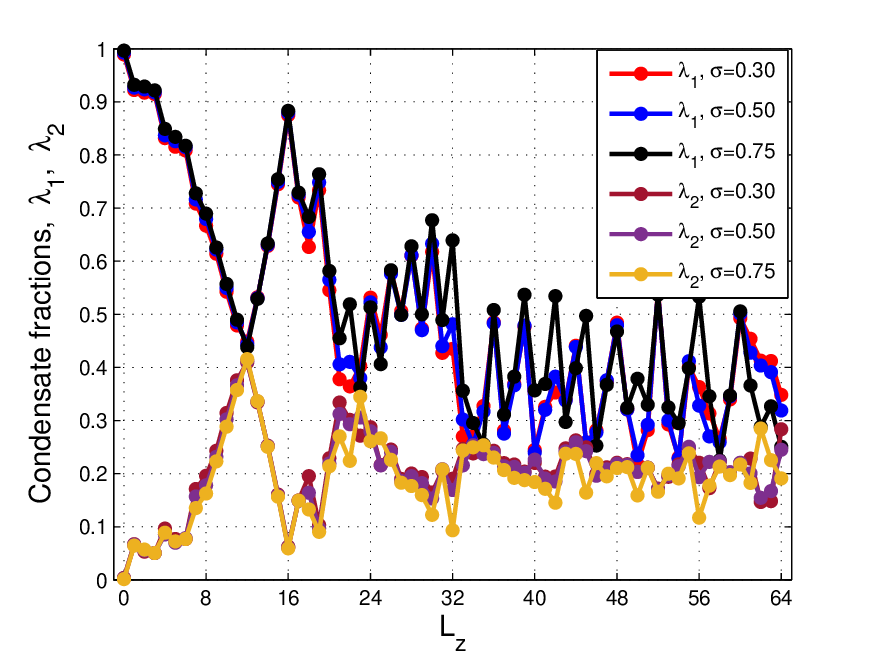}
\caption{\label{fig:condfra1}(Color online) The largest two  condensate fractions $\lambda_{1}>\lambda_{2}$ {\it vs} the total angular momentum in the regime $0\le L_{z}\le 4N$ for $N=16$ bosons with interaction strength ${U_{0}}=0.2171$ and three values of interaction range $\sigma= 0.30,0.50,0.75$.
For $\sigma=0.30$, the largest two eigenvalues of OPRDM become comparable {\it i.e.} $\lambda_{1}\approx \lambda_{2}$ at angular momenta $L_{z}=12,23,34,40,45,50,54,58$. For other values of $\sigma$, the corresponding angular momenta are listed in Table~\ref{tab:lamb_comp}.
}
\end{figure}
\\
\indent In Fig.~\ref{fig:condfra1}, we observe that the largest condensate fraction $\lambda_{1}$ of OPRDM, defined in Eq.~(\ref{eq:rdm}), is maximum for the non-rotating $L_{z}=0$ state and decreases with increase in angular momentum $L_{z}$. 
However, the second largest condensate fraction $\lambda_{2}$ which is minimum for the non-rotating state, increases with increase in angular momentum and becomes comparable to the first one {\it i.e.}
$\lambda_{2} \approx \lambda_{1}$ for the angular momentum $L_{z}=12$ state in addition to several other angular momentum states, in slowly to moderately rotating regime for the three values of interaction range $\sigma$ as given in Table~\ref{tab:lamb_comp}.
\begin{table}[!htb]
\caption{For $N=16$, values of total angular momentum in the regime $0\leq L_{z}\leq4N$ for which the largest two eigenvalues of OPRDM become comparable {\it i.e. } $\lambda_{2} \approx \lambda_{1}$ as seen in Fig.~\ref{fig:condfra1}, for interaction strength ${U_{0}}=0.2171$ and interaction range $\sigma=0.30,0.50,0.75$.}
\begin{tabular}{|c|ccccccccc|}
	\hline\hline
	Interaction range $\sigma$& \multicolumn{9}{c|}{Total angular momentum $L_{z}$}\\ \hline
	$\sigma=0.30$ &12 &23 &34 &40&45&50&54&58&  \\ \hline
	$\sigma=0.50$ &12 &23 &34 &40&45&50&54&57&58  \\ \hline
	$\sigma=0.75$ &12 &23 &34 &35&43&58&62&& \\ \hline
	\end{tabular}
	\label{tab:lamb_comp}
\end{table}
\subsection{von Neumann entropy \label{sec:von_S1}}
\indent For an $N$-body system, the quantum-entanglement between $n$ and the remaining $(N-n)$ particles is measured~\cite{hengfan09} by the von Neumann entropy
$S_{n}=-\mbox{Tr} \left(\hat{\rho}^{(n)} \ln \hat{\rho}^{(n)}\right)$
where $\hat{\rho}^{(n)}$ is the $n$-particle reduced density matrix. 
In the present study, we calculate the one-particle von Neumann entropy 
for the ground state of the condensate, 
defined in terms of OPRDM $\hat{\rho}^{(1)}$ as ~\cite{Horodecki09,hengfan09,JEisert10,Julia11},
\begin{eqnarray}
S_{1} &=-\mbox{Tr}& \left(\hat{\rho}^{(1)} \ln \hat{\rho}^{(1)}\right)
=-\sum_{{\mu}=1}^{{\cal M}}\lambda_{{\mu}} \ln \lambda_{{\mu}} ~~~~~~~
\label{entp_gen}
\end{eqnarray}
where ${\mu}$ labels the symmetry-broken one-particle eigenstates $\chi_{\mu;m^{\mu}}$ of OPRDM, defined in Eq.(\ref{eq:rdm}).
\begin{figure}[!htb]
\centering
\includegraphics[width=1.10\linewidth]{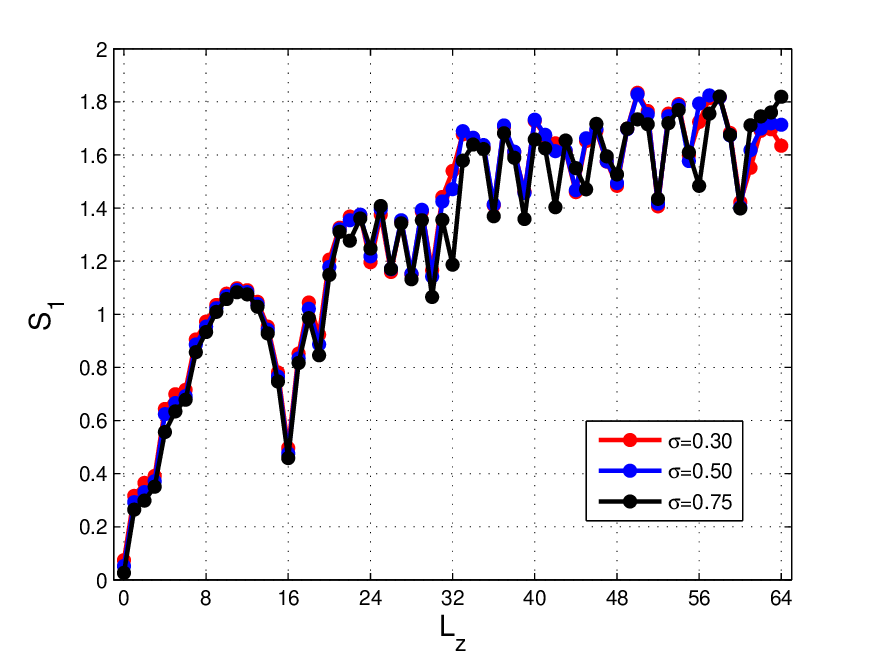}
\caption{(Color online) The von Neumann entropy $S_{1}\left(L_{z},\{\sigma\}\right)$ {\it vs} the total angular momentum in the regime $0\le L_{z}\le 4N$ for $N=16$ bosons with interaction strength ${U_{0}}=0.2171$ and three values of interaction range $\sigma= 0.30,0.50,0.75$. Decrease in the value of von Neumann entropy $S_{1}$ is observed with an increase in the interaction range $i.e.$ $S_{1}\left(\sigma^{\prime}\right)< S_{1}\left(\sigma\right)$ for $\sigma^{\prime}>\sigma$.
\label{fig:S_1}}
\end{figure}
\begin{figure}[!htb]
\centering 
\includegraphics[width=1.10\linewidth]{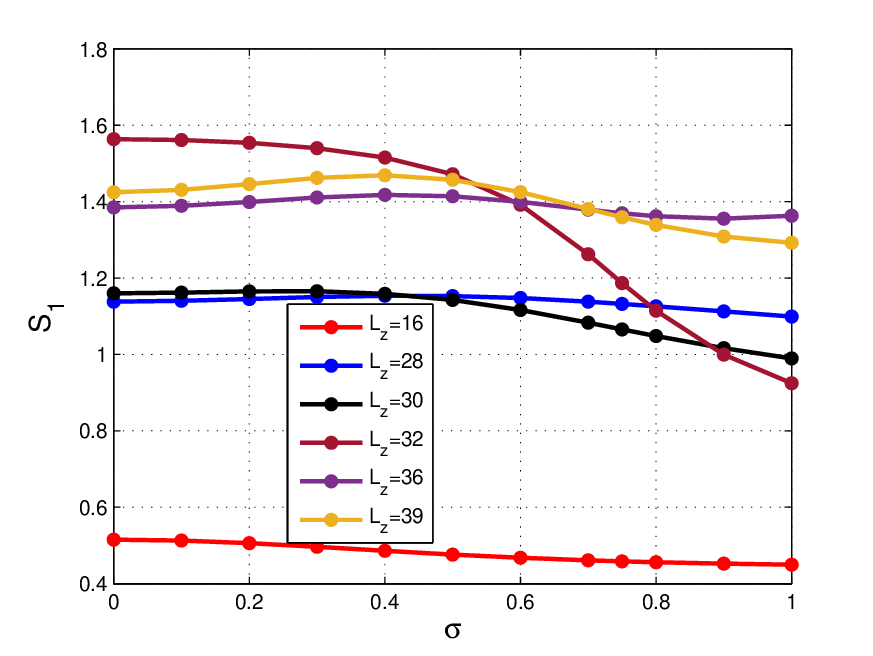}
\caption{(Color online) The von Neumann entropy $S_{1}\left(L_{z},\{\sigma\}\right)$ {\it vs} the interaction range in the regime $0\leq \sigma \leq 1$ with interaction strength ${U_{0}}=0.2171$, for $N=16$ bosons for some of the total angular momentum states, namely, $L_{z}=16,28,30,32,36,39$, listed in Table ~\ref{tab:stb_point}. 
We observe several crossings in the von Neumann entropy $S_{1}\left(L_{z},\{\sigma\}\right)$ of different angular momentum states as interaction range is varied over $0\leq \sigma \leq 1$.
\label{fig:S1vsS}}
\end{figure}
\begin{figure}[!htb] 
	\centering
\includegraphics[width=1.10\linewidth]{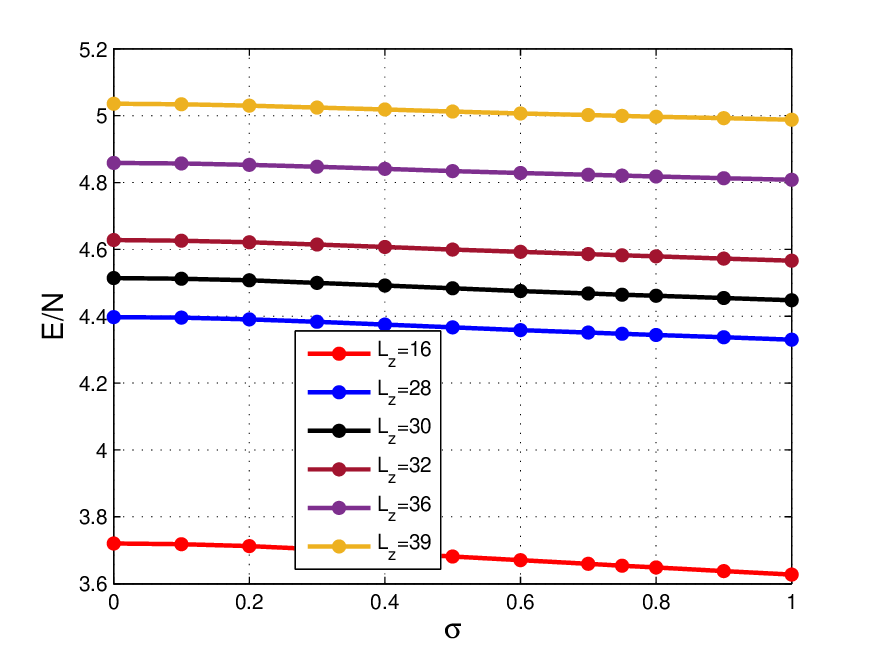}
\caption{(Color online) The energy per particle $E/N$ {\it vs} the interaction range $\sigma$ with interaction strength ${U_{0}}=0.2171$, for $N=16$ bosons for some of the total angular momentum states, namely, $L_{z}=16,28,30,32,36,39$. We observe a decrease in ground state energy per particle with increase in interaction range $\sigma$.\label{fig:envss}}
\end{figure}
\\
\indent 
In Fig.~\ref{fig:S_1}, we present von Neumann entropy for given interaction range $\sigma=0.30,0.50,0.75$ in the angular momentum regime $0\leq L_{z}\leq 4N$.
We observe that for $N=16$, the non-rotating ($L_{z}=0$) condensate has the lowest von Neumann entropy for the three values of
interaction range $\sigma=0.30,0.50,0.75$, implying that quantum mechanically it is the most ordered state of all the angular momentum states for a given given number of particle $N$. It is further observed that for a given angular momentum $L_{z}$ state, the von Neumann entropy is lower for higher values of $\sigma$ $i.e.$ $S_{1}(L_{z},\{\sigma^{\prime}\})<S_{1}(L_{z},\{\sigma\})$ for $\sigma^{\prime}>\sigma$~\cite{Imran_20}. Hence, an increase in interaction range leads to a more quantum-mechanically-ordered state. \\
\indent In Table~\ref{Tab:Local_min_s1}, we list the angular momentum states for which von Neumann entropy has a local minima, for three values of interaction
range $\sigma = 0.30, 0.50, 0.75$. The angular momentum states in bold-face are the stable states and others are the meta-stable states, as observed from the $L_{z}-\Omega$ stability graph in Fig.~\ref{fig:stbn16}. The angular momentum states $L_{z}=30,39$ which are meta-stable states for interaction range $\sigma=0.30,0.50$, become stable states with increase in interaction range to $\sigma=0.75$.
\begin{table*}[!htb]
\caption{For $N=16$ bosons, values of total angular momentum $L_{z}$ corresponding to minima in the von Neumann entropy $S_{1}\left(L_{z},\{\sigma\}\right)$, in Fig.~\ref{fig:S_1}, for the interaction strength ${U_{0}}=0.2171$ and different values of the interaction range $\sigma$ in Eq.~(\ref{eq:int_pot}) in the total angular momentum regime $0\leq L_{z}\leq4N$. The angular momentum states in bold-face are the stable states, corresponding to plateaus in $L_{z}-\Omega$ graph in Fig.~\ref{fig:stbn16}, and other states are the meta-stable states.}
\label{Tab:Local_min_s1}
\begin{tabular}{|c|ccccccccccccccccc|}
\hline\hline
Interaction range $\sigma$& \multicolumn{17}{c|}{Total angular momentum $L_{z}$}\\ \hline
$\sigma=0.30$ &{\bf 16} &19 &24 &26&{\bf 28}&30&{\bf 32}&{\bf 36}&39& &44&&{\bf 48}& &{\bf 52}&55&{\bf 60}  \\ \hline
$\sigma=0.50$ &{\bf 16} &19 &24 &26&{\bf 28}&30&{\bf 32}&{\bf 36}&39 &42& 44&&{\bf 48} &&{\bf 52}&55&{\bf 60} \\ \hline
$\sigma=0.75$ &{\bf 16} &19 &24 &26&{\bf 28}&{\bf 30}&{\bf 32}&{\bf 36}&{\bf 39}&42&&45&{\bf 48} &50&{\bf 52}&55&{\bf 60} \\ \hline
\end{tabular}
\label{tab:stb_point}
\end{table*}
\\
\indent 
In Fig.~\ref{fig:S1vsS}, we present von Neumann entropy for several angular momentum $L_{z}$ states  listed in Table~\ref{Tab:Local_min_s1}, in the interaction regime $0\leq \sigma \leq 1$. It is observed that the single-vortex state $L_{z}=N=16$ has minimum value of $S_{1}\left(L_{z},\{\sigma\}\right)$ among all the angular momentum $L_{z}=16,28,30,32,36,39$ states, $i.e.$ it is quantum-mechanically the most ordered state, well isolated from other angular momentum states listed in Table~\ref{Tab:Local_min_s1}. We further observe that there are crossings in the von Neumann entropy $S_{1}\left(L_{z},\{\sigma\}\right)$, in the regime $0\leq  \sigma\leq 1$. Among the angular momentum states $L_{z}=28,30,32,36,39$, we observe that for the two-vortex angular momentum $L_{z}=2N=32$ state, the von Neumann entropy $S_{1}\left(L_{z},\{\sigma\}\right)$ is maximum at $\sigma=0$, decreases with increase in $\sigma$ and takes the least value at $\sigma=1$, crossing the angular momentum states $L_{z}=28,30,36,39$ at the values of interaction range $\sigma = 0.80,0.90,0.60,0.50$, respectively. It is to be noted that the two-vortex $L_{z}=2N=32$ state is unique in that it exhibits merging of cores of vortices for values of interaction range beyond $\sigma>0.75$, as seen in Fig.~\ref{fig:sp3L32A}, where we present the conditional probability distribution(CPD) plots for the condensate. 
We also observe that there is bunching of two-vortex states
$L_{z}=28$ and $L_{z}=30$ with crossing in von Neumann entropy $S_{1}\left(L_{z},\sigma\right)$ at the interaction range $\sigma=0.40$ {\ i.e.} $S_{1}\left(L_{z}=28,\sigma<0.4\right) <  S_{1}\left(L_{z}=30,\sigma<0.4\right)$ and $S_{1}\left(L_{z}=28,\sigma>0.4\right) >  S_{1}\left(L_{z}=30,\sigma>0.4\right)$. Similarly, there is bunching of three-vortex states $L_{z}=36$ and $L_{z}=39$ with crossing in von Neumann entropy $S_{1}\left(L_{z},\sigma\right)$ at the interaction range $\sigma=0.70$
{\ i.e.} $S_{1}\left(L_{z}=36,\sigma<0.7\right) <  S_{1}\left(L_{z}=39,\sigma<0.7\right)$ and $S_{1}\left(L_{z}=36,\sigma>0.7\right) >  S_{1}\left(L_{z}=39,\sigma>0.7\right)$.
We further observe that for the angular momentum $L_{z}=30,39$ states, the von Neumann entropy $S_{1}\left(L_{z},\{\sigma\}\right)$ remains more or less constant upto the interaction range $\sigma=0.40$ and then decreases with increase in $\sigma$. This decrease in $S_{1}\left(L_{z},\{\sigma\}\right)$ with increase in $\sigma$ leads to an increase in quantum entanglement between the particles in the condensate. It is to be noted that the angular momentum states $L_{z}=30$ and $L_{z}=39$ which are meta-stable states for $\sigma=0.30,0.50$, become stable states for 
$\sigma=0.75$ as shown in Table~\ref{Tab:Local_min_s1}. This observation can be attributed to enhancement in quantum entanglement with increase in interaction range $\sigma$.
\\
\indent In Fig.~\ref{fig:envss}, we present variation of the ground state energy per particle $E/N$ with interaction range $\sigma$ for some of the stable total angular momentum states, listed in Table~\ref{Tab:Local_min_s1}. 
The energy per particle decreases with increase in interaction range $\sigma$ whereas increases with increase in angular momentum state $L_{z}$.  
It is observed that $E\left(L_{z},\sigma'\right)>E\left(L_{z},\sigma\right)$ for $\sigma' > \sigma$ and 
$E\left(L'_{z},\sigma\right)>E\left(L_{z},\sigma\right)$ for $L_{z}' > L_{z}$ in the interaction range regime $0\leq \sigma \leq 1$. There are no crossings of energy levels as $\sigma$ is varied.
\subsection{Internal structure of the condensate: conditional probability distribution \label{sec:cpd}}
\begin{figure*}
		\subfigure[]{
      \hspace{-5mm}
\begin{tikzpicture}[spy using outlines={magnification=10.0, circle, size=1.50cm, black, connect spies}]	\node (n1) at (0,0)
   {\includegraphics[width=0.40\linewidth]{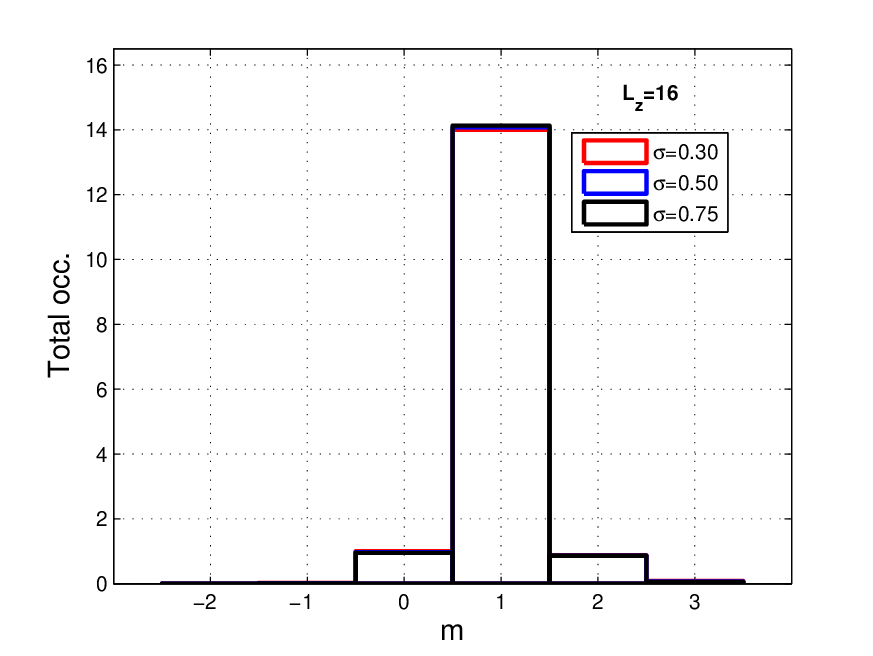}};
  \spy on (0.40,1.390) in node at (-0.9,0.0);
\end{tikzpicture}
		\hspace{-8.mm}
		\includegraphics[width=0.30\linewidth]       {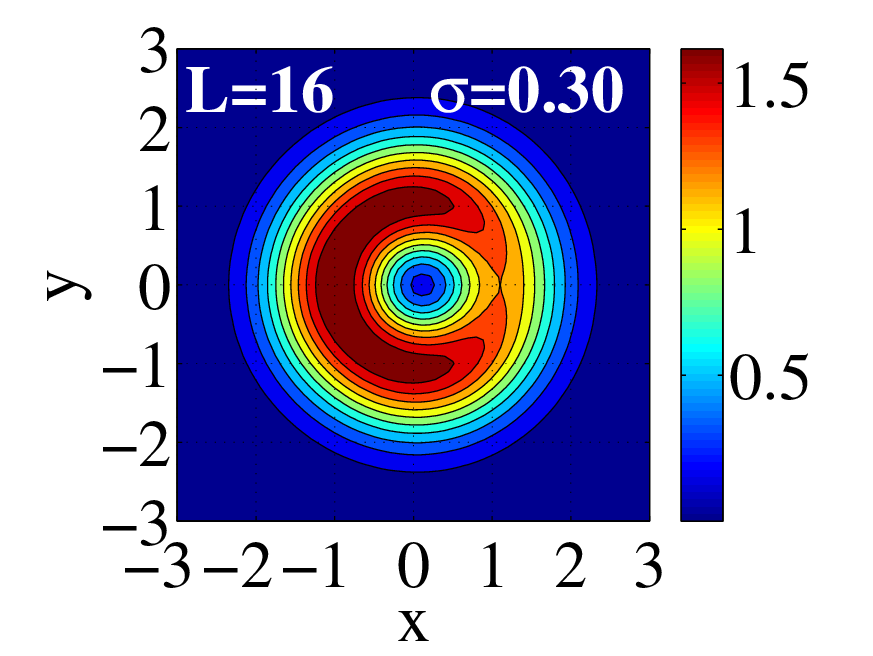}\label{fig:l16s3}
			\hspace{-5mm}
		\includegraphics[width=0.30\linewidth]   {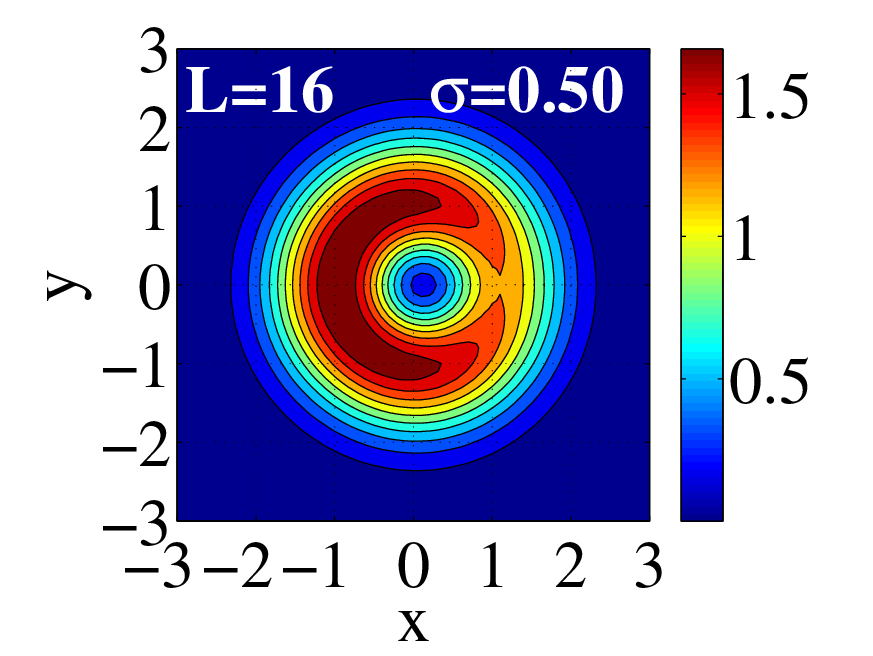}\label{fig:l16s5}
			\hspace{-5mm}
		\includegraphics[width=0.30\linewidth]{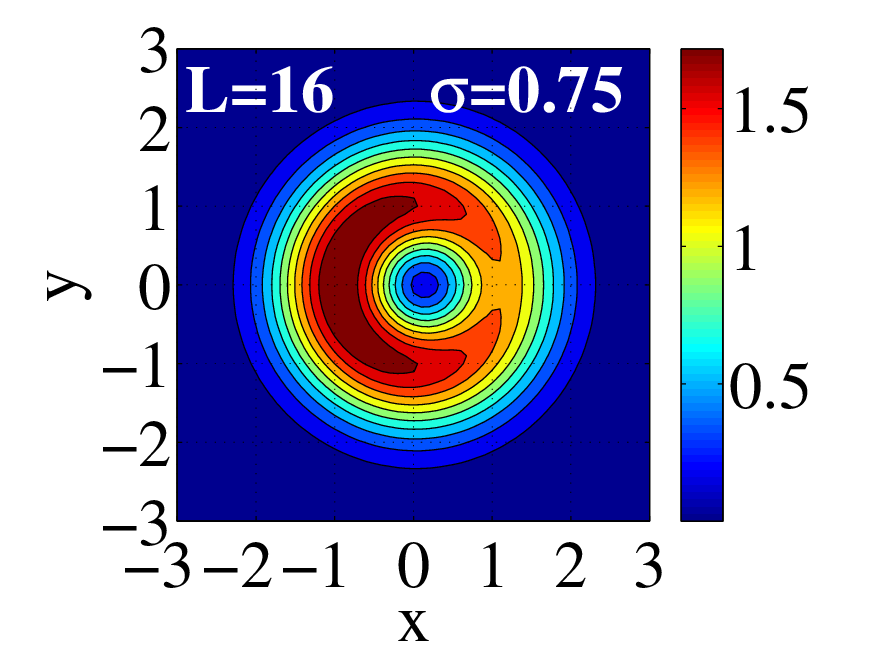}\label{fig:l16s75}}
\caption{(Color online) The macroscopic occupations of one-particle angular momentum states of OPRDM {\it vs} one-particle angular momentum $m$ and the corresponding CPD plots depicting the entry of the first vortex for $N=16$ with interaction strength $U_{0}=0.2171$ and interaction range $\sigma=0.30,0.50,0.75$. The reference point has been chosen at $\mathbf{r}_{0}=(3,0)$. In (a)-(d), with increase in angular momentum $L_{z}$, there is increase in the macroscopic occupation of one-particle angular momentum $m=1$ state and decrease in the macroscopic occupation of $m=0$ state of OPRDM. In the CPD plots, with increase in interaction range $\sigma$, the value of probability density at the peak increases. (e) for $L_{z}=N=16$ state, the macroscopic occupation of maximally occupied one-particle angular momentum $m=1$ state of OPRDM becomes maximum for states $0\leq L_{z}\leq 16$ in (a)-(e) and singly quantized vortex appears for all three values $\sigma$. In CPD plots, with increase in interaction range $\sigma$, the values of probability density increases at the peaks and decreases at the cores.\label{fig:b4_vortx}}
\end{figure*}

\begin{figure*}
	\subfigure[]{
  \hspace{-6.0mm}
	\begin{tikzpicture}[spy using outlines={magnification=10.0, circle, size=1.50cm, black, connect spies}]\node (n1) at (0,0)
   {\includegraphics[width=0.4\linewidth]{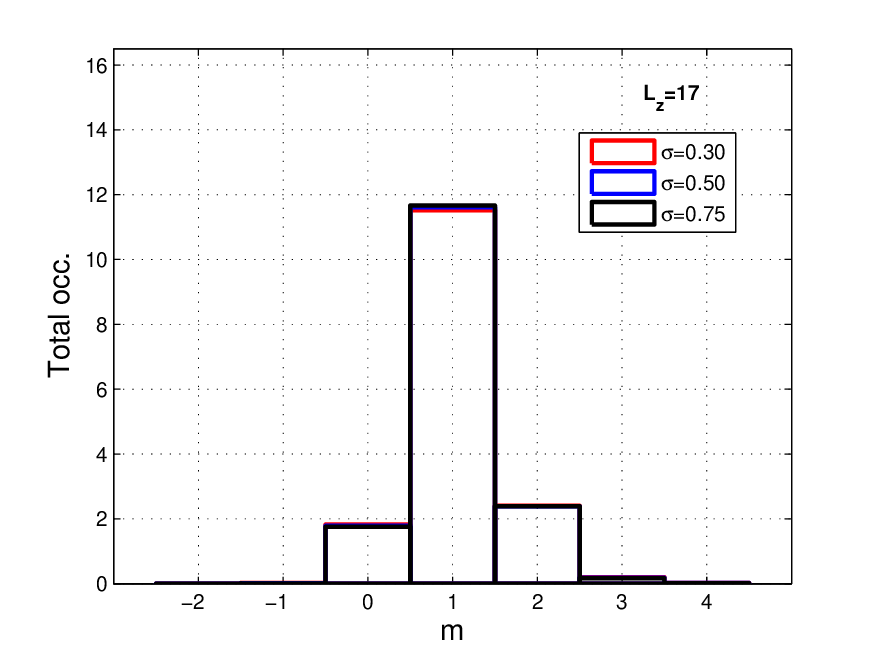}};
		\spy on (0.15,0.85) in node at (-1.350,0.0);
		\end{tikzpicture}
		\hspace{-8.0mm}
		\includegraphics[width=0.30\linewidth]{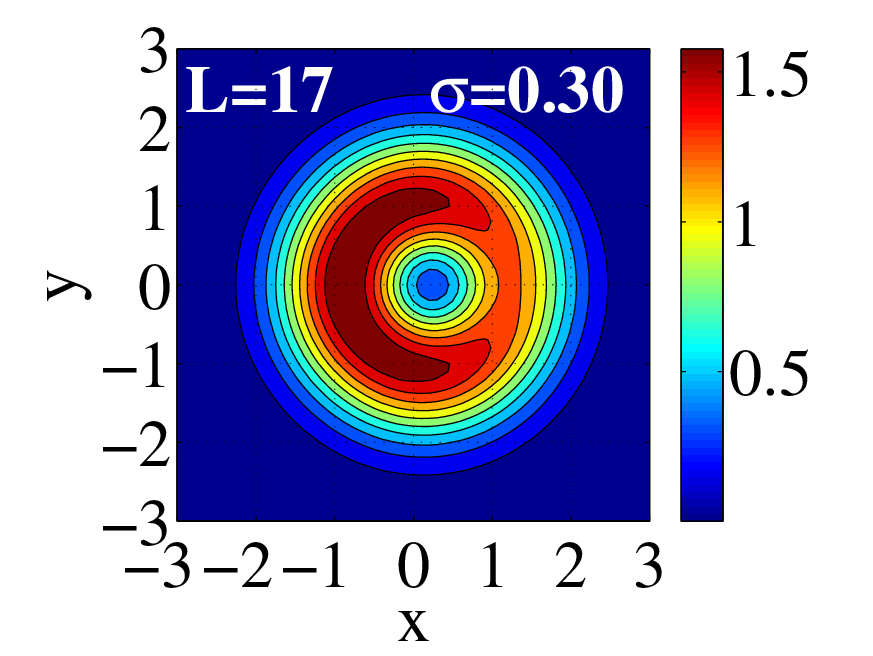}\label{fig:l17s3}
		\hspace{-5.0mm}
		\includegraphics[width=0.30\linewidth]{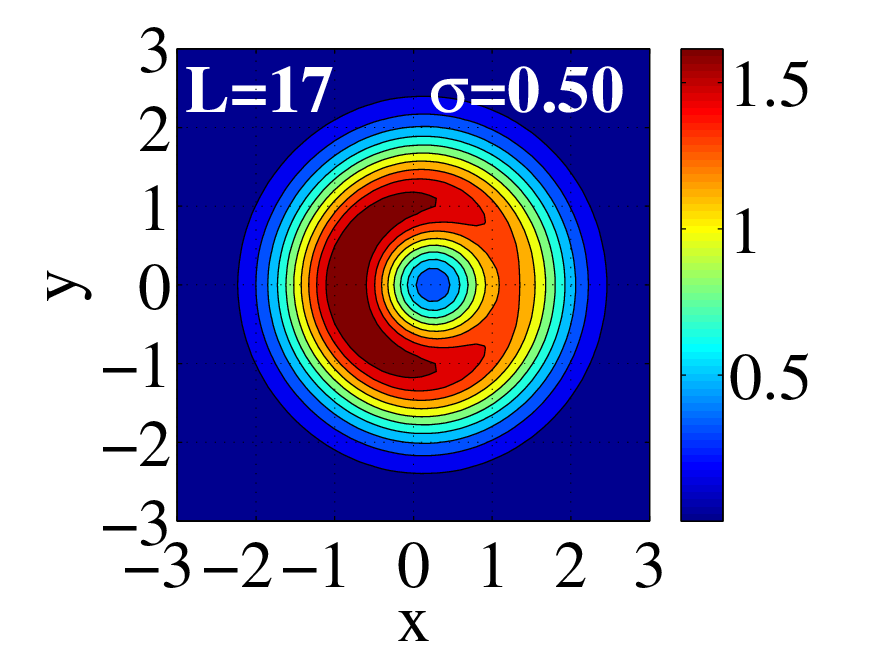}\label{fig:l17s5}
		\hspace{-5mm}
		\includegraphics[width=0.30\linewidth]{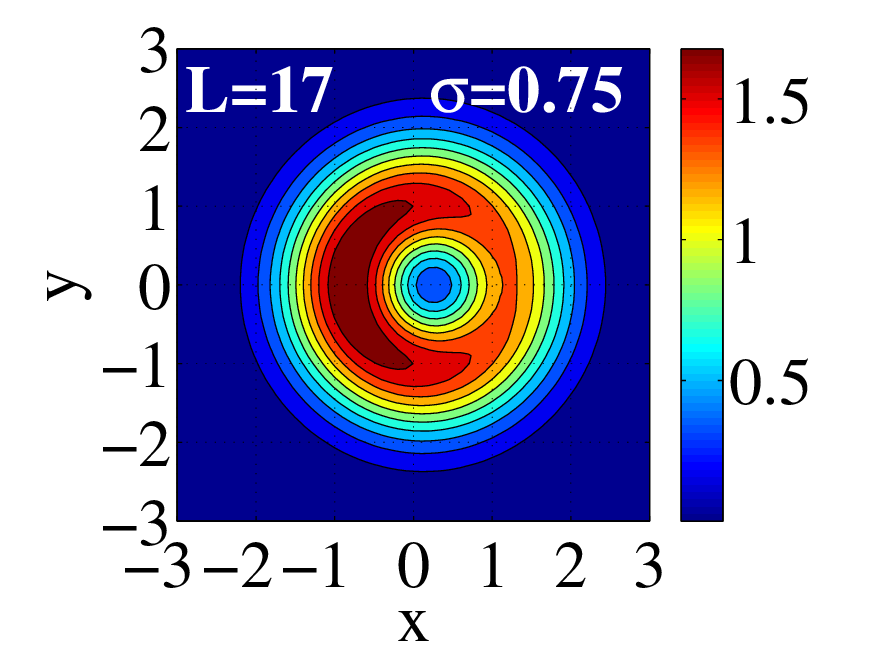}\label{fig:l17s75}}
     \vspace{-0.2in}
\caption{(Color online)	\label{fig:lz17} The macroscopic occupation of one-particle angular momentum states of OPRDM {\it vs} one-particle angular momentum $m$ and the corresponding CPD plots depicting the single-vortex for $N=16$ bosons with interaction strength $U_{0}=0.2171$ and three values of interaction range $\sigma=0.30,0.50,0.75$. The reference point has been chosen at $\mathbf{r}_{0}=(3,0)$. For the angular momentum $L_{z}=17$, the macroscopic occupation one-particle angular momentum state $m=1$ of OPRDM decreases with respect to $L_{z}=N$ state as shown in Fig.~\ref{fig:l16s3}. However, with increase in interaction range $\sigma$, the macroscopic occupation of maximally occupied one-particle angular momentum state $m=1$ of OPRDM increases whereas, in the CPD plots, the values of probability density at the peaks increases and at the cores decrease.}
\end{figure*}

\begin{figure*}
	\subfigure[]{
     \hspace{-5.0mm}
     	\begin{tikzpicture}[spy using outlines={magnification=10.0, circle, size=1.50cm, black, connect spies}]\node (n1) at (0,0)
  {\includegraphics[width=0.35\linewidth]         {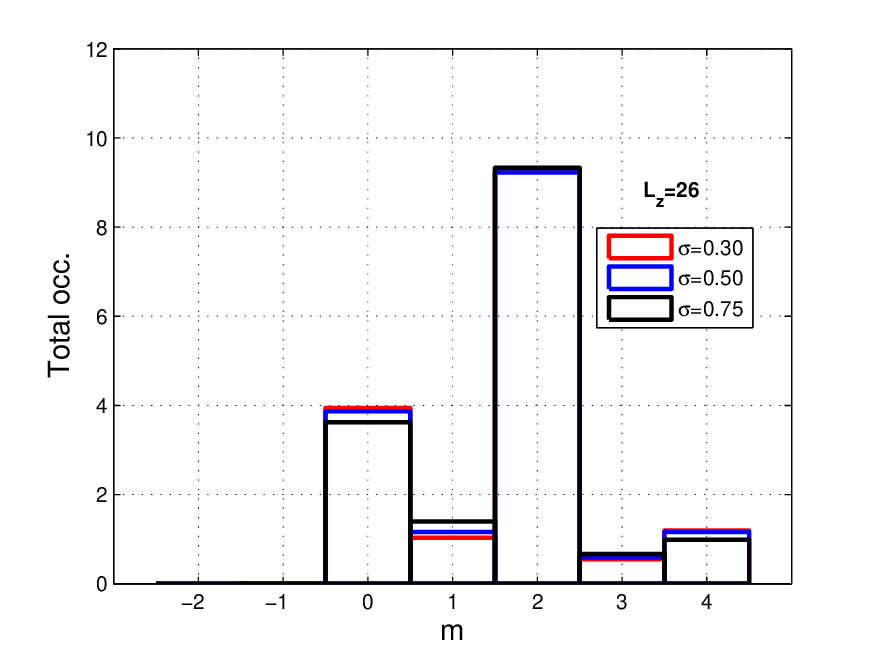}};
		\spy on (0.6,0.95) in node at (-1.0,0.880);
		\end{tikzpicture}
	   	  \hspace{-6.0mm}
		 \includegraphics[width=0.3\linewidth] {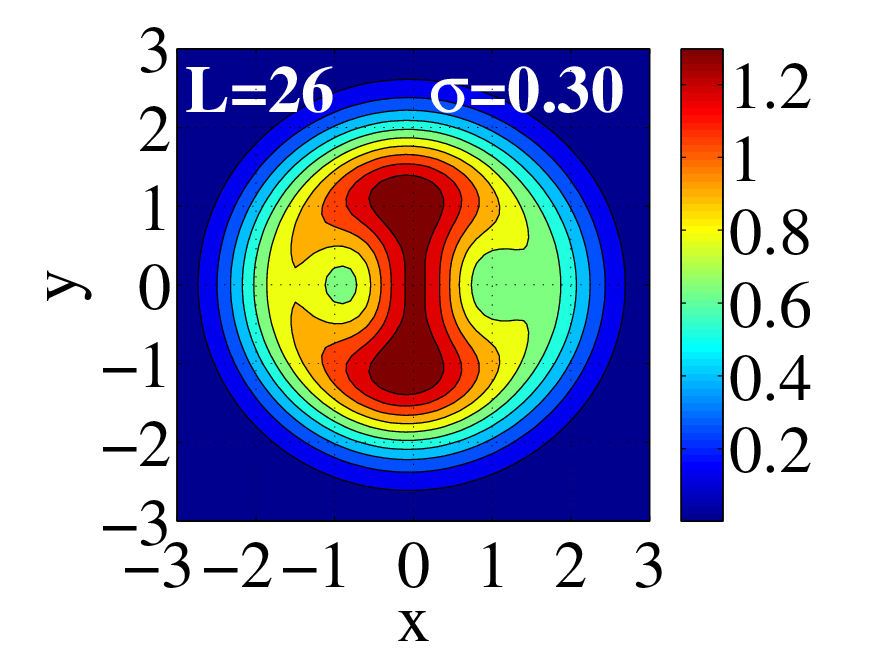}\label{fig:sp3_L26}
		\hspace{-5.0mm}
		 \includegraphics[width=0.3\linewidth]{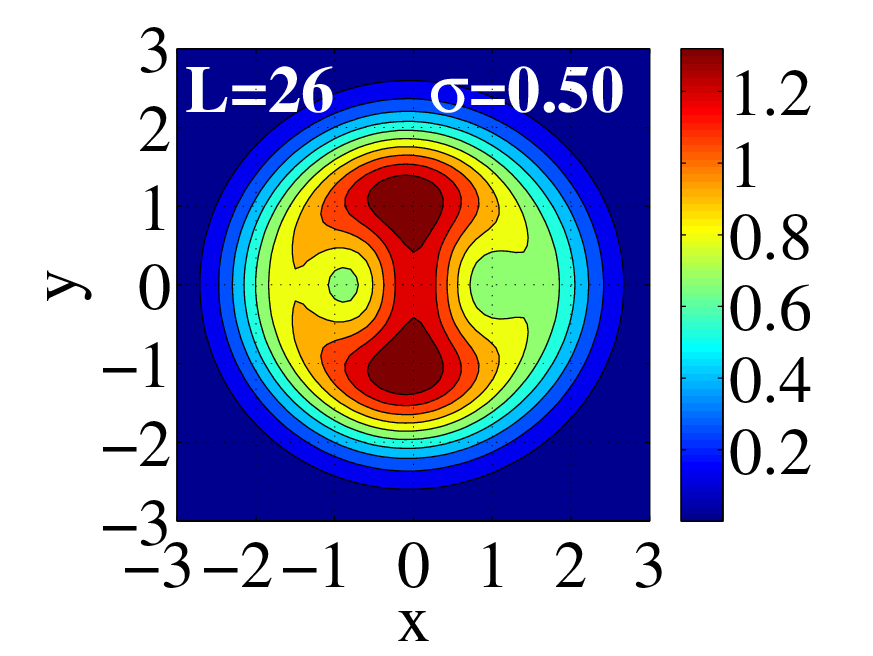}\label{fig:sp5_L26}
		\hspace{-5.0mm}
		 \includegraphics[width=0.3\linewidth]{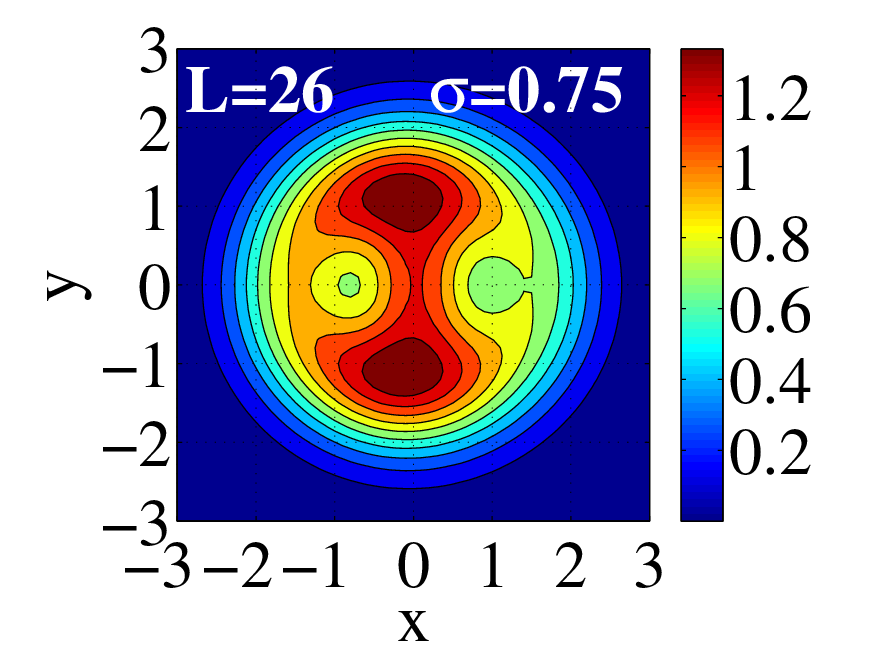}\label{fig:sp75_L26}}
	\subfigure[]{ 
   \hspace{-5.0mm}
	   \includegraphics[width=0.35\linewidth]{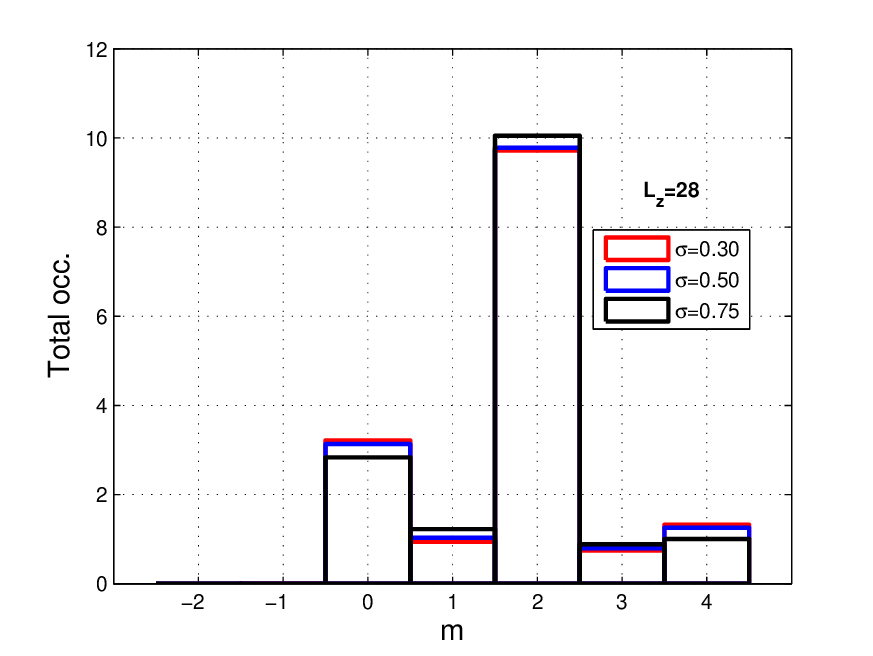}
   \hspace{-6.0mm}
        \includegraphics[width=0.3\linewidth]{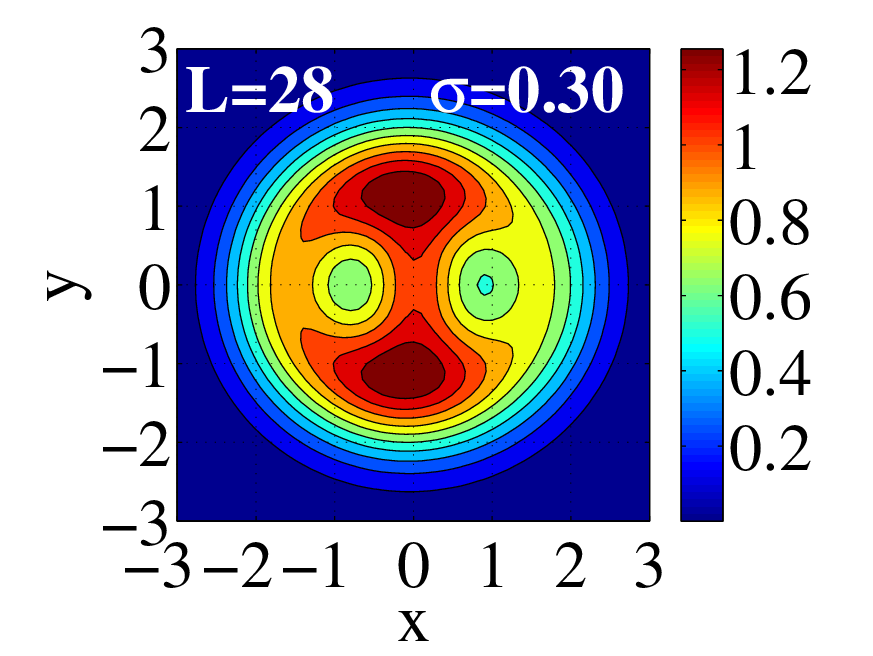}\label{fig:sp3_L28}
	\hspace{-5.0mm}
		\includegraphics[width=0.3\linewidth]{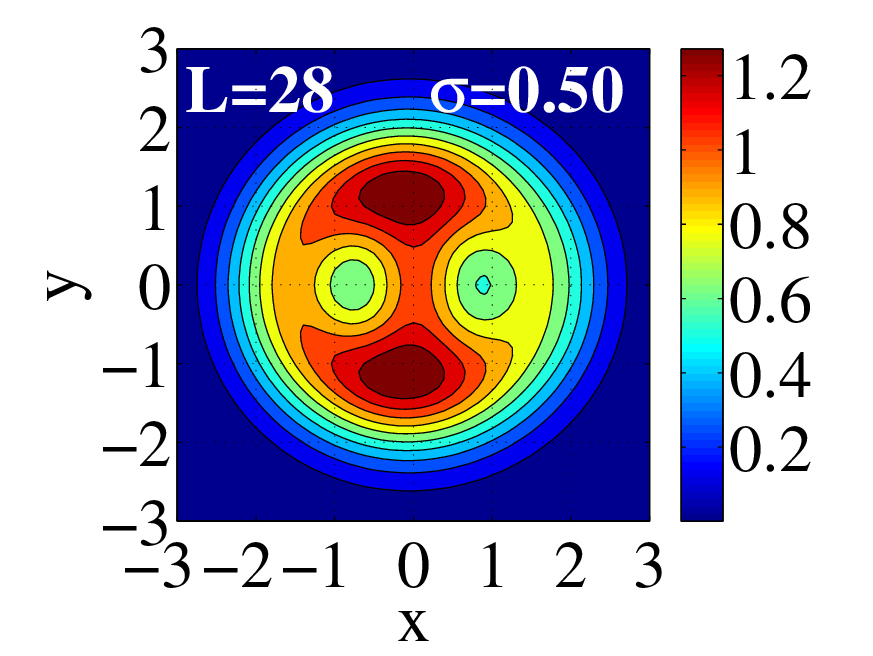}\label{fig:sp5_L28}
	\hspace{-5.0mm}
		\includegraphics[width=0.3\linewidth]{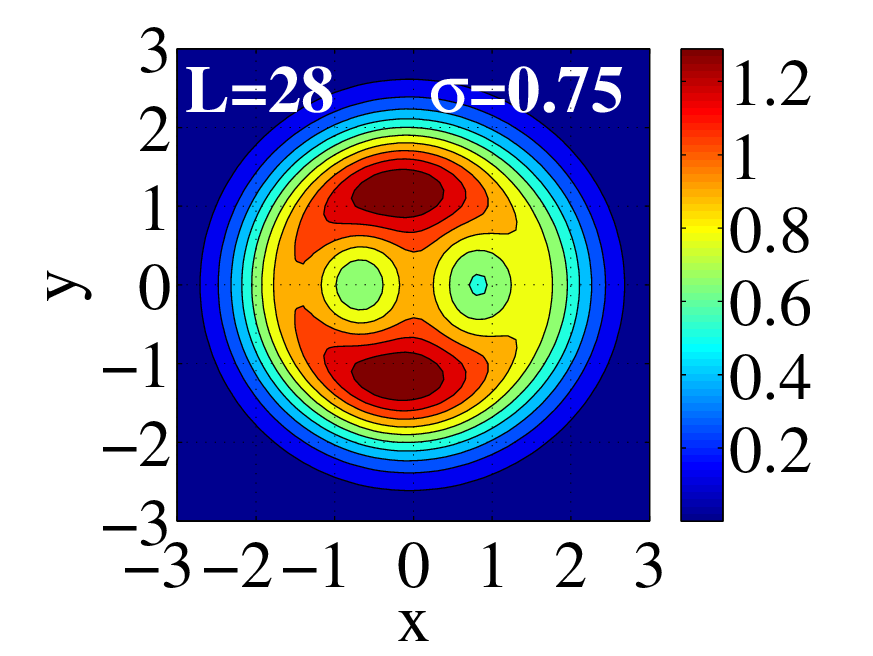}\label{fig:sp75_L28}}
	\subfigure[]{
		\includegraphics[width=0.35\linewidth]{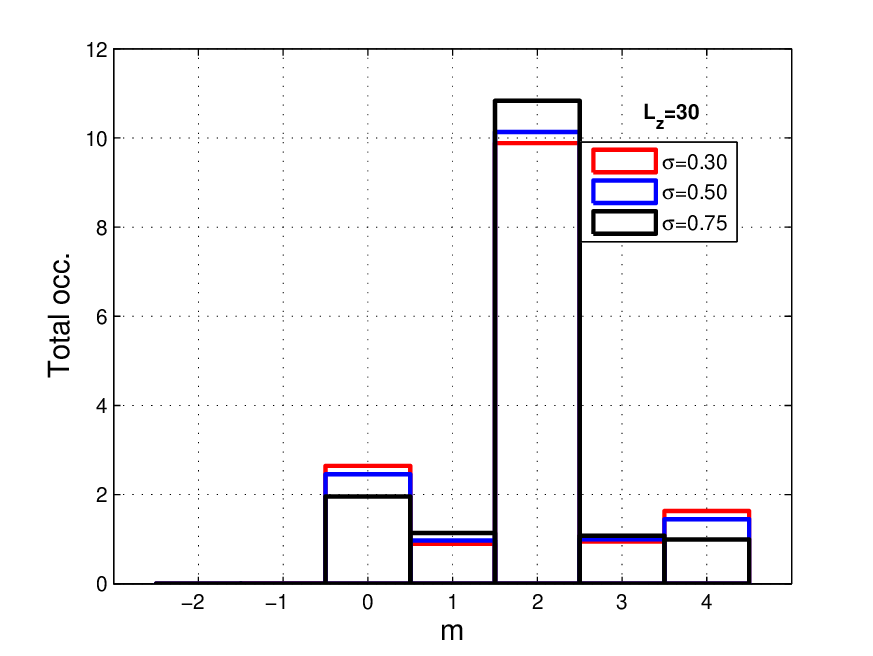}
		\hspace{-6.0mm}
		\includegraphics[width=0.3\linewidth]{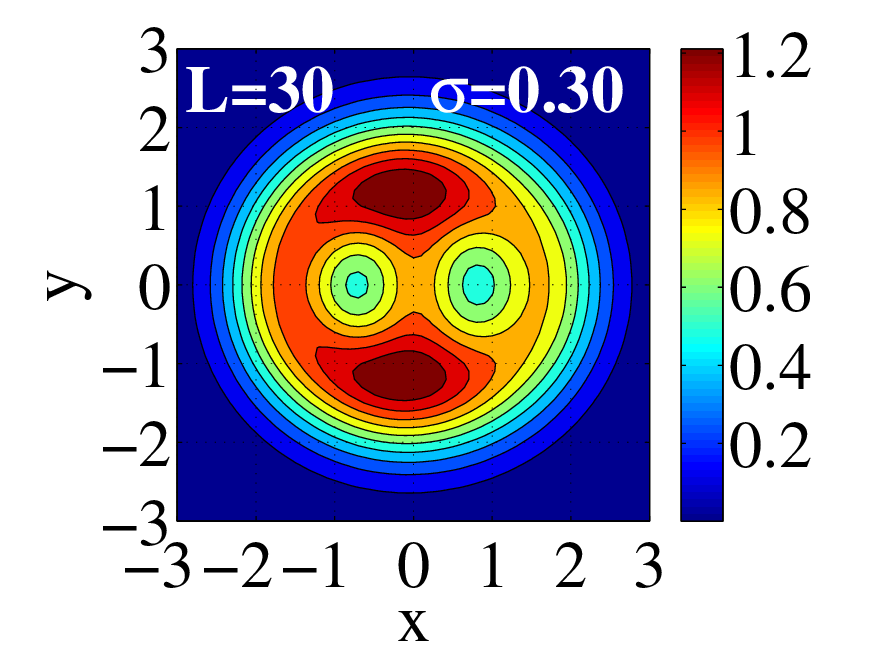}\label{fig:sp3_L30}
		\hspace{-5.0mm}
		\includegraphics[width=0.3\linewidth]{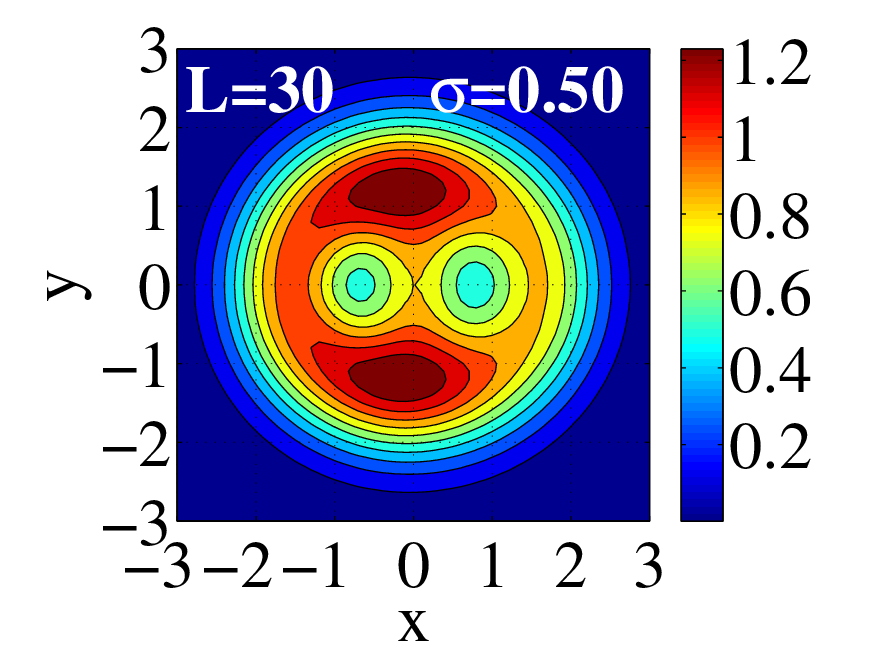}\label{fig:sp5L30A}
		\hspace{-5.0mm}
		\includegraphics[width=0.3\linewidth]{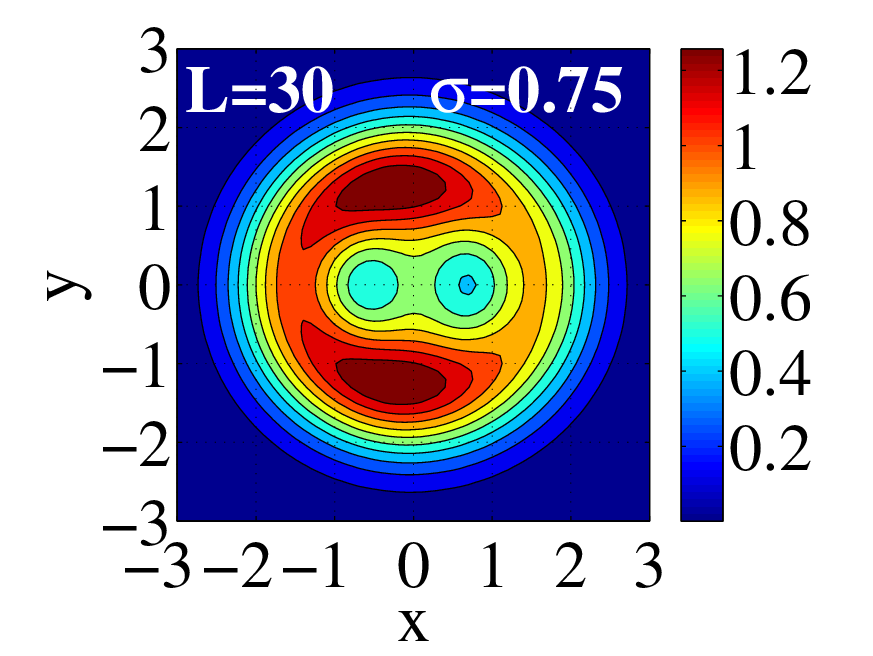}\label{fig:sp75L30A}}
	\subfigure[]{
		\includegraphics[width=0.35\linewidth]{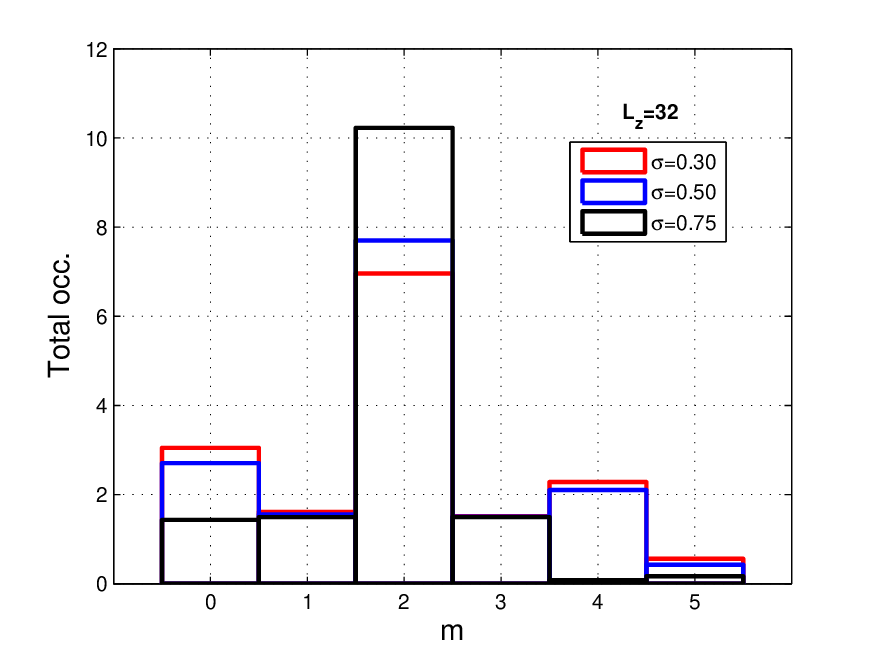}
		\hspace{-6.0mm}
		\includegraphics[width=0.3\linewidth]{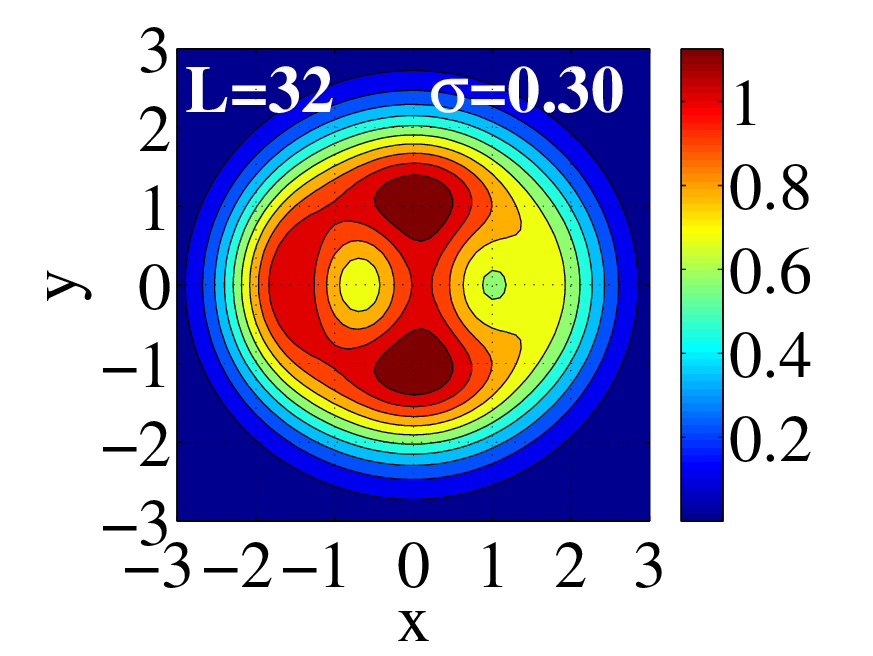}\label{fig:sp3L32A}
		\hspace{-5.0mm}
		\includegraphics[width=0.3\linewidth]{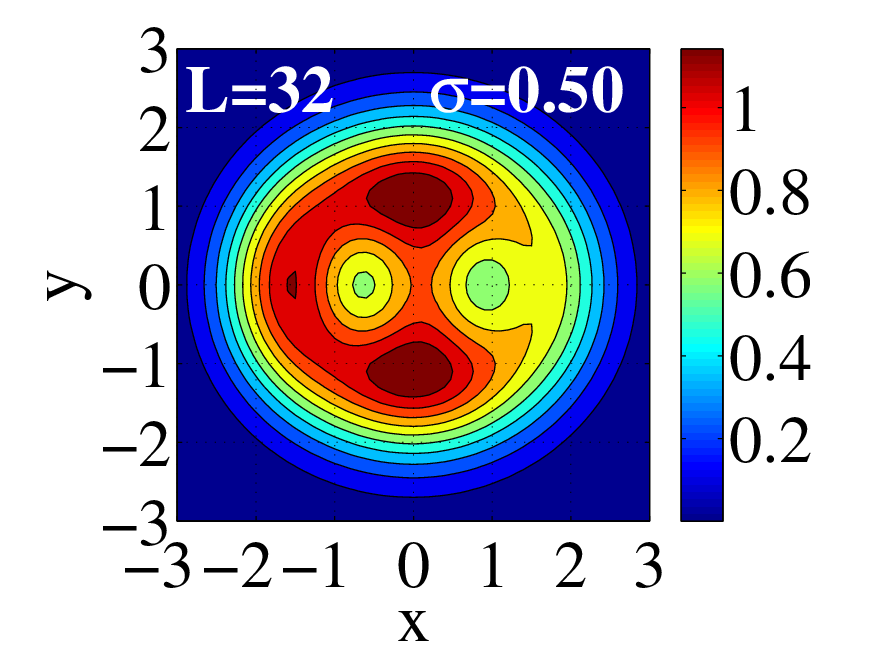}\label{fig:sp5L32A}
		\hspace{-5.0mm}
		\includegraphics[width=0.3\linewidth]{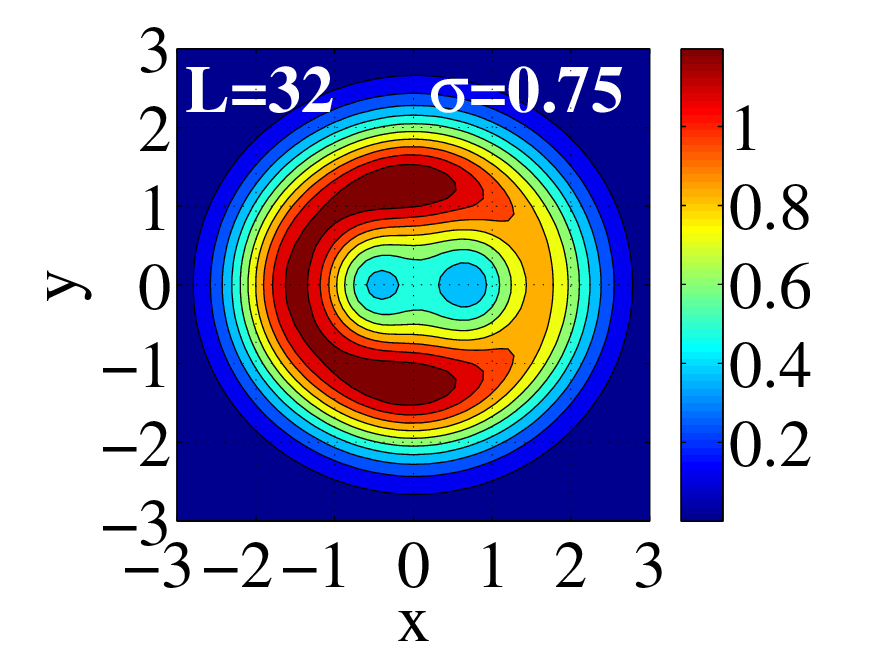}\label{fig:sp75L32A}}
\caption{\label{fig:twovortx2630}(Color online) The macroscopic occupations of one-particle angular momentum states of OPRDM {\it vs} one-particle angular momentum $m$ and the corresponding CPD plots depicting diagonally symmetric two-vortex state for $N=16$ bosons with the interaction strength ${U_{0}}=0.2171$ and the interaction range $\sigma=0.30,0.50,0.75$. The reference point has been chosen at $\mathbf{r}_{0}=(3,0)$. In (a), for the two-vortex $L_{z}=26$ states, with increase in interaction range $\sigma$, the macroscopic occupation of maximally occupied one-particle angular momentum $m=2$ state of OPRDM first 
decreases then increases whereas, in the CPD plots, the values of probability density increases at the peaks and decreases at the cores. 
In (b)-(d) for $L_{z}=28,30,32$ states, with increase in interaction range $\sigma$, the macroscopic occupation of maximally occupied one-particle angular momentum $m=2$ state of OPRDM increases.}
\end{figure*}

\begin{figure*}[t]
	\centering
	\vspace{-5mm}
	\subfigure[$N=10$, $L_{z}=20$]
	{\hspace{-4mm}
		\includegraphics[width=0.225\linewidth]{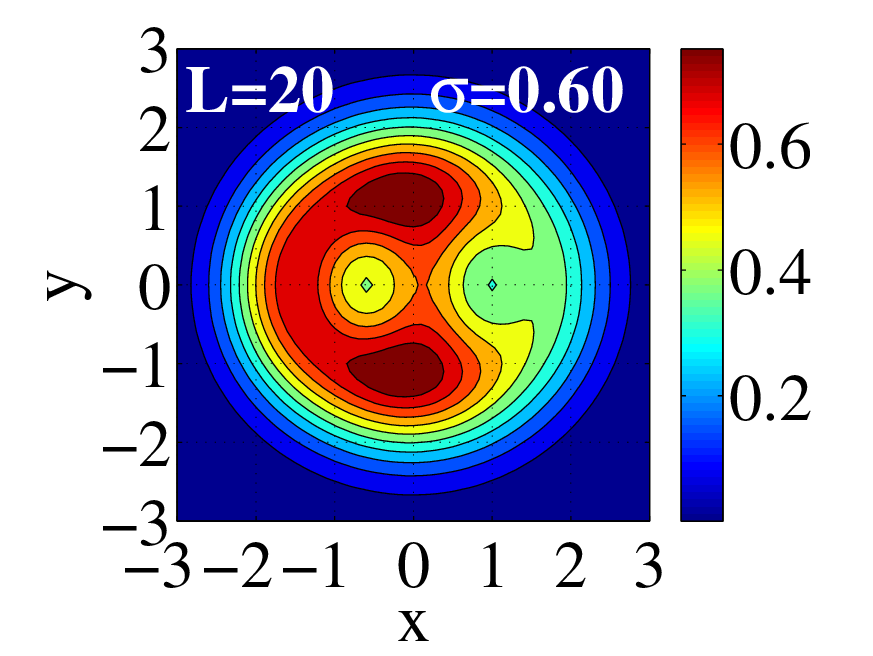}\label{fig:sp6_L20N10}
		\hspace{-4mm}
		\includegraphics[width=0.225\linewidth]{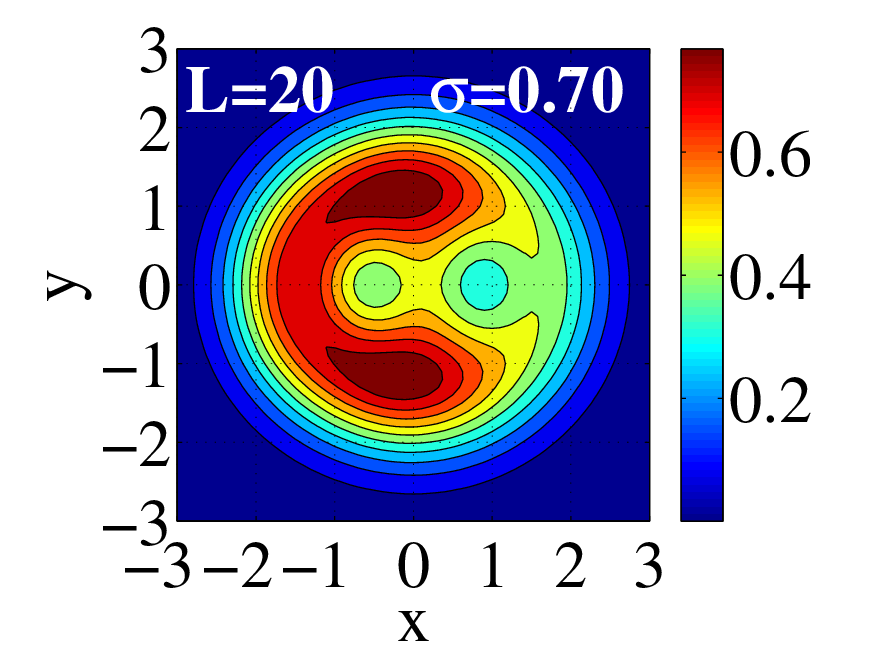}\label{fig:sp7_L20N10}
		\hspace{-4mm}
		\includegraphics[width=0.225\linewidth]{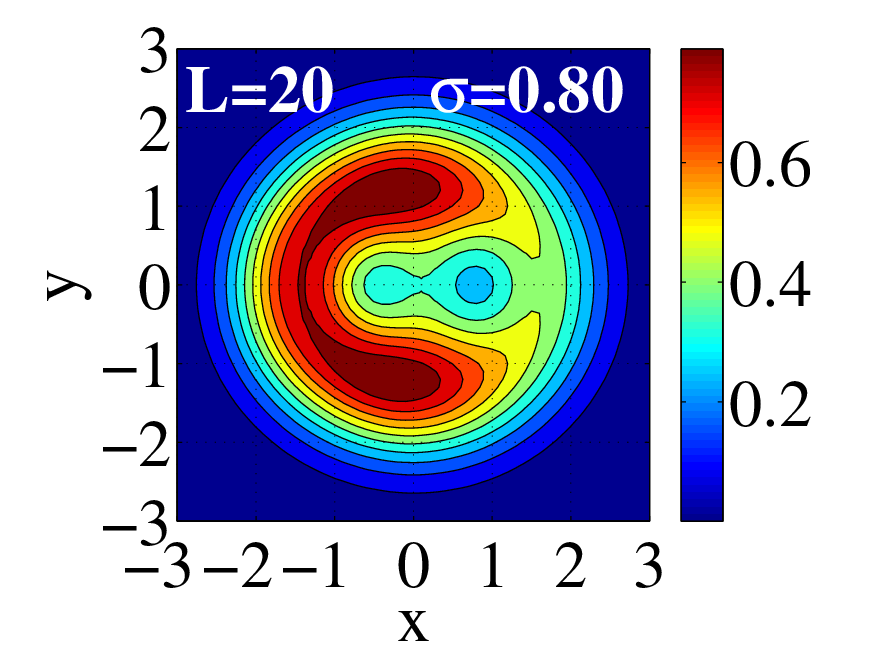}\label{fig:sp8_L20N10}
		\hspace{-4mm}
		\includegraphics[width=0.225\linewidth]{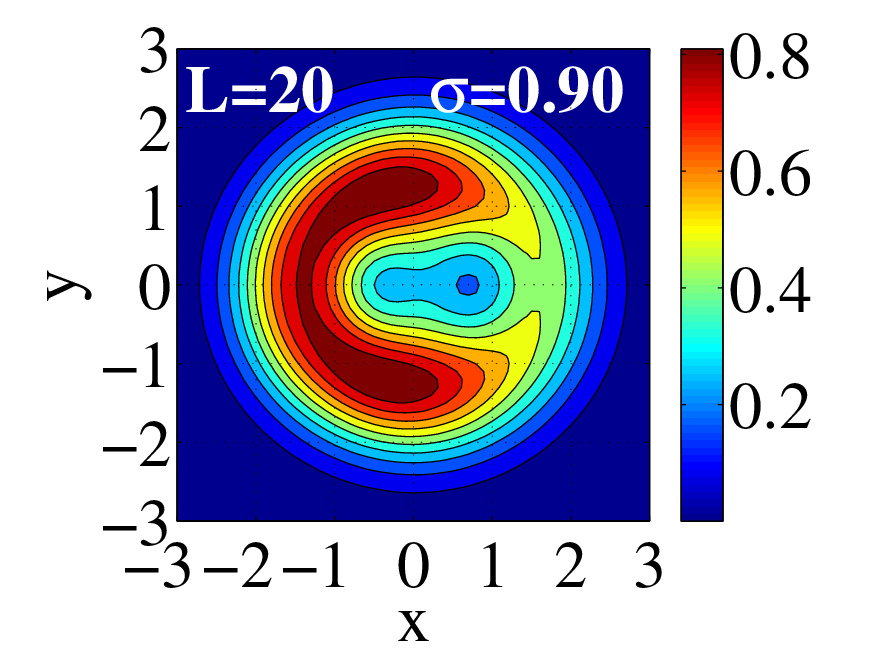}\label{fig:sp9_L20N10}
		\hspace{-4mm}
		\includegraphics[width=0.225\linewidth]{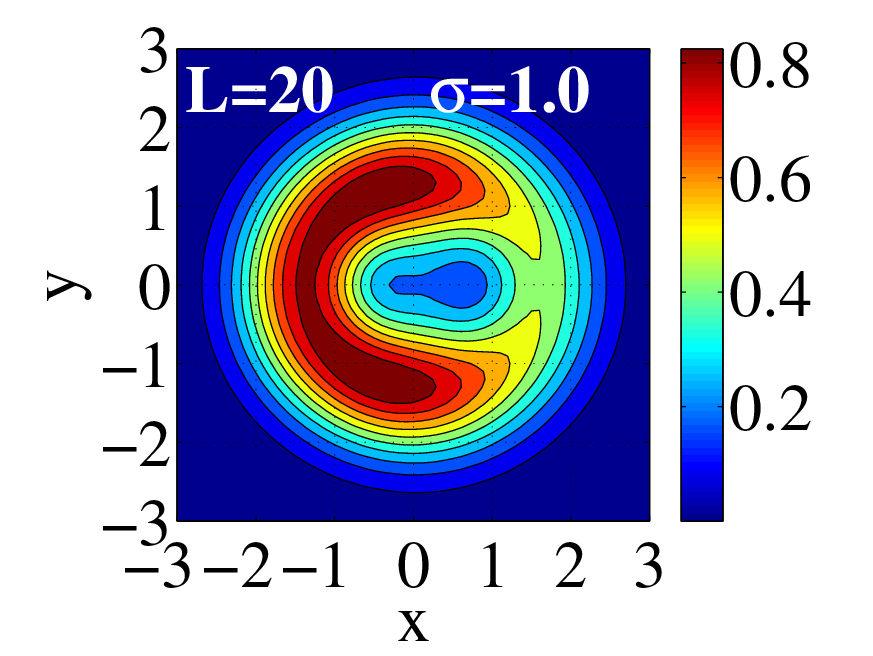}\label{fig:s1_L20N10}
	}
	\centering
	\vspace{-3mm}
	\subfigure[$N=12$, $L_{z}=24$]
	{\hspace{-4mm}
		\includegraphics[width=0.225\linewidth]{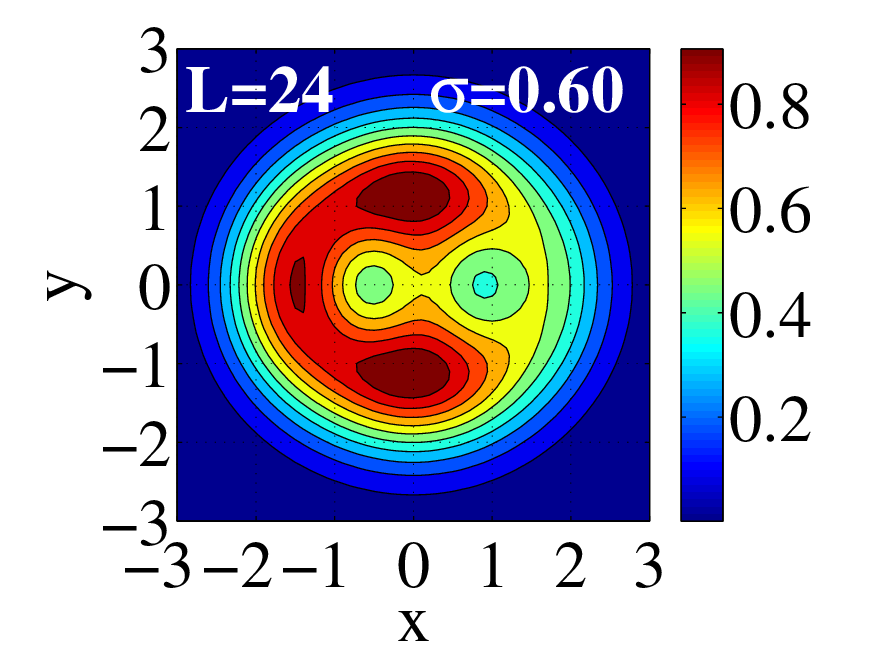}
		\hspace{-4mm}
		\includegraphics[width=0.225\linewidth]{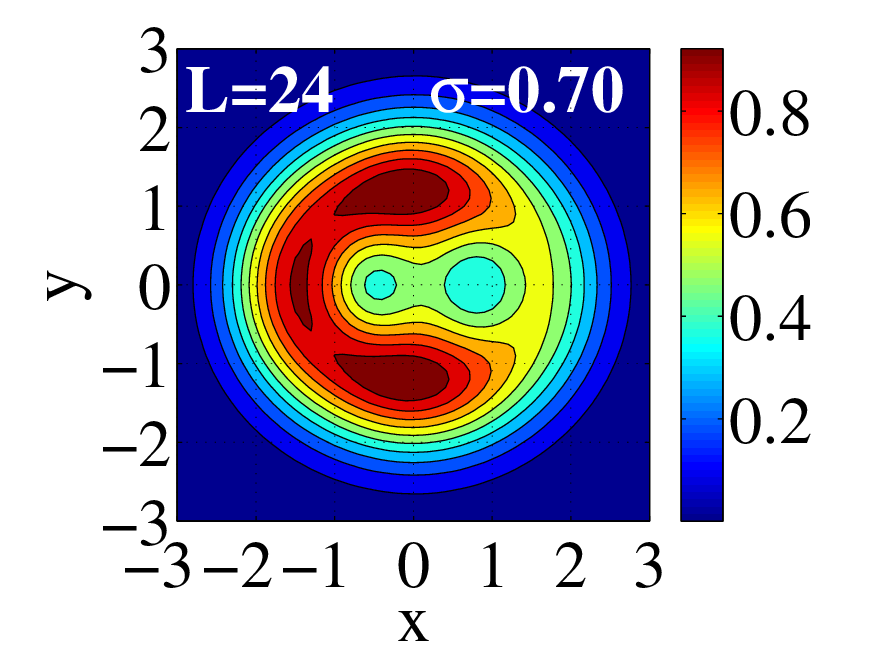}
		\hspace{-4mm}
		\includegraphics[width=0.225\linewidth]{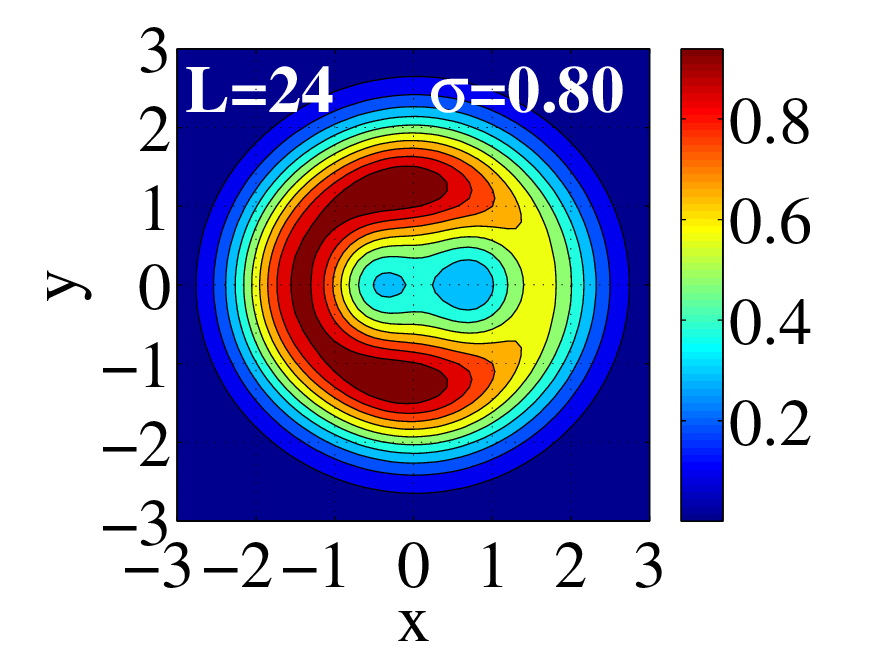}
		\hspace{-4mm}
		\includegraphics[width=0.225\linewidth]{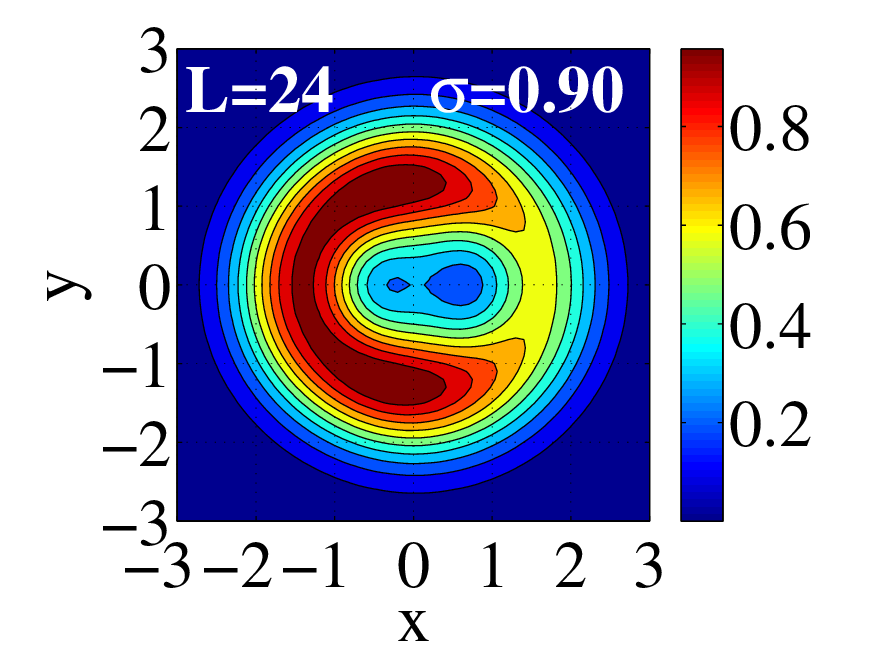}
		\hspace{-4mm}
		\includegraphics[width=0.225\linewidth]{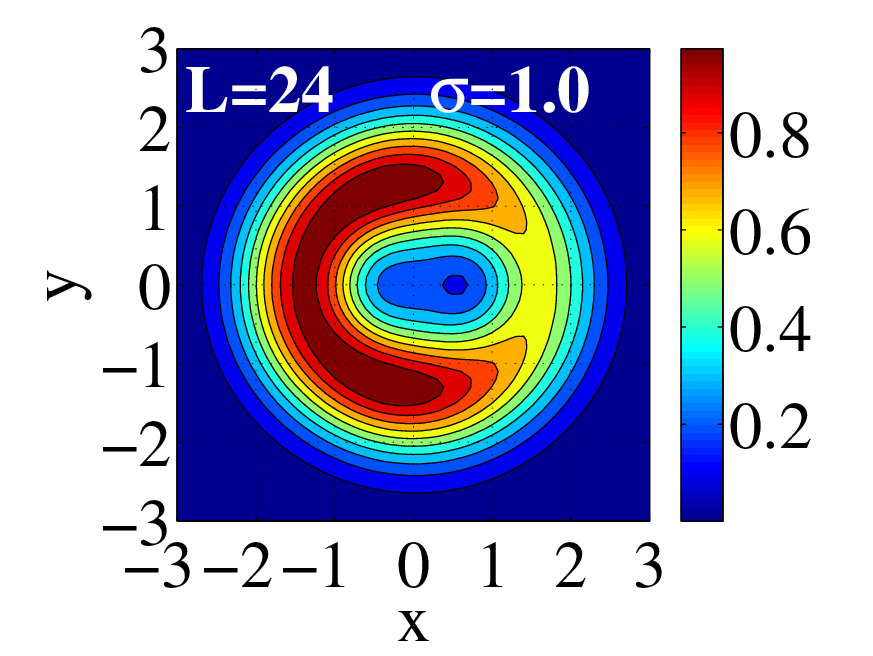}
	}
	\centering
	\vspace{-3mm}
\caption{\label{fig:spl32}(Color online) Merging of vortex cores with increase in interaction range: CPD plots depicting diagonally symmetric two-vortex state for $N=10,12,16$ bosons
with interaction strength ${U_{0}}=0.2171$ and interaction range $\sigma=0.60-1.0$ in the beyond lowest Landau level approximation. The reference point has been chosen at $\mathbf{r}_{0}=(3,0)$. (a) for $L_{z}=2N=20$, with increase in interaction range beyond $\sigma=0.6$, the cores begin to merge and at $\sigma=0.80$ the overlap of cores becomes significant.
(b) similarly for $L_{z}=2N=24$ state, with increase in interaction range beyond $\sigma=0.7$, the cores begin to merge and at $\sigma=0.80$ the overlap of cores becomes significant.}
\end{figure*}

\begin{figure*}[!htb]
	\subfigure[$N=16$, $L_{z}=32$]{\hspace{-3mm}
		\includegraphics[width=0.225\linewidth]{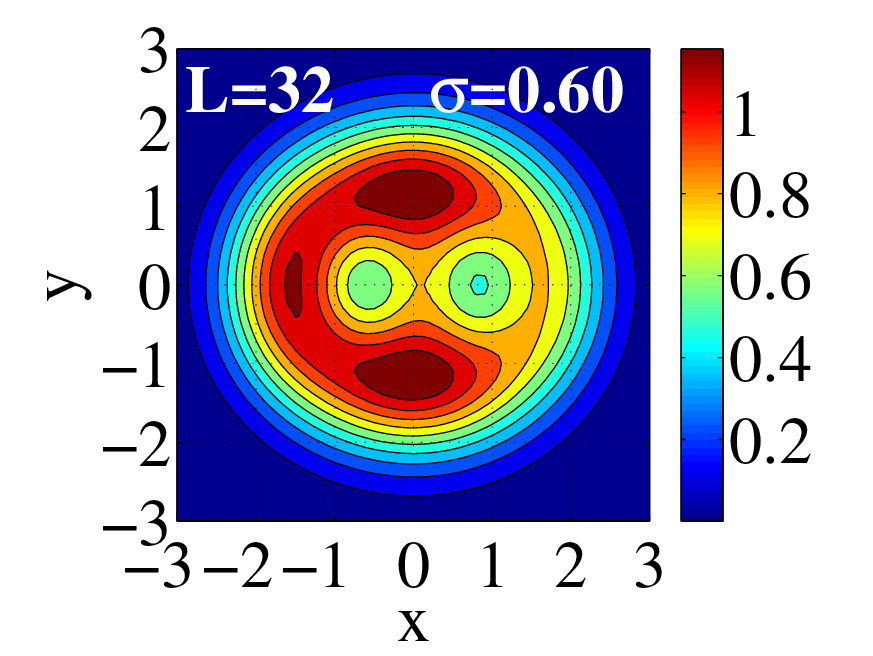}\label{fig:sp6_L32}
		\hspace{-4mm}
		\includegraphics[width=0.225\linewidth]{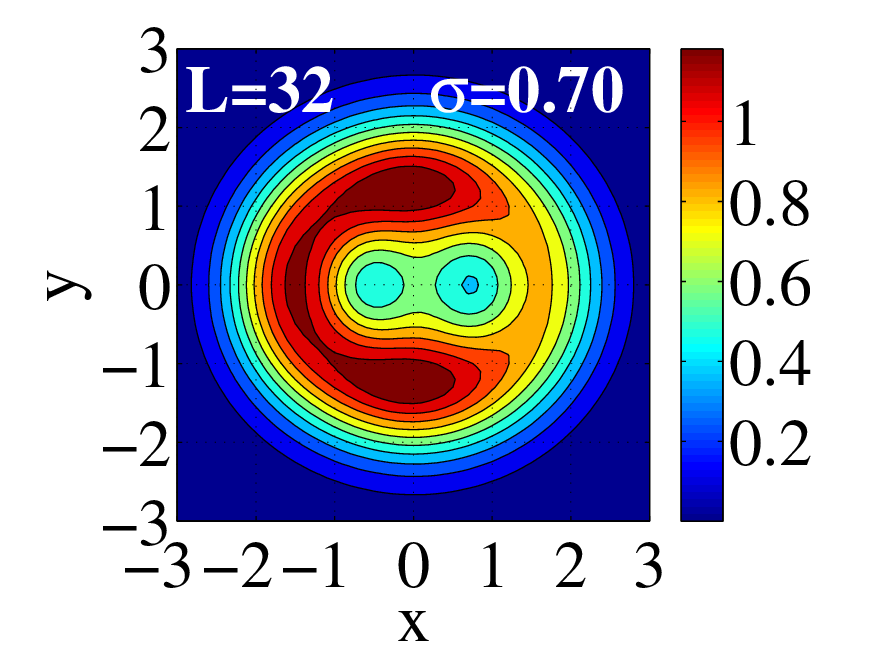}\label{fig:sp7_L32}
		\hspace{-4mm}
		\includegraphics[width=0.225\linewidth]{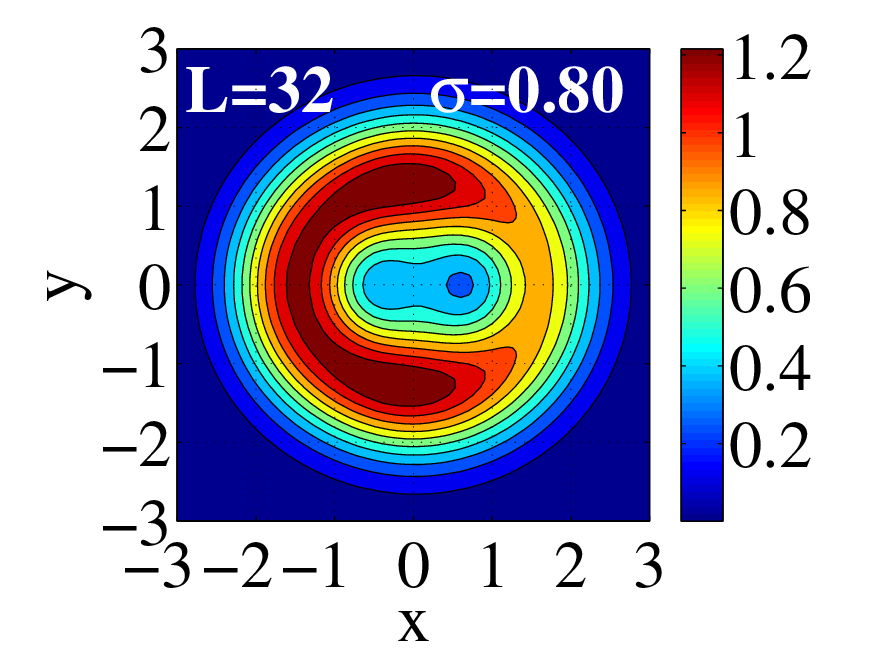}\label{fig:sp8_L32}
		\hspace{-4mm}
		\includegraphics[width=0.225\linewidth]{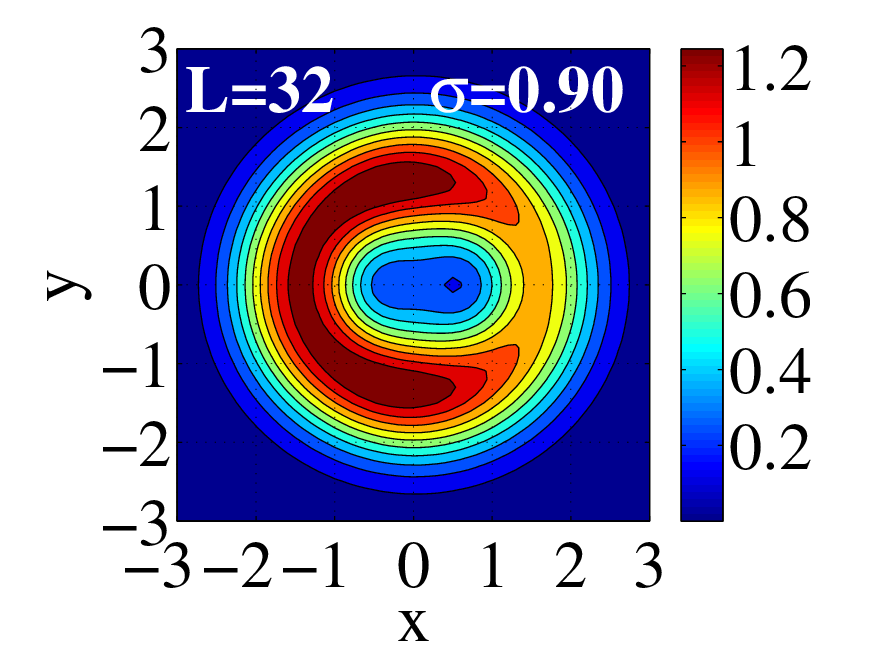}\label{fig:sp9_L32}
		\hspace{-4mm}
		\includegraphics[width=0.225\linewidth]{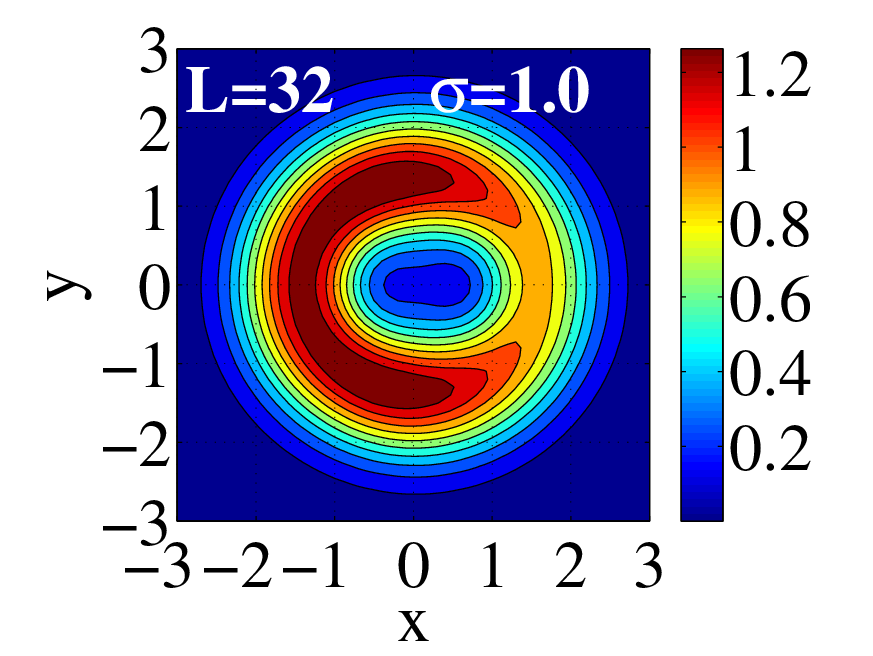}\label{fig:s1_L32}
       }
	\begin{minipage}{0.5\textwidth}
	\subfigure[]{\hspace{-3mm}
		\includegraphics[width=1.0\linewidth ]{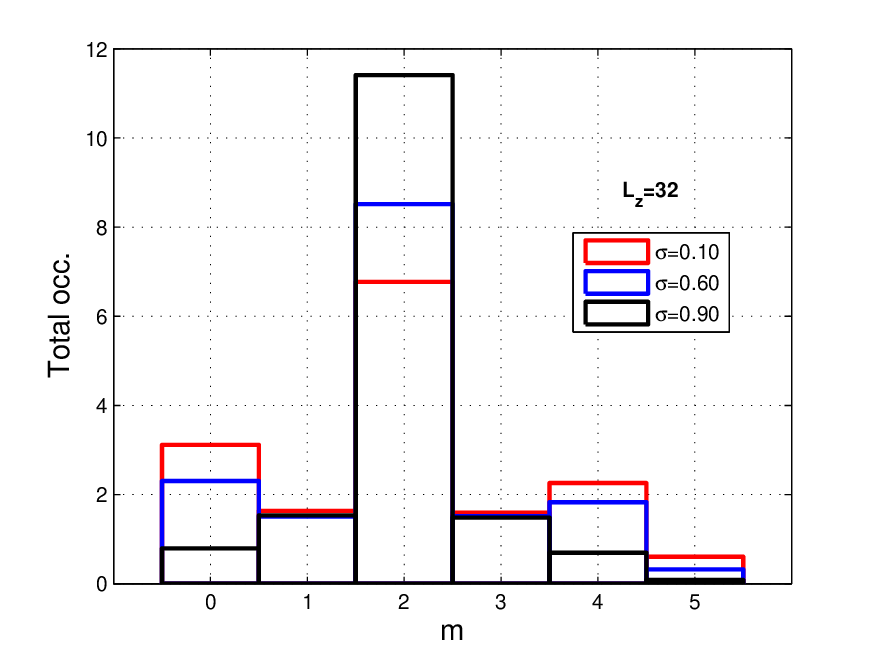}
		}
	\end{minipage}
\hspace{0.325in}
	\begin{minipage}{0.35\textwidth}
\caption{\label{fig:spN16l32}(Color online) The macroscopic occupations of one-particle angular momentum states of OPRDM {\it vs} one-particle angular momentum $m$ and the corresponding 
CPD plots depicting diagonally symmetric two-vortex state 
with ${U_{0}}=0.2171$ and $0.6\le \sigma \le 1.0$ in the beyond lowest Landau level approximation. The reference point has been chosen at $\mathbf{r}_{0}=(3,0)$. (a) for $L_{z}=2N=32$, with increase in interaction range beyond $\sigma=0.7$, the cores begin to merge and at $\sigma=0.80$ the overlap of cores become significant. (b) with increase in interaction range $\sigma$, the macroscopic occupation of maximally occupied one-particle angular momentum $m=2$ state of OPRDM increases.}
	\end{minipage}
\end{figure*}

\begin{figure*}[t]
	\centering
	\vspace{-4.877mm}
	\subfigure[$N=20$, $L_{z}=40$]
	{
		\hspace{-4mm}
		\includegraphics[width=0.225\linewidth]{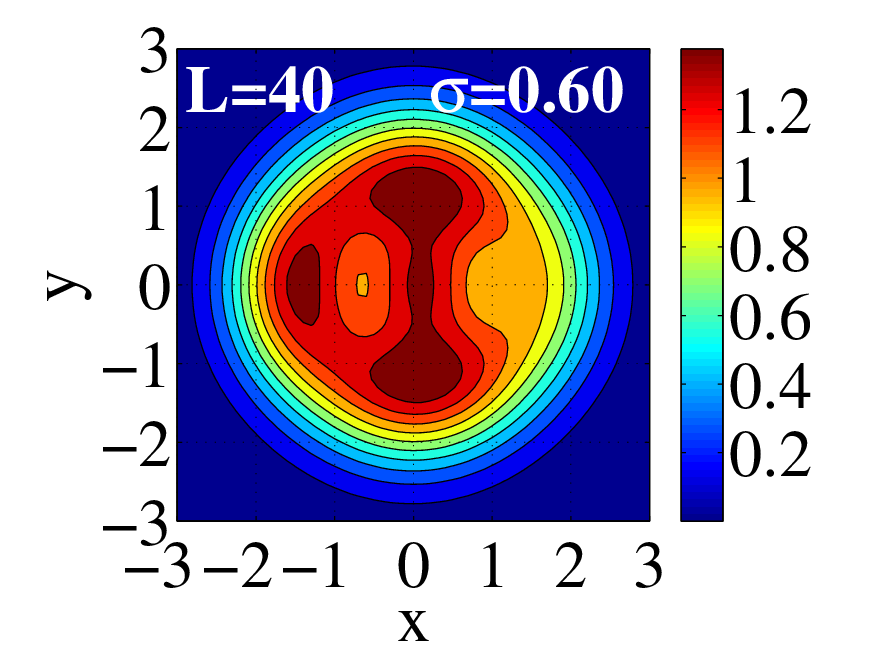}
		\hspace{-4mm}
		\includegraphics[width=0.225\linewidth]{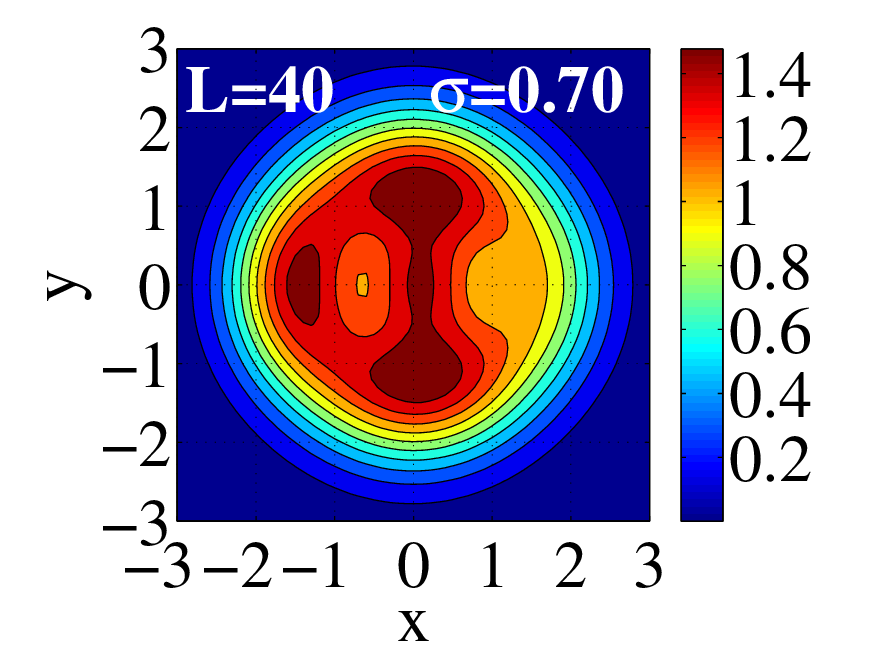}
		\hspace{-4mm}
		\includegraphics[width=0.225\linewidth]{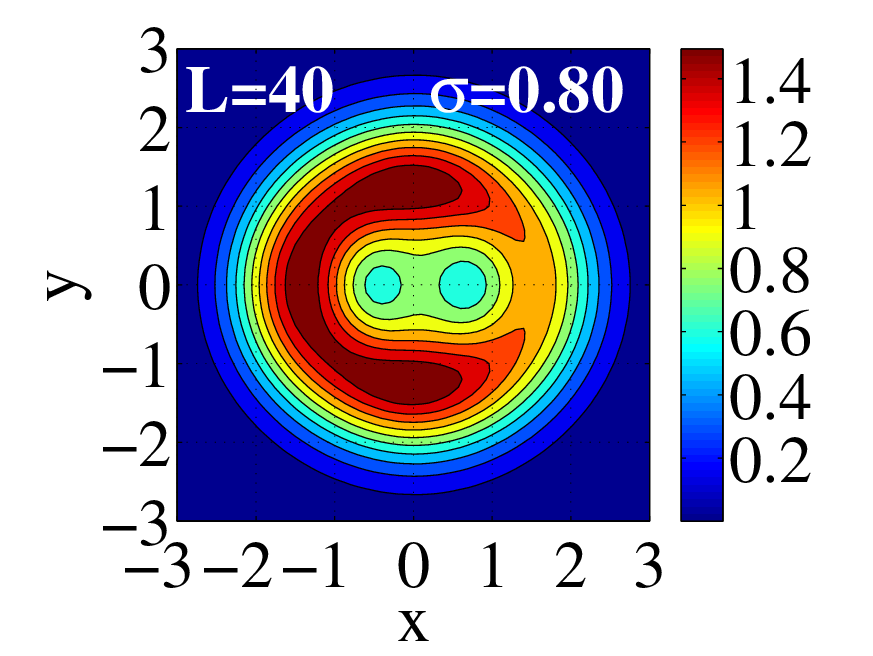}
		\hspace{-4mm}
		\includegraphics[width=0.225\linewidth]{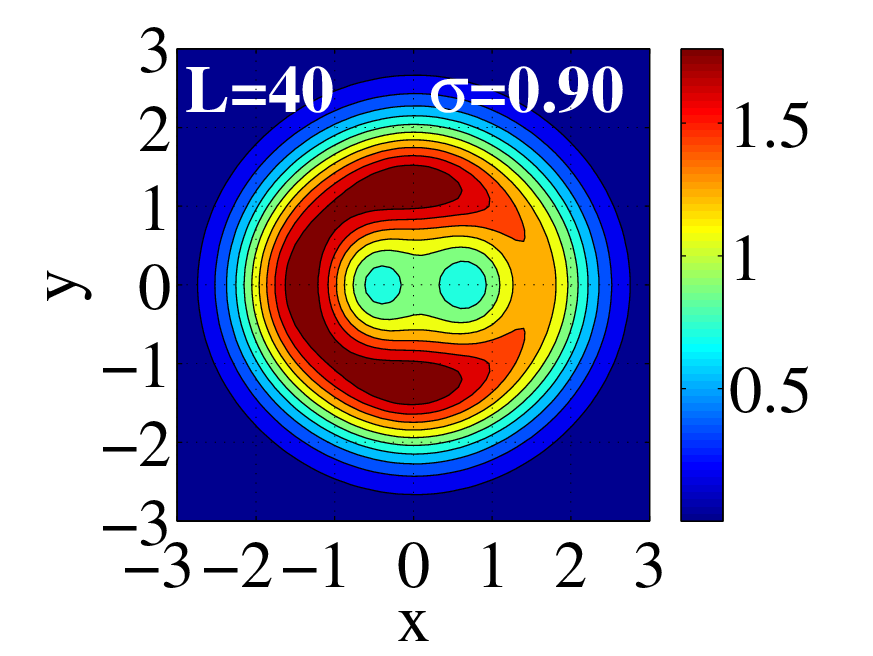}
		\hspace{-4mm}
		\includegraphics[width=0.225\linewidth]{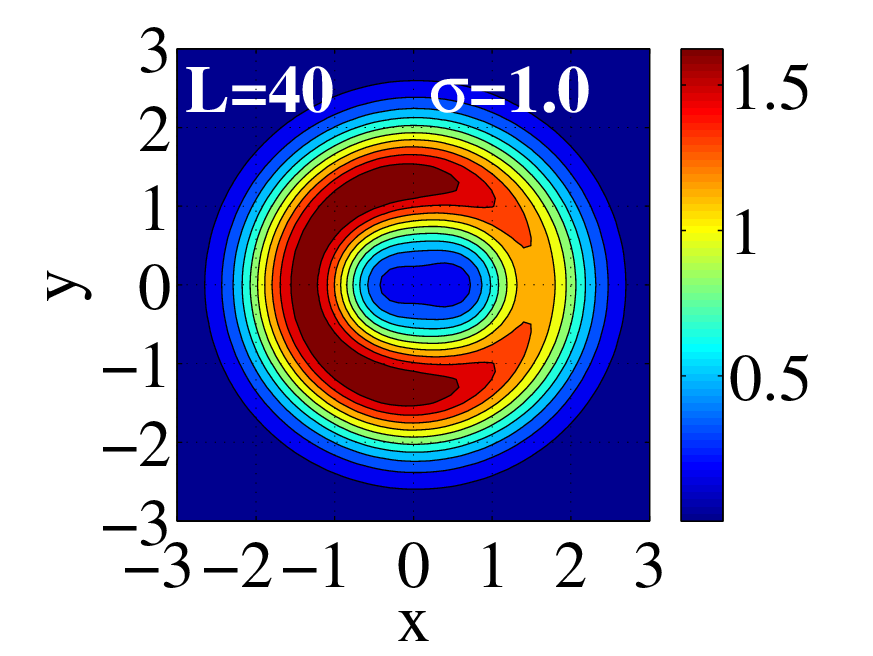}
	}
	\vspace{-4.5mm}
	\subfigure[$N=24$, $L_{z}=48$]
	{
		\hspace{-4mm}
		\includegraphics[width=0.225\linewidth]{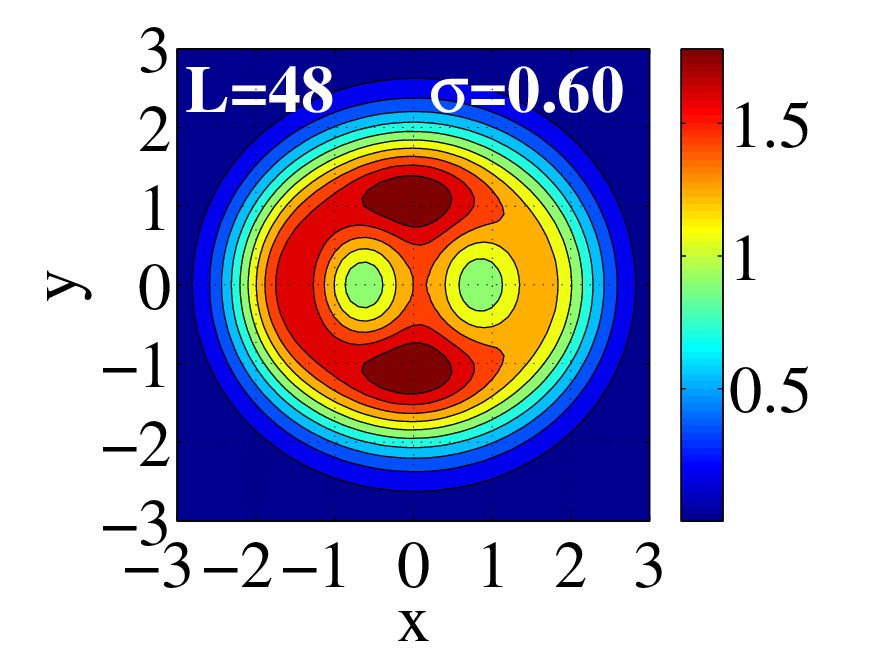}
		\hspace{-4mm}
		\includegraphics[width=0.225\linewidth]{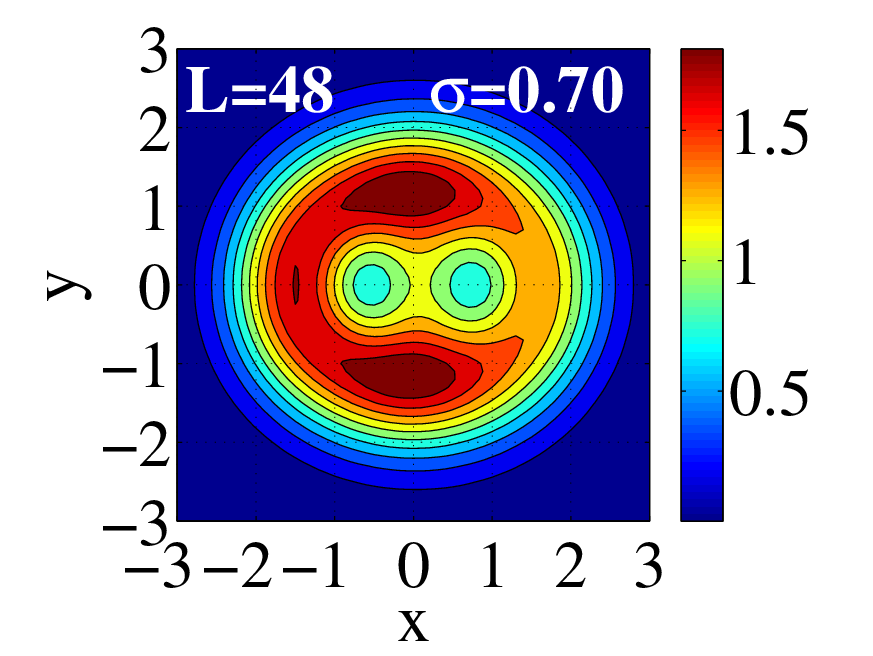}
		\hspace{-4mm}
		\includegraphics[width=0.225\linewidth]{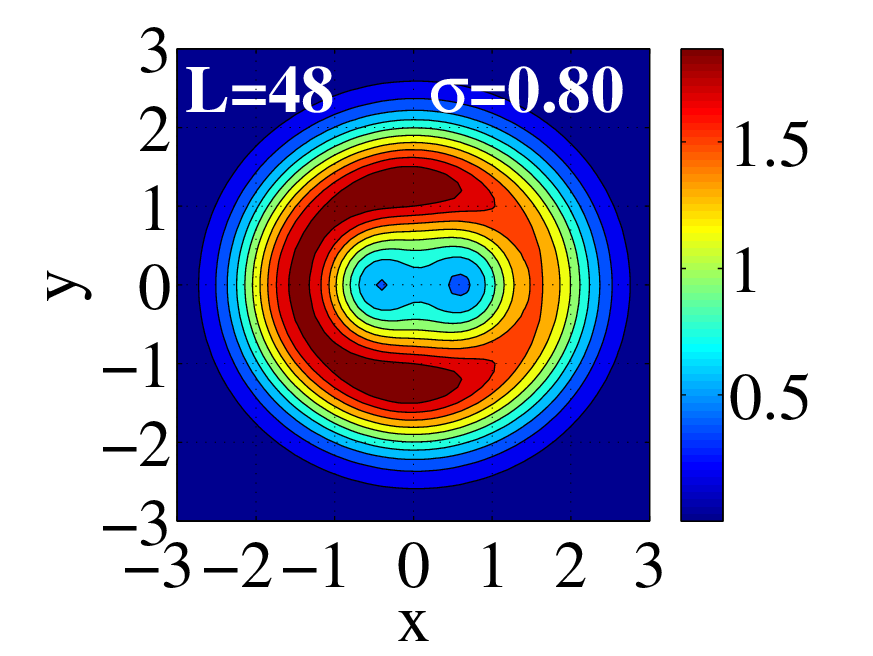}
		\hspace{-4mm}
		\includegraphics[width=0.225\linewidth]{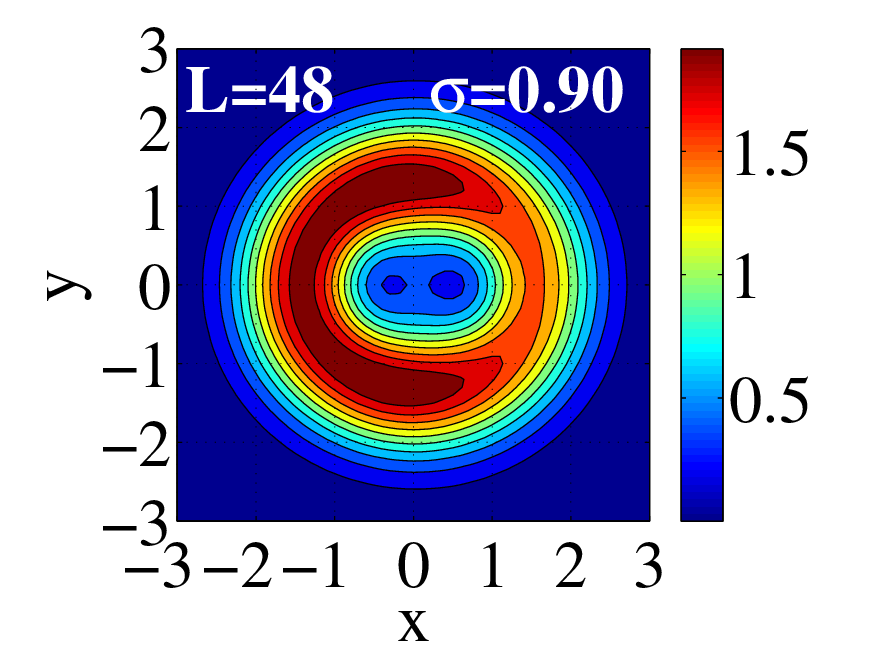}
		\hspace{-4mm}
		\includegraphics[width=0.225\linewidth]{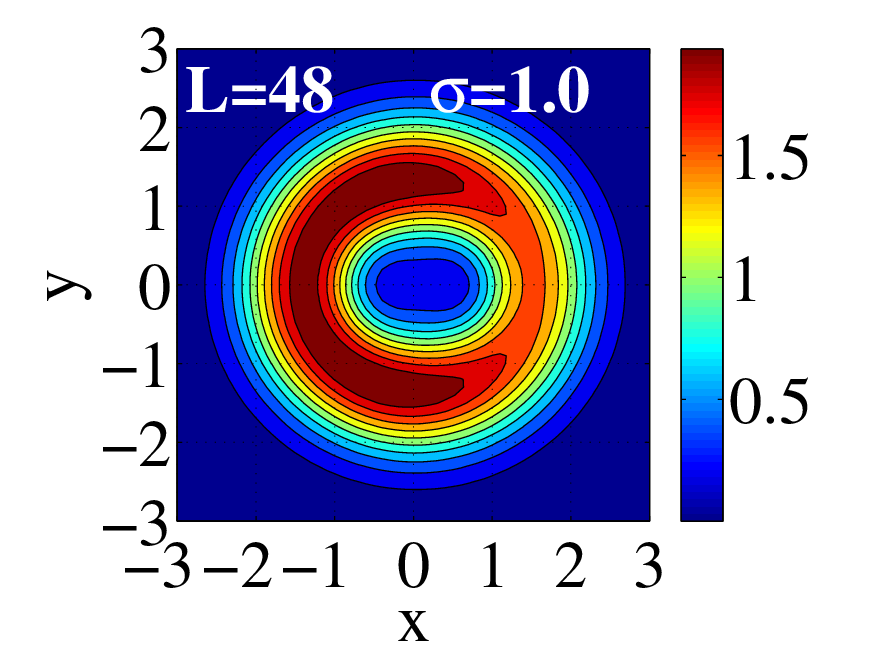}}
	\centering
\caption{\label{fig:spl40N20}(Color online) Merging of vortex cores with increase in interaction range: CPD plots depicting diagonally symmetric two-vortex state with $U_{0}=0.2171$ and $\sigma=0.60-1.0$ in the lowest Landau level approximation. The reference point has been chosen at $\mathbf{r}_{0}=(3,0)$. In CPD plots, red represents the highest probability density region and blue the least probability density region as shown on the adjoining color bar: (a) for $L_{z}=2N=40$, with increase in interaction range beyond $\sigma=0.8$, the merging of vortex-cores take place and at $\sigma=0.90$ the overlap of cores becomes significant. (b) similarly for $L_{z}=2N=48$ state, with increase in interaction range beyond $\sigma=0.8$, the vortex cores begin to merge and at $\sigma=0.90$ the overlap of cores becomes significant.}
\end{figure*}

\indent To examine the internal structure of the ground state of the condensate, we calculate the conditional probability distribution(CPD)\cite{ChienNanLiu12,LeiGeng20,JerzyCioslowski17,Xia01_ro} which is a measure of the density distribution in the body-fixed frame, defined as the probability of finding a boson at $\mathbf{r}$ given that the other is at $\mathbf{r}_{0}$~\cite{UziLandman00,srh10}:
\begin{eqnarray}
\cal{P}
(\mathbf{r,r_{0}}) &=& \frac{\langle\Psi_{0} |\sum_{i\neq j}\delta(\mathbf{r}-\mathbf{r}_{i})\delta(\mathbf{r}_{0}-\mathbf{r}_{j})|\Psi_{0}\rangle}{(N-1)\sum_{j}\langle\Psi_{0} |\delta(\mathbf{r}_{0}-\mathbf{r}_{j})|\Psi_{0}\rangle}\nonumber\\
\label{eq:cpd}
\end{eqnarray}
where $|\Psi_{0}\rangle$ represents the many-body ground-state obtained through exact diagonalization and $\mathbf{r}_{0}=(x_{0},y_{0})$ is the reference point in the $x$-$y$ plane (in units of $a_{\perp}$). For small values of angular momentum $0\leq L_{z}\leq 2N$, the reference point is chosen to take significantly large value {\it i.e.} $\mathbf{r}_{0}=(3,0)$,
whereas in the angular momentum regime $2N< L_{z}\le 4N$, the reference point is chosen to be ${\bf r}_{0}=(1.5,0)$~\cite{Imran_EJPD23}.
For the chosen reference point, the line joining ($0,0$) and   $(x_{0},0)$ is the reference line. For multi-vortex states, the nucleating vortices arrange themselves symmetrically about the reference line. The CPD is depicted through contour plot which is the iso-surface density profile viewed along z-axis in the body-fixed frame. In CPD plots presented in the following, red represents the highest probability density region and blue the lowest probability density region as shown on the adjoining color bar in Figures.~\ref{fig:b4_vortx}-\ref{fig:fivs}.\\
\indent In the following, we discuss the variations in the internal structure of the single-vortex state $L_{z}=N$ followed by the multi-vortex states as the interaction range $\sigma$ is varied. 
The number of vortices in the condensate is proportional to the vorticity $m^{1}$ which is the one-particle angular momentum quantum number corresponding to the largest eigenvalue $\lambda_{1}$ in OPRDM in Table.~{\ref{tab:Oc_table1_sp3}(column 2)}. Further, to study the evolution of the macroscopic occupation of one-particle angular momentum state $m^{\mu}$ with rotation, we present $(N\lambda_{\mu})$ {\it vs} $m^{\mu}$ bar-graph with $\mu=1,\cdots {\cal M}$, for given values of angular momentum $L_{z}=\sum_{\mu=1}^{\cal M}(N\lambda_{ \mu})m^{\mu}$ in Eq.~(\ref{eq:eval_norm}). 
\begin{figure*}
	\subfigure[]{
\begin{tikzpicture}[spy using outlines={magnification=10.0,         circle, size=1.50cm, black, connect spies}]\node (n1)           at (0,0)
    {\includegraphics[width=0.40\linewidth]{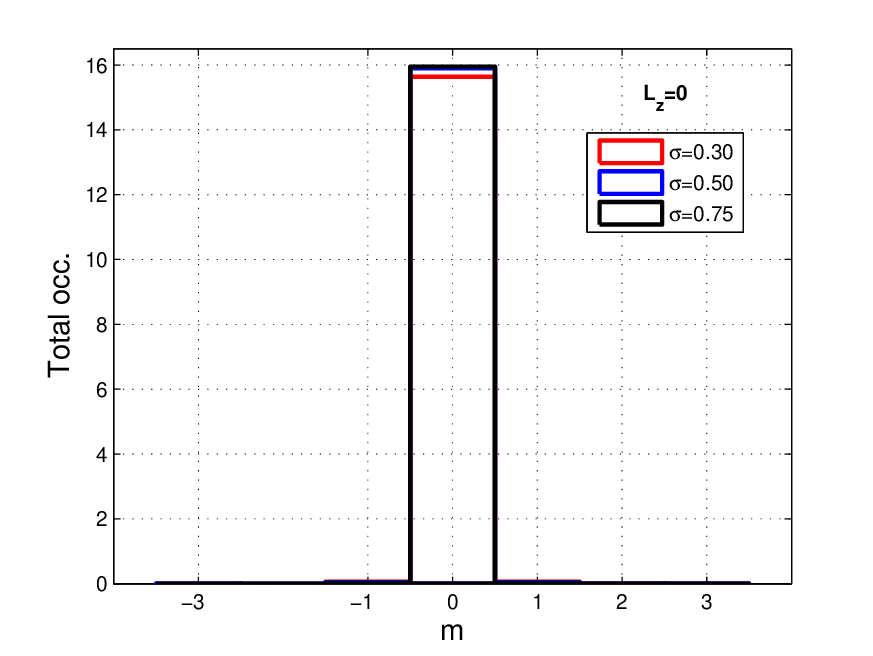}};
	\spy on (0.150,1.780) in node at (-1.20,0.0);
\end{tikzpicture}
		\vspace{-0.25in}	\hspace{-3.mm}
		\includegraphics[width=0.27\linewidth]{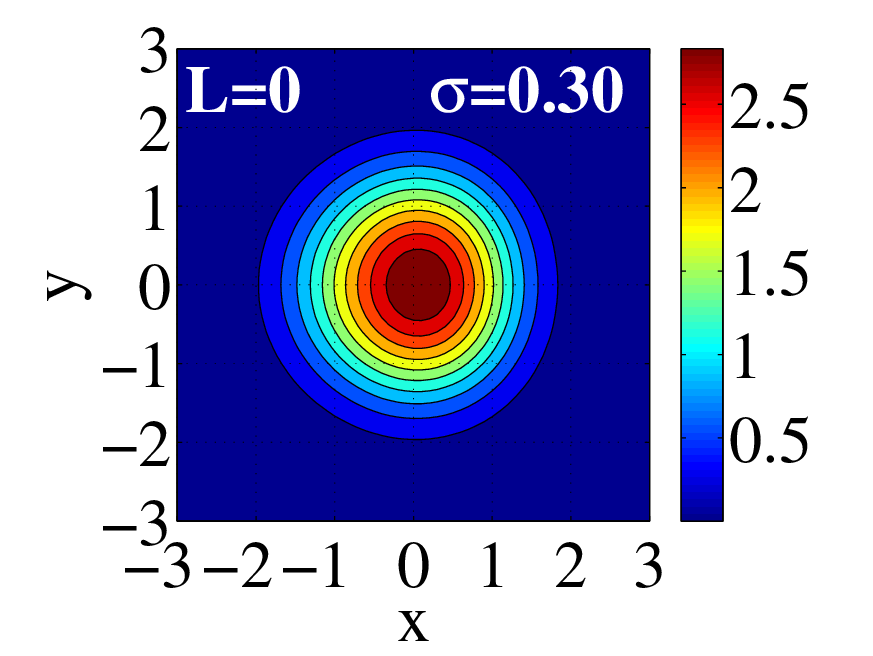}\label{fig:l0s3}
		\vspace{-0.25in}	\hspace{-5mm}
		\includegraphics[width=0.27\linewidth]{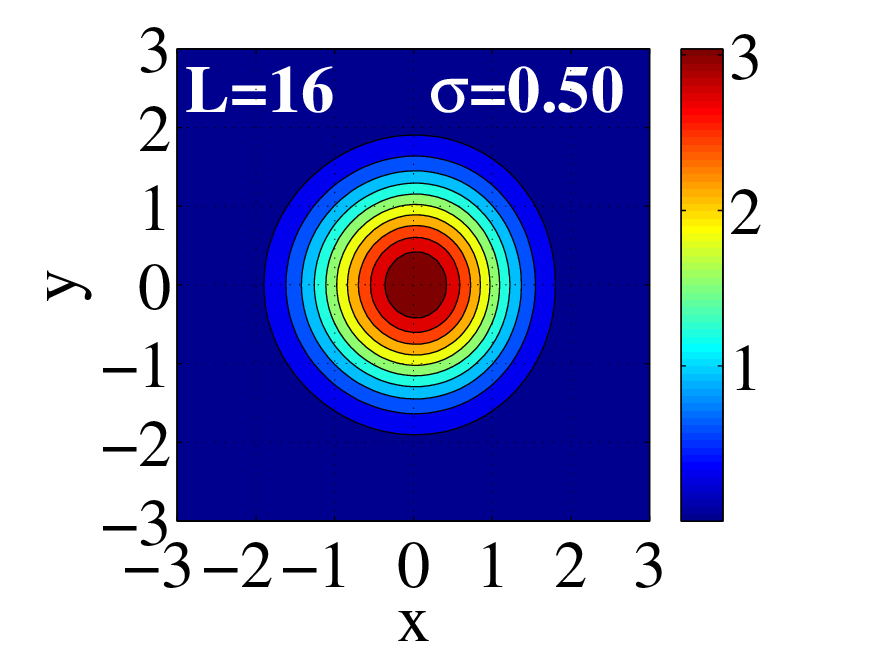}\label{fig:l0s5}
		\vspace{-0.25in}	\hspace{-5mm}
		\includegraphics[width=0.27\linewidth]{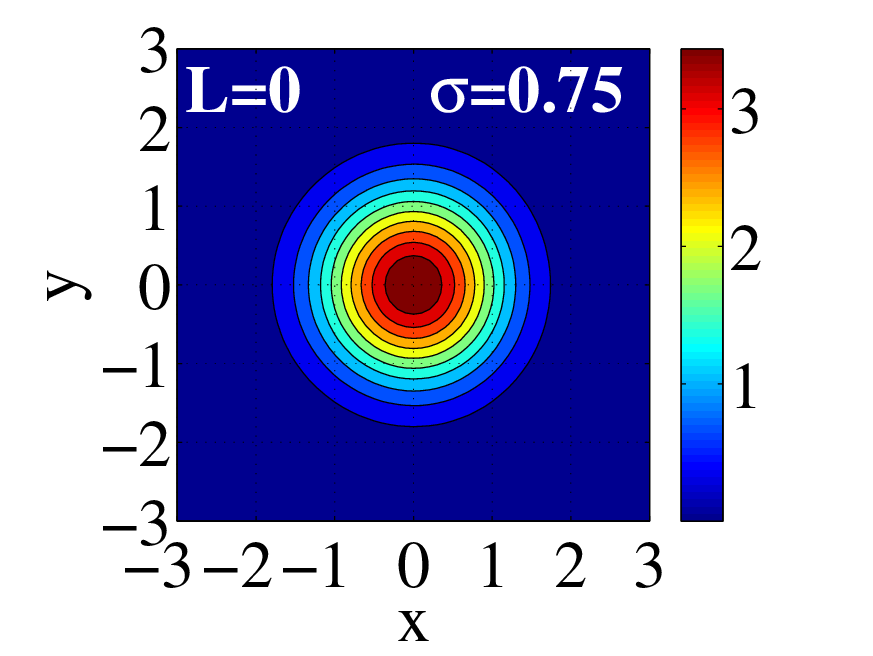}\label{fig:l0s75}}
	\subfigure[]{
\begin{tikzpicture}[spy using outlines= {magnification=10.0, circle, size=1.50cm, black, connect spies}]	\node (n1) at (0,0)
    {\includegraphics[width=0.4\linewidth]{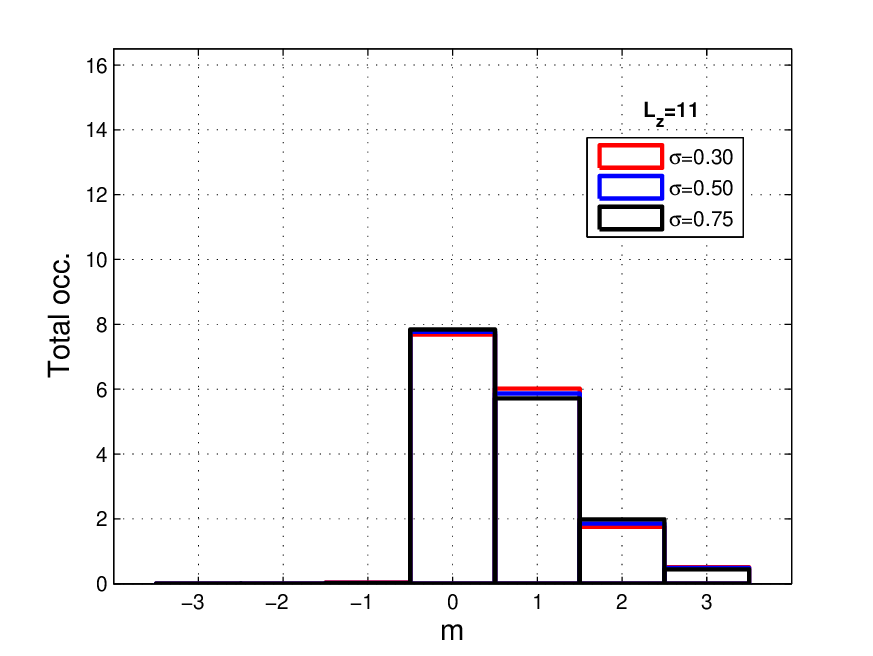}};
\spy on (0.2,-0.01) in node at (-1.20,-0.20);
\end{tikzpicture}
		\vspace{-0.25in}	\hspace{-3.mm}
		\includegraphics[width=0.27\linewidth]{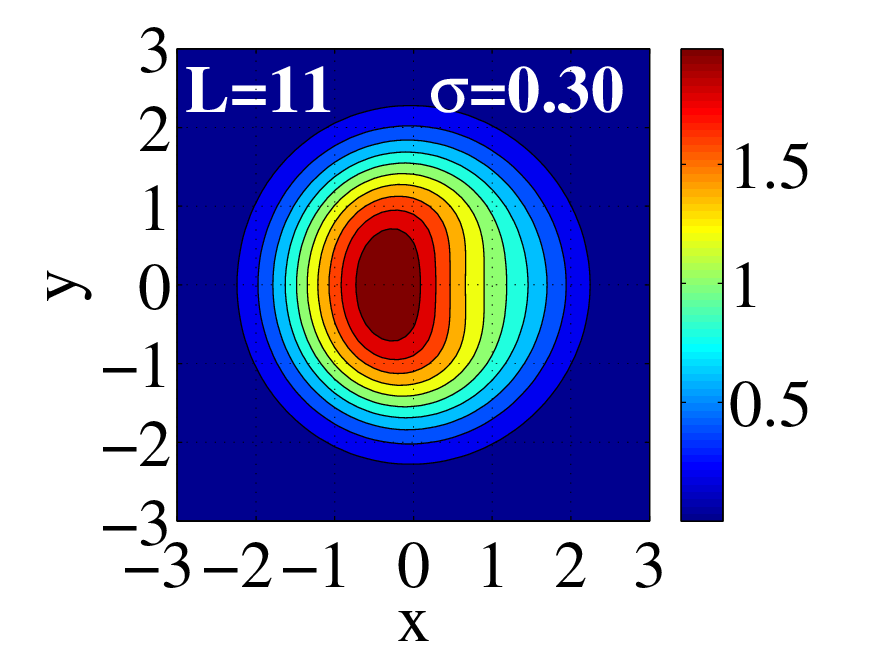}\label{fig:l11s3}
		\hspace{-4mm}
		\includegraphics[width=0.27\linewidth]{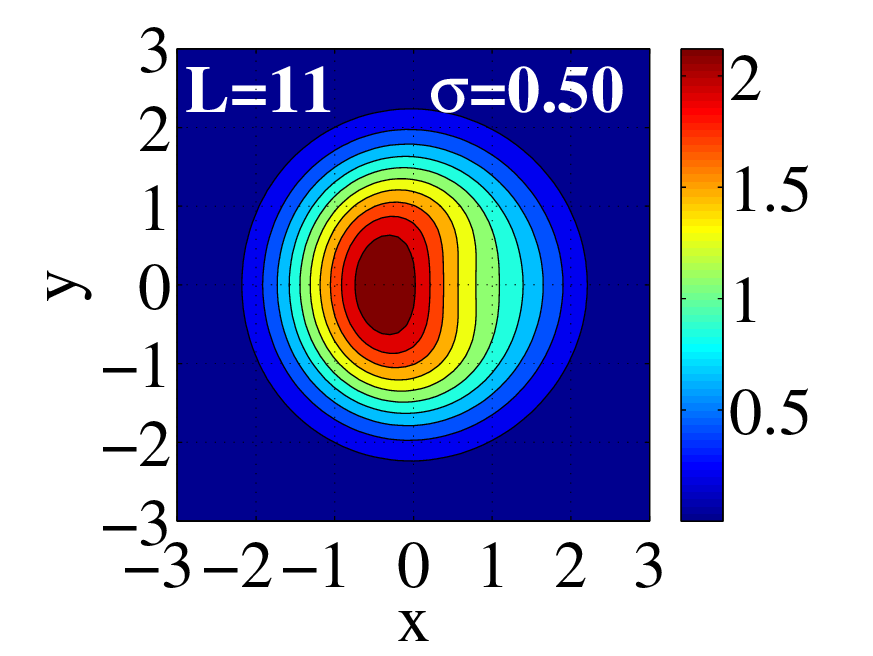}\label{fig:l11s5}
		\vspace{-0.25in}	\hspace{-4mm}
		\includegraphics[width=0.27\linewidth]{{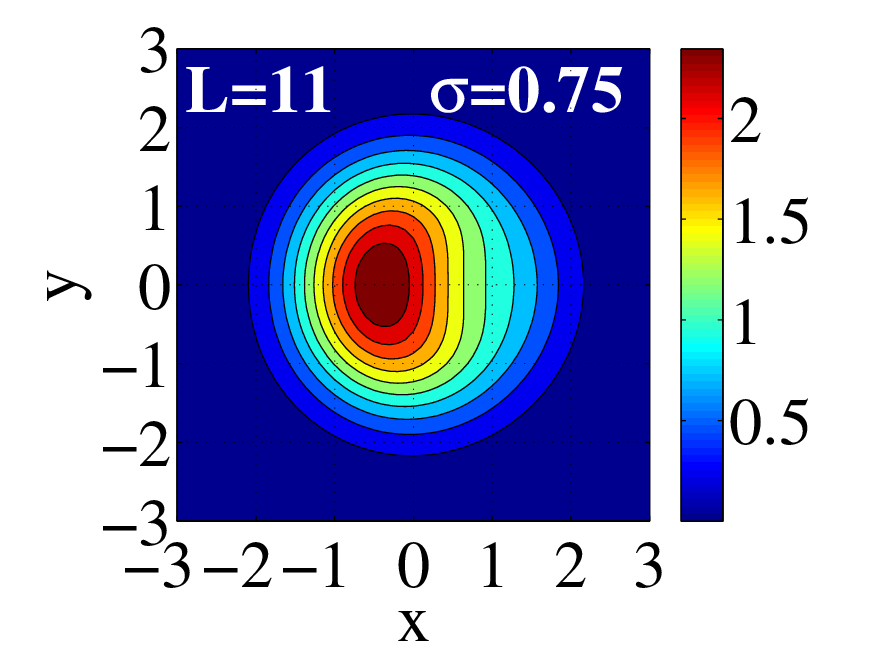}}\label{fig:l11s75}}
\subfigure[]{
   \begin{tikzpicture}[spy using outlines={magnification=10.0, circle, size=1.50cm, red, connect spies}]\node (n1) at (0,0)
   {\includegraphics[width=0.4\linewidth]{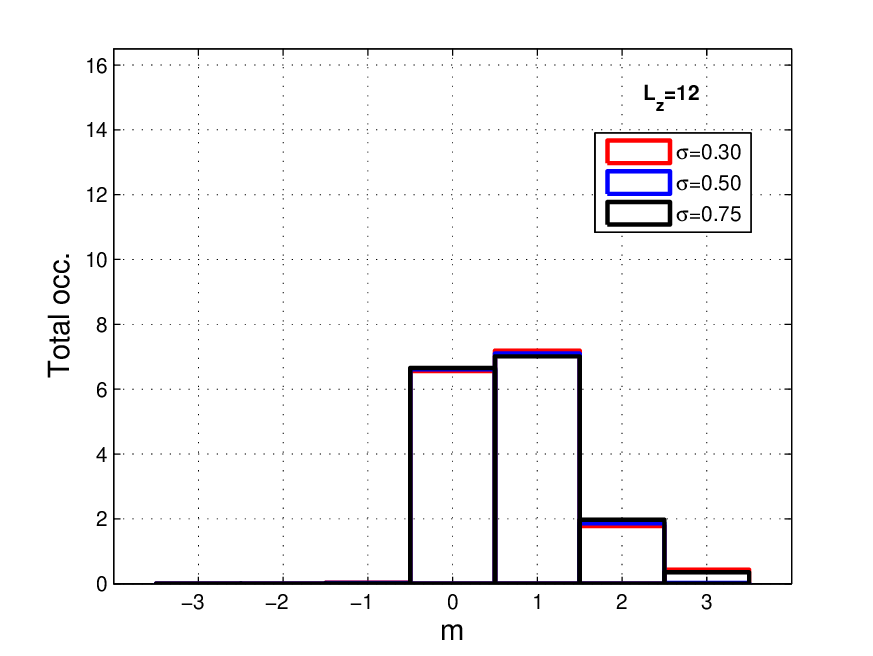}};
  \spy on (0.0,-0.290) in node at (-1.20,-0.50);
\end{tikzpicture}
		\vspace{-0.25in}	\hspace{-3.4mm}
		\includegraphics[width=0.27\linewidth]{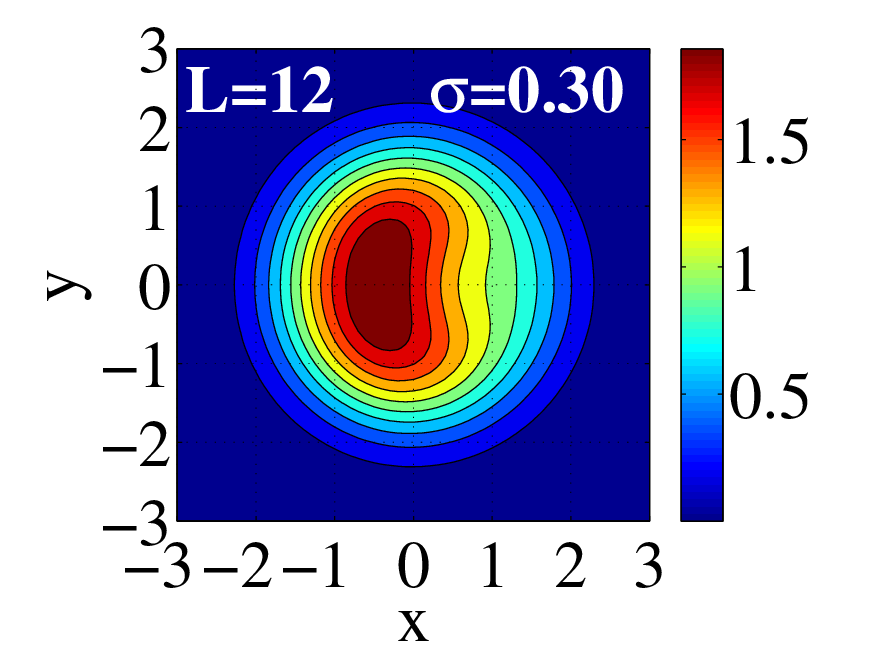}\label{fig:l12s3}
		\vspace{-0.25in}	\hspace{-5mm}
		\includegraphics[width=0.27\linewidth]{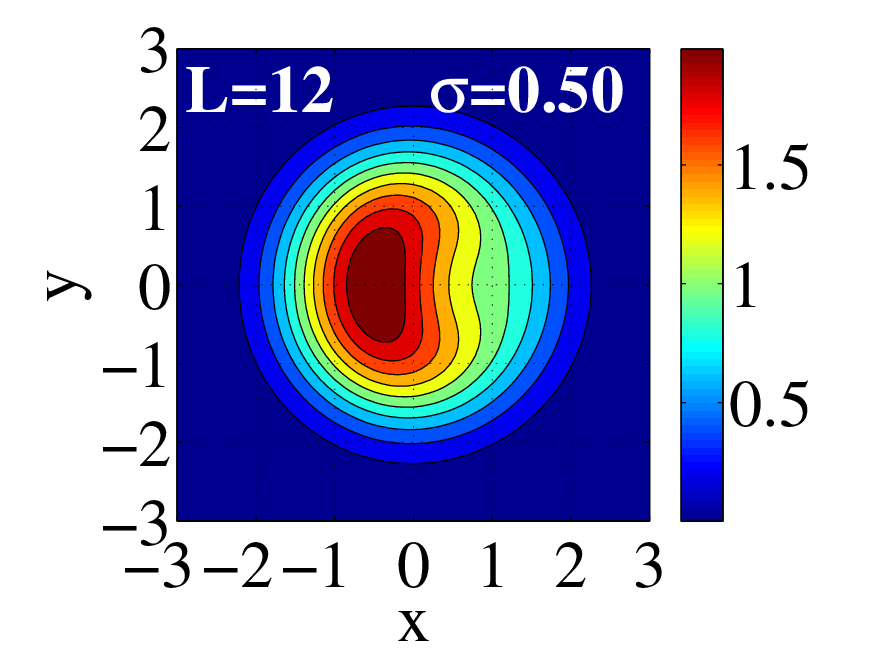}\label{fig:l12s5}
		\vspace{-0.25in}	\hspace{-5mm}
		\includegraphics[width=0.27\linewidth]{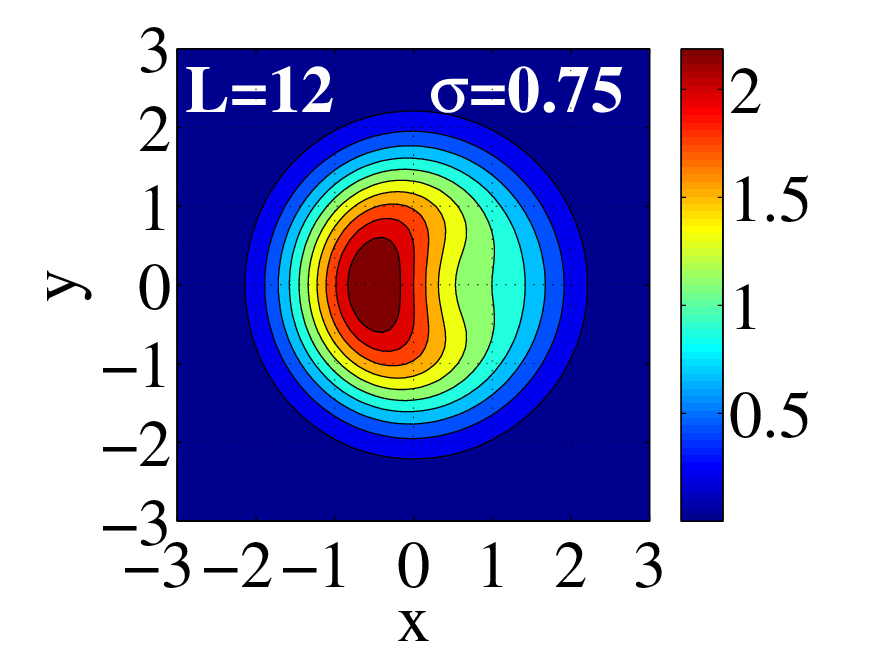}\label{fig:l12s75}}
\subfigure[]{
  \begin{tikzpicture}[spy using outlines={magnification=10.0, circle, size=1.50cm, black, connect spies}]	\node (n1) at (0,0)
 {\includegraphics[width=0.4\linewidth]{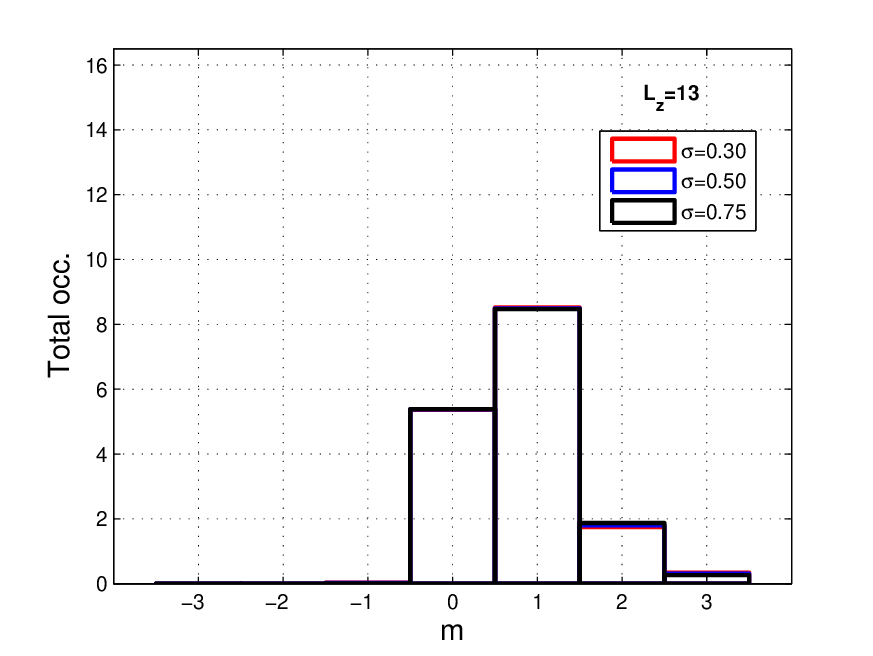}};
     \spy on (0.0,-0.570) in node at (-1.0,0.50);
\end{tikzpicture}
   		\vspace{-0.25in}	\hspace{-3.0mm}
	\includegraphics[width=0.27\linewidth]{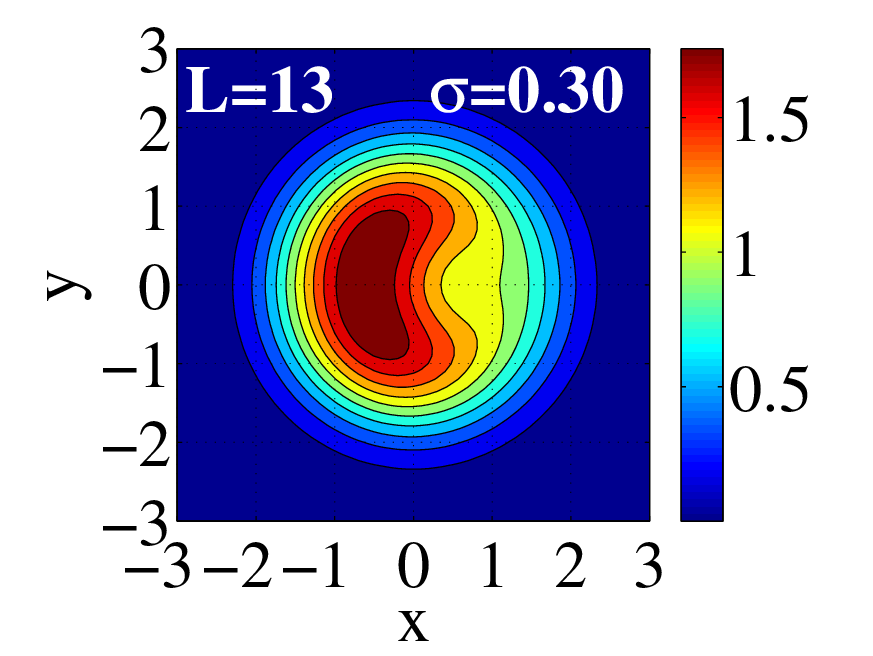}\label{fig:l13s3}
		\vspace{-0.25in}	\hspace{-5mm}
		\includegraphics[width=0.27\linewidth]{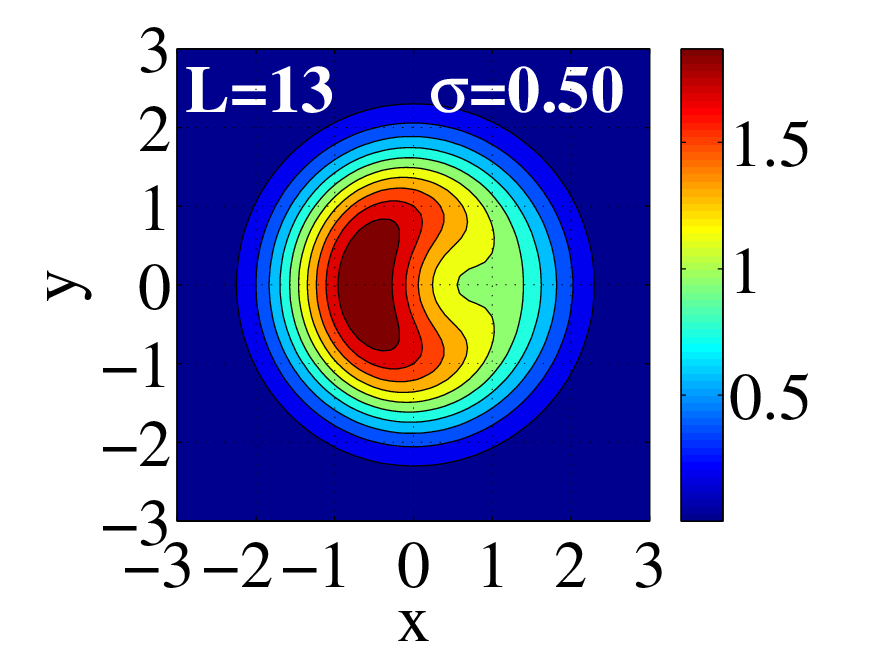}\label{fig:l13s5}
		\vspace{-0.25in}	\hspace{-5mm}
		\includegraphics[width=0.27\linewidth]{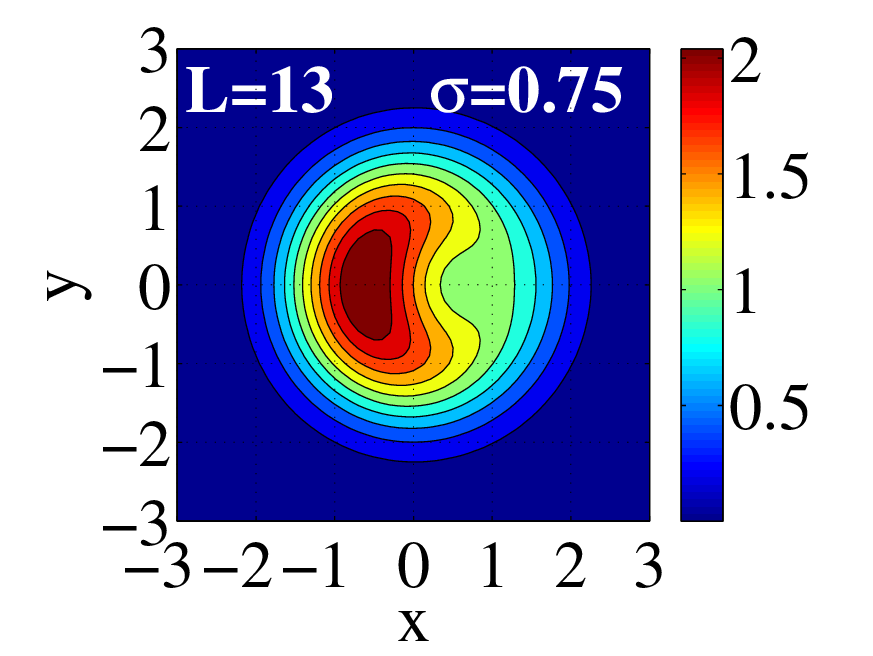}\label{fig:l13s75}}
	\end{figure*}
\subsection{Single-vortex state}
\indent To study the nucleation of the single-vortex state $L_{z}=N=16$ in the regime $0\le L_{z}\le N$ 
for different values of interaction range $\sigma=0.30,0.50,0.75$, we present contour plots (isodensity lines) for conditional probability distribution (CPD) along with the macroscopic occupation ($N\lambda_{\mu}$) of one-particle angular momentum state $m^{\mu}$ of OPRDM as shown in Fig.~\ref{fig:b4_vortx}. 
For the non-rotating $L_{z}=0$ state in Fig.~\ref{fig:l0s3}, the macroscopic occupation of maximally occupied one-particle angular momentum $m=0$ state of OPRDM increases with increase in interaction range $\sigma$. Further, In CPD plots, the density profile is symmetric Gaussian which becomes narrower in width and increases in height with increase in interaction range $\sigma$.
\\ 
\indent For $L_{z}=11$ angular momentum state, we observe in Fig.~\ref{fig:l11s3}, that the macroscopic occupation of $m=0$ state increases while that of $m=1$ state decreases with increase in interaction range $\sigma$. The isodensity lines (contour) of the nucleating vortex are symmetric about the reference line and the reference point $(x_{0},0)$.
In the CPD plot, the isodensity lines (contours) shrink along the reference line $(x_{0},0)$ and the probability density increases at the peak with increase in interaction range $\sigma$. 
\\
\indent For the angular momentum $L_{z}=12$ state, we observe a maximum in the von Neumann entropry $S_{1}$ as shown in Fig.~\ref{fig:S_1} and a minimum in the largest condensate fraction $\lambda_{1}$ of OPRDM as shown in Fig.~\ref{fig:condfra1}.
\\
\indent For $L_{z}=12$ angular momentum state in Fig.~\ref{fig:l12s3}, the macroscopic occupation of one-particle angular momentum $m=1$ state of OPRDM becomes greater than the macroscopic occupation of one-particle angular momentum $m=0$ state 
and the macroscopic occupation of $m=0$ state increases while that of $m=1$ state decreases with increase in interaction range $\sigma$. Further, with increase in interaction range $\sigma$, the isodensity lines (contours) at the peak of the density profile gets closer and merge into each other to form a sharper peak. On the other hand, at the core of the nucleating vortex, the isodensity lines move away from each other and the dip in the density profile flattens.
For $L_{z}=13$ state as shown in Fig.~\ref{fig:l13s3}, the macroscopic occupation of both $m=0,1$ states become comparable with increase in $\sigma$, whereas in CPD, similar behavior is exhibited to $L_{z}=12$ state.\\ 
\indent The angular momentum $L_{z}=N=16$ state appears as 
the single-vortex state on the $L_{z}-\Omega$ stability graph in Fig.~\ref{fig:stbn16}. The macroscopic occupation of maximally occupied one-particle angular momentum $m=1$ state of OPRDM increases with increase in interaction range $\sigma$ as shown in Fig.~\ref{fig:l16s3}. The appearance of single-vortex state corresponds to a maxima in the largest condensate fraction $\lambda_{1}$ of OPRDM and a minima in the von Neumann entropy $S_{1}\left(L_{z},\sigma\right)$  as shown in Figs.~\ref{fig:condfra1} and \ref{fig:S_1}, respectively. As the reference point is chosen at $(x_{0},0)$~\cite{Imran_EJPD23,Xia01_ro,Riemann02}, the density peak appears as crescent symmetric about the reference line and facing towards the reference point $({x}_{0},0)$. With increase in interaction range $\sigma$, the crescent opens further along the reference line and shrink in size as the peak becomes narrower with increase in height as shown on the adjoining colour-bar. The single-vortex state emerges through the breaking of Gaussian-symmetry of the density profile as the total angular momentum increases in the regime $0\leq L_{z}\leq N$.
\\ 
\indent With further increase in angular momentum in the regime $N< L_{z}\leq 4N$, multiple vortices appear~\cite{Xia01_ro,jmf_gunn00,hengfan09}. These vortices arrange themselves in geometries like, a diagonally opposite, 
an equilateral triangle, a square and a pentagon for two, three, four and five vortices, respectively, symmertic about the reference line~\cite{Imran_EJPD23,Rokhsar_99,Madison_00}. These vortex configurations are stable due to the effective repulsion experienced by the vortices rotating in the same direction carrying the same `charge'~\cite{Rokhsar_99}.
\subsection{Two-vortex states: merging of vortex-cores \label{sec:twovortx26}}
\begin{figure}
\subfigure[$U_{0}=0.02171$]{
\includegraphics[width=0.90\linewidth]{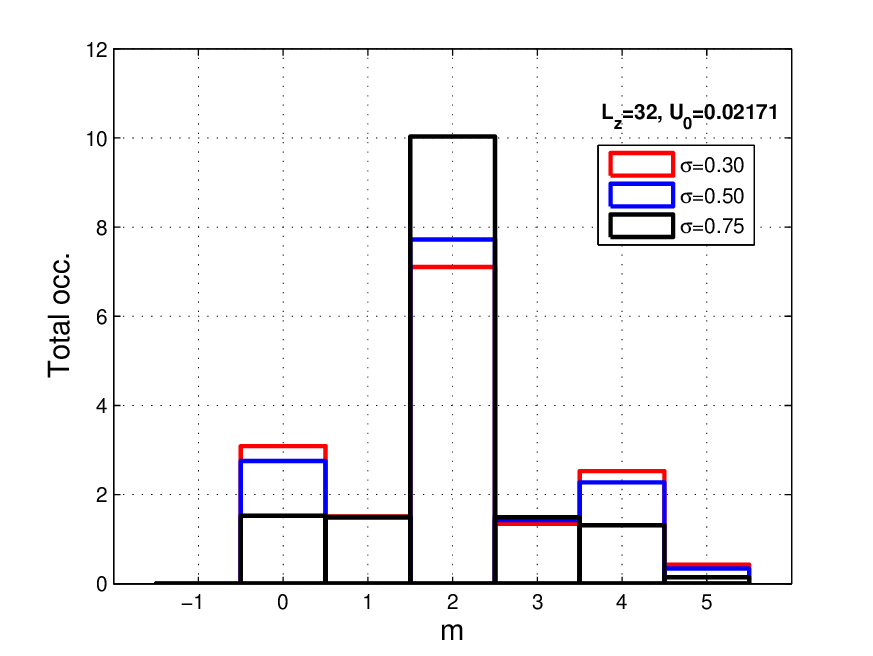}\label{fig:occl32g09s}}
\subfigure[$U_{0}=2.171$]{
\includegraphics[width=0.90\linewidth]{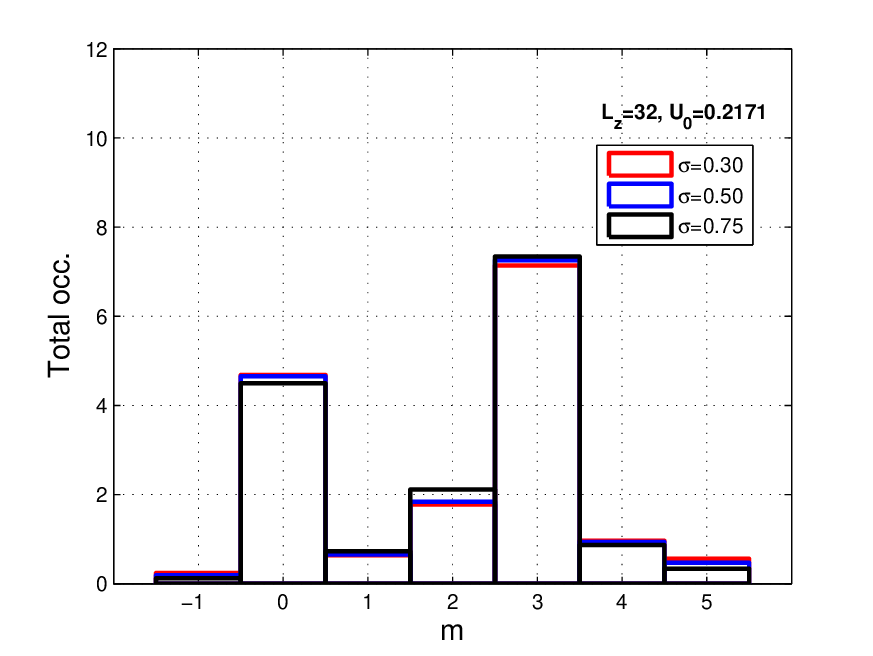}\label{fig:occl32g9s}}
\caption{(Color online) For the angular momentum $L_{z}=32$, the macroscopic occupations of one-particle angular momentum states of OPRDM {\it vs} one-particle angular momentum states $m$. (a) with increase in interaction range $\sigma$, the macroscopic occupation of maximally occupied one-particle angular momentum state $m=2$ of OPRDM increases. (b) with increase in interaction strength $U_{0}$, the maximally occupied one-particle angular momentum state changes to $m+1=3$.\label{fig:g2m}}
\end{figure}

\indent 
To study the nucleation of two-vortex state in the angular momentum regime $N\leq L_{z} \leq 2N$, we present CPD plots along with the macroscopic occupations ($N\lambda_{\mu}$) of one-particle angular momentum state $m^{\mu}$ of OPRDM.  
Beyond the single-vortex $L_{z}=N=16$ state, the macroscopic occupation of one-particle angular momentum state $m=1$ of OPRDM begins to decrease, whereas that of state $m=2$
begins to increase, as shown in Fig.~\ref{fig:lz17} for angular momentum $L_{z}=17$ state. 
Further, the macroscopic occupation of one-particle angular momentum state $m=1$ increases with increase in interaction range $\sigma$. In CPD, the crescent corresponding to density peak shrinks around the core at $(0,0)$ with increase in interaction range $\sigma$.
\\
\indent The two-vortex state begins to nucleate at angular momentum $L_{z}=26$. It is observed that for the angular momentum $L_{z}=26$ state, the macroscopic occupation of the one-particle angular momentum $m=2$ state first decrease and then increases with increase in interaction range $\sigma$, as shown in Fig.~\ref{fig:sp3_L26}. However, for the angular momentum $L_{z}=28,30,32$ states, the macroscopic occupation of the one-particle angular momentum $m=2$ state of OPRDM increases with increase in interaction range $\sigma$, as shown in Fig.~\ref{fig:sp3_L28}-\ref{fig:sp3L32A}. It is to be noted that the angular momentum $L_{z}=26,28$ are the meta-stable sates whereas $L_{z}=30,32$ are the stable states on $L_{z}-\Omega$ stability graph in Fig.~\ref{fig:stbn16}.
In the CPD plots, the contours of vortex become more pronounced with increase in interaction range $\sigma$.
\\
\indent For the two-vortex state with angular momentum $L_{z}=2N$, the core of vortices are placed symmetrically about the center of the trap $(0,0)$ on the reference line. With increase in interaction range $\sigma$, we observe merging of the cores of two-vortex $L_{z}=2N$ state along the reference line~\cite{GIMenon96,Cooper01} forming effectively a single-vortex state for $N=10,12,16$ bosons in the beyond lowest Landau level approximation as shown in Fig.~\ref{fig:spl32}-\ref{fig:spN16l32} and for $N=20,24$ bosons in the lowest Landau level approximation as shown in Fig.~\ref{fig:spl40N20}. For the vortex states comprising of more than two vortices, we do not observe merging of the vortex cores. This may be due to the presence of the density peak at the centre of the trap which prevents the cores of multi-vortex states (states with more than two vortices) from merging. 
For the total angular momentum $L_{z}=2N=32$ state, the macroscopic occupation of one-particle angular momentum $m=2$ state of OPRDM increases with increase in interaction range $\sigma$ as shown in Fig.~\ref{fig:spN16l32} which may lead to the merger of cores of two-vortex state due to quantum fluctuation.
\\
\indent It is further observed that for $L_{z}=2N=32$ state
in the weakly to moderately interacting regime with $U_{0}=0.02171$ and $U_{0}=0.2171$ as shown in  Fig.~\ref{fig:occl32g09s} and Fig.~\ref{fig:occl32g9s} respectively, there is an increase in the macroscopic occupation of the maximally occupied one-particle angular momentum $m=2$ state of OPRDM with increase in $\sigma$. Whereas, in the strongly interacting regime with $U_{0}=2.171$, instead of $m=2$ state, the macroscopic occupation of $m=3$ state becomes the maximally occupied one-particle angular momentum state of OPRDM
as shown in Fig.~\ref{fig:occl32g9s}.
\\
\indent Thus, for a particular angular momentum $L_{z}$ state, an increase in interaction strength $U_{0}$ beyond a certain value changes the macroscopic occupation of maximally occupied one-particle angular momentum state $m$ of OPRDM to $m'=m+1$ state.
\begin{figure*}
	\hspace{-5.0mm}
	\subfigure[]{\includegraphics[width=0.35\linewidth]{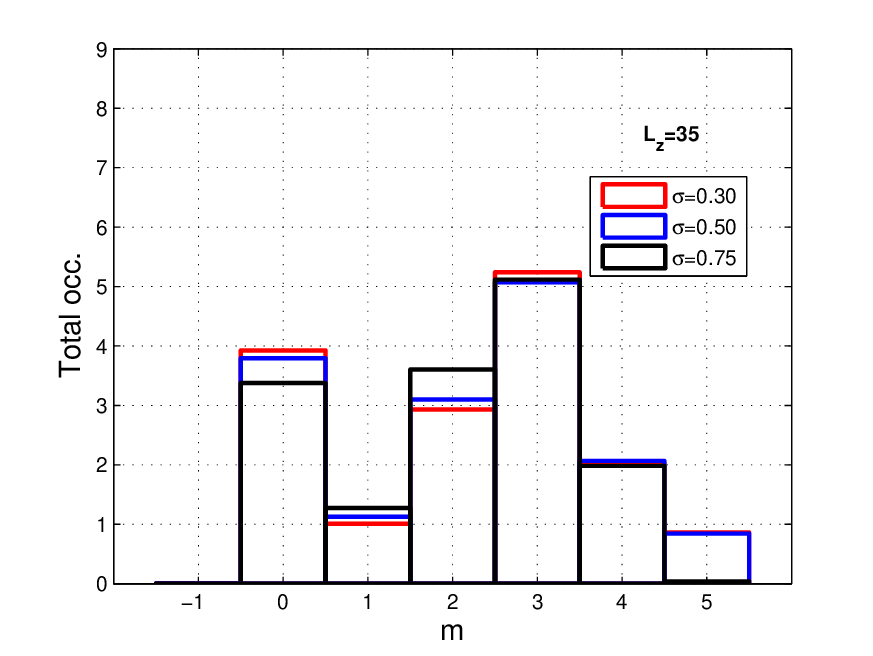}\label{fig:sp3_L35}
		\hspace{-6.0mm}
		\includegraphics[width=0.3\linewidth]{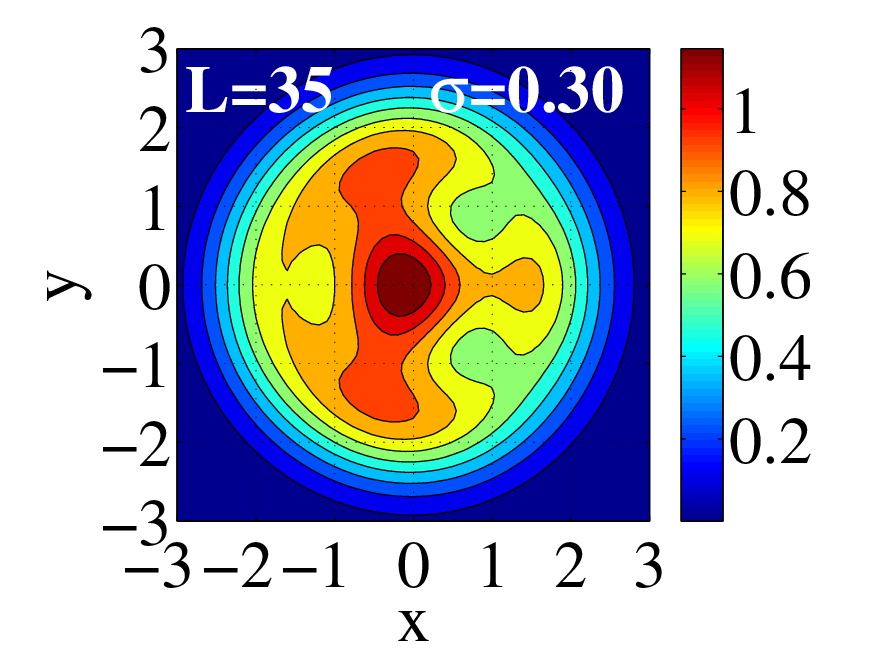}\label{fig:occL35s}
		\hspace{-5.0mm}
		\includegraphics[width=0.3\linewidth]{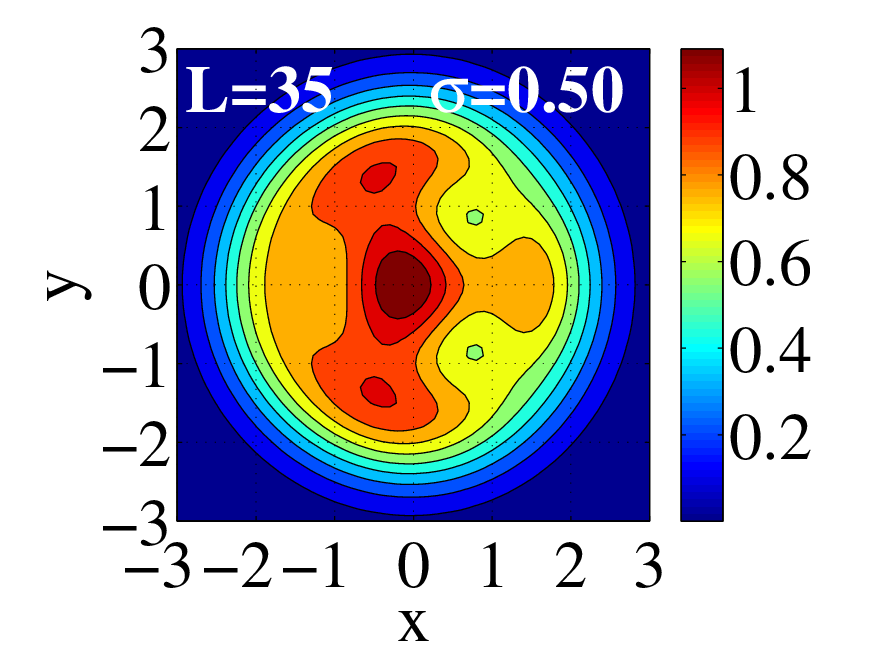}\label{fig:sp5_L35}
		\hspace{-5.0mm}
		\includegraphics[width=0.3\linewidth]{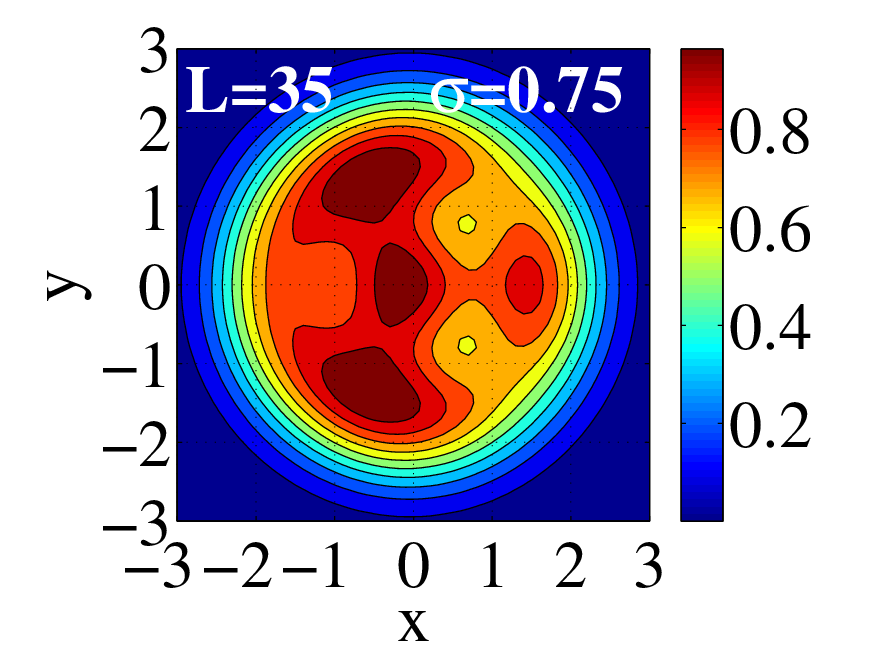}\label{fig:sp75_L35}}
	\subfigure[]{ \hspace{-5.0mm}
    \includegraphics[width=0.35\linewidth]{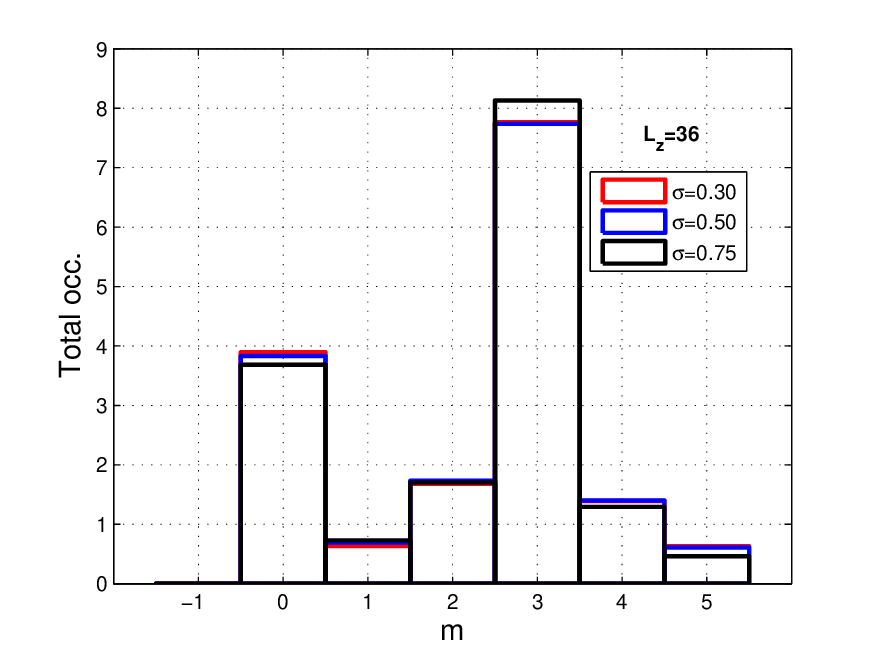}
		\hspace{-6.0mm}
		\includegraphics[width=0.3\linewidth]{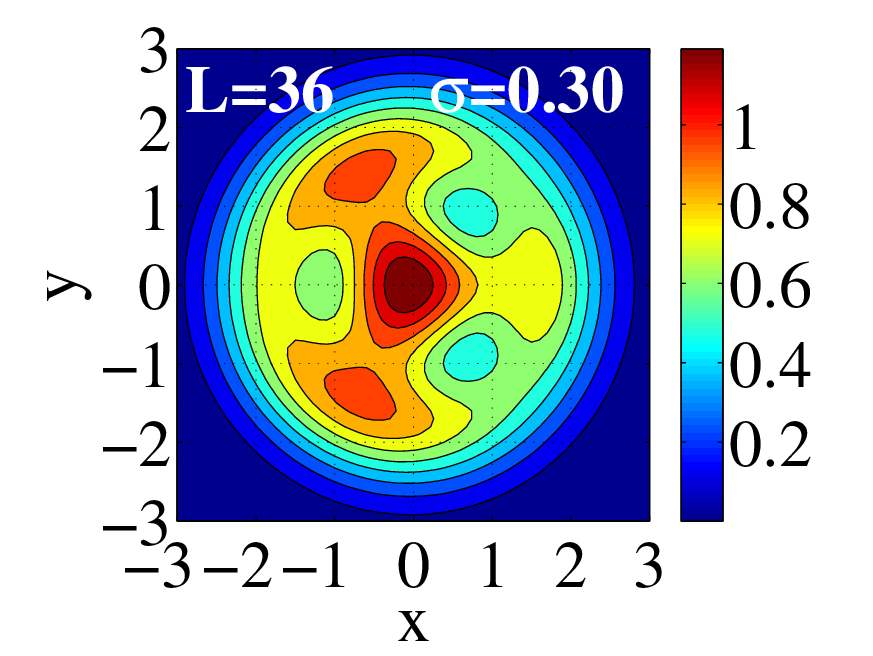}\label{fig:sp3_L36}
		\hspace{-5.0mm}
		\includegraphics[width=0.3\linewidth]{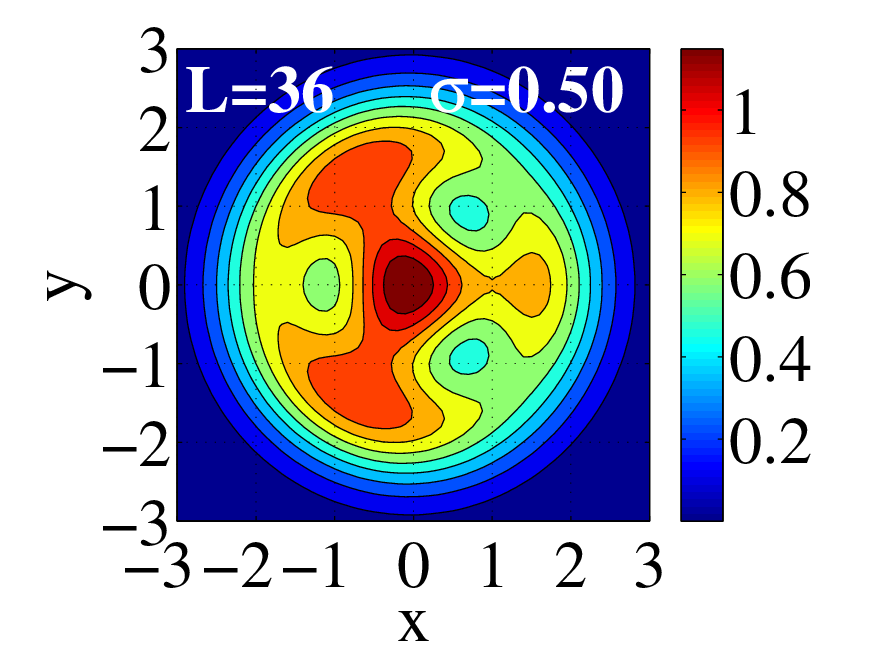}\label{fig:sp5_L36}
		\hspace{-5.0mm}
		\includegraphics[width=0.3\linewidth]{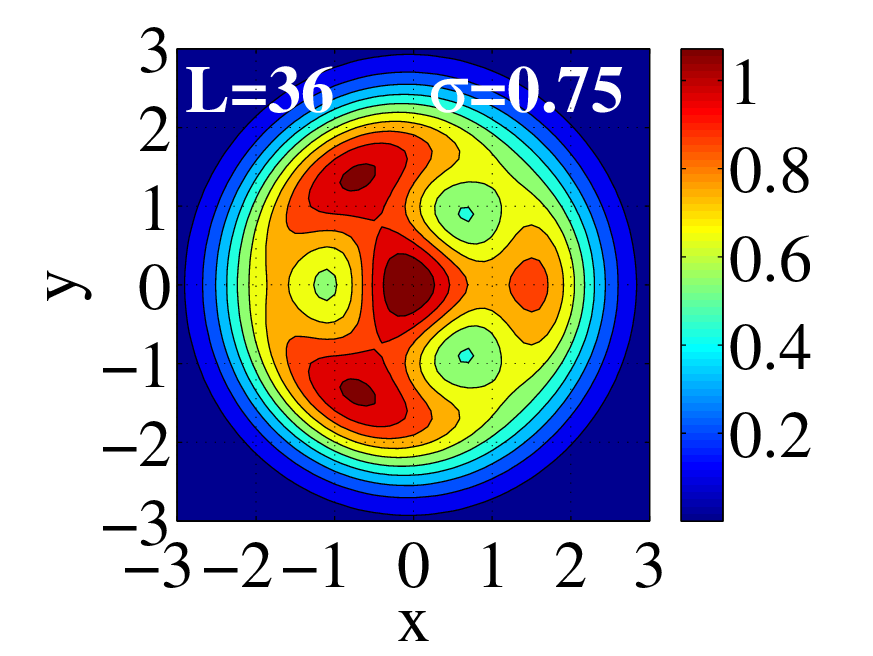}\label{fig:sp75_L36}}
	\subfigure[]{\hspace{-5.0mm}
		\includegraphics[width=0.35\linewidth]{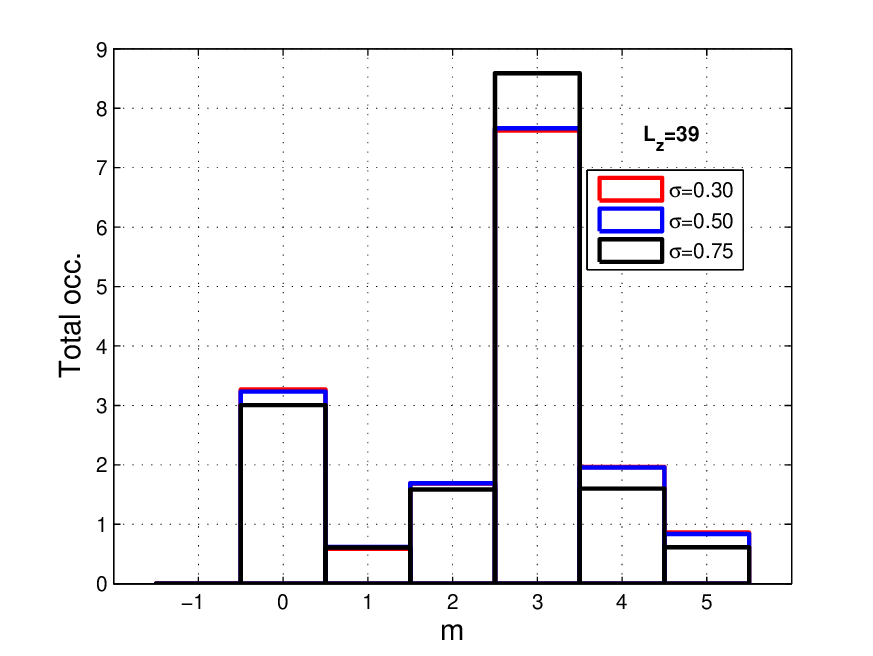}
		\hspace{-6.0mm}
		\includegraphics[width=0.3\linewidth]{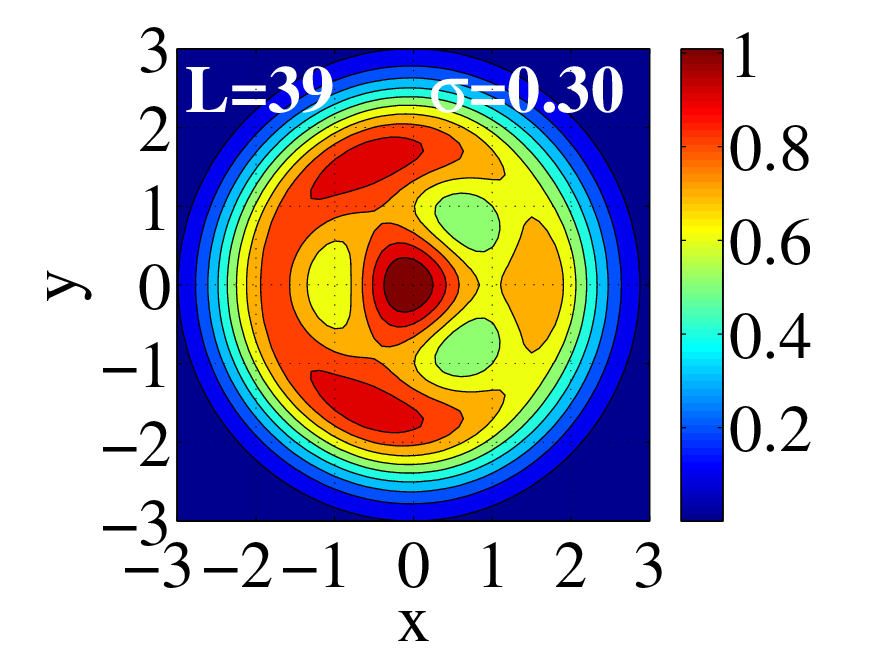}\label{fig:sp3_L39}
		\hspace{-5.0mm}
		\includegraphics[width=0.3\linewidth]{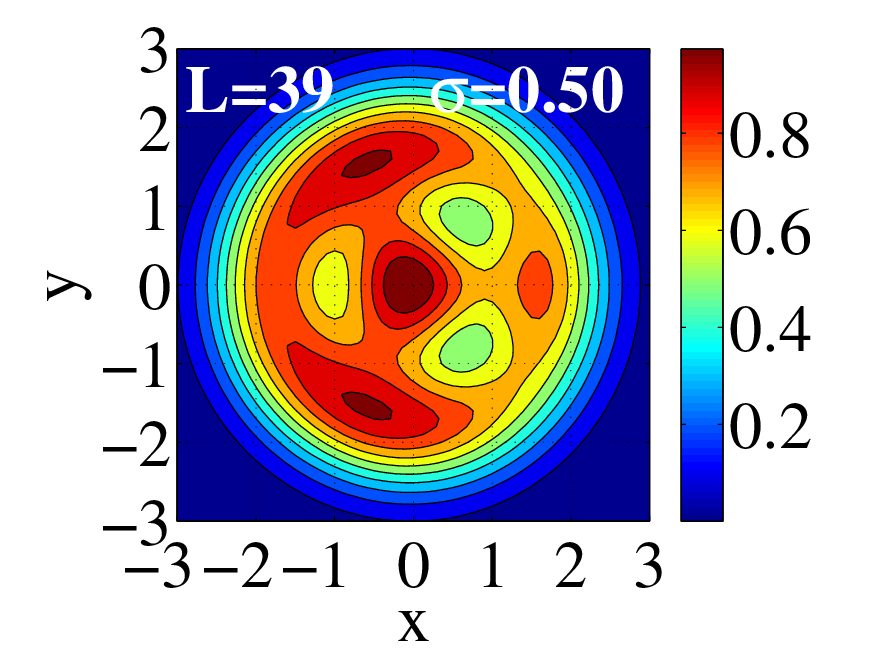}\label{fig:sp5_L39}
		\hspace{-5.0mm}
		\includegraphics[width=0.3\linewidth]{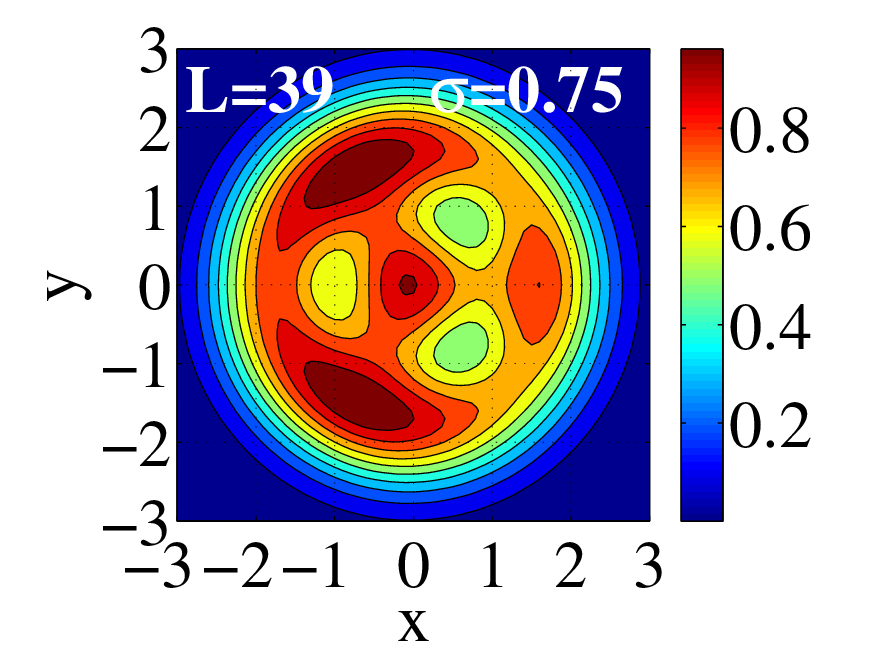}\label{fig:sp75_L39}}
	\centering
\caption{\label{fig:tvs}(Color online) The macroscopic occupations of one-particle angular momentum states of OPRDM {\it vs} one-particle angular momentum $m$ and the corresponding CPD plots depicting the three-vortex state for $N=16$ with interaction strength ${U_{0}}=0.2171$ and interaction range $\sigma=0.30,0.50,0.75$. The reference point has been chosen at $\mathbf{r}_{0}=(1.5,0)$. (a) for the meta-stable three-vortex state $L_{z}=35$, with increase in interaction range $\sigma$, the macroscopic occupation of maximally occupied one-particle angular momentum state $m=3$ of OPRDM decreases, whereas, in the CPD plots, the values of probability density decreases at the central peak and increases at the off-center peaks due to redistribution with increase in interaction range $\sigma$. In (b)-(c), for stable three-vortex $L_{z}=36,39$ states, with increase in interaction range $\sigma$, the macroscopic occupation of maximally occupied one-particle angular momentum state $m=3$ of OPRDM increases.}
\end{figure*}
\begin{figure*}
\hspace{-5.0mm}
	\subfigure[]{\hspace{-5.0mm}
    \begin{tikzpicture}[spy using outlines={magnification=10.0, circle, size=1.50cm, black, connect spies}]\node (n1) at (0,0)
      {\includegraphics[width=0.35\linewidth]{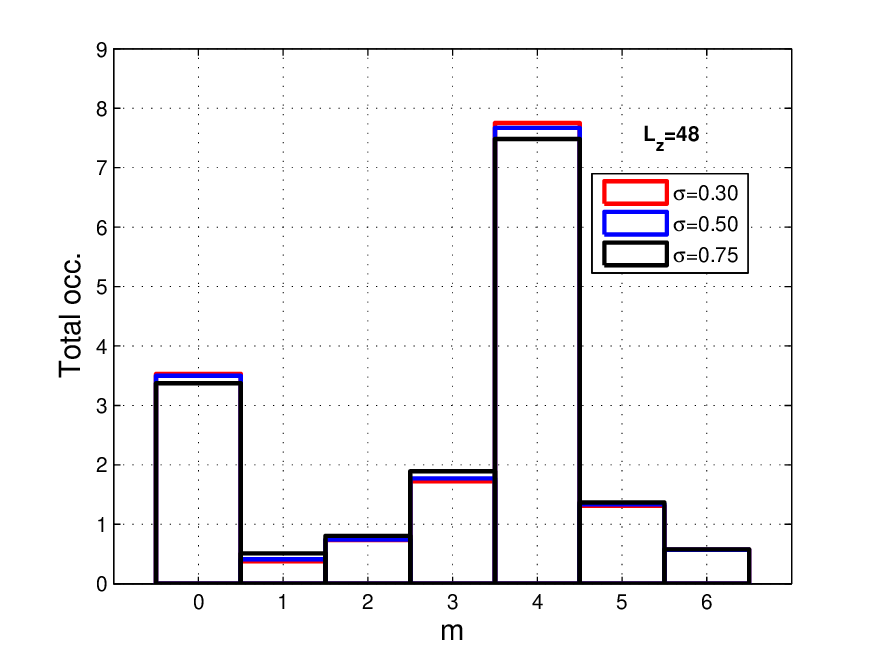}};
	\spy on (0.50,1.20) in node at (-1.0,0.50);
	\end{tikzpicture}
		\hspace{-6.0mm}
		\includegraphics[width=0.3\linewidth]{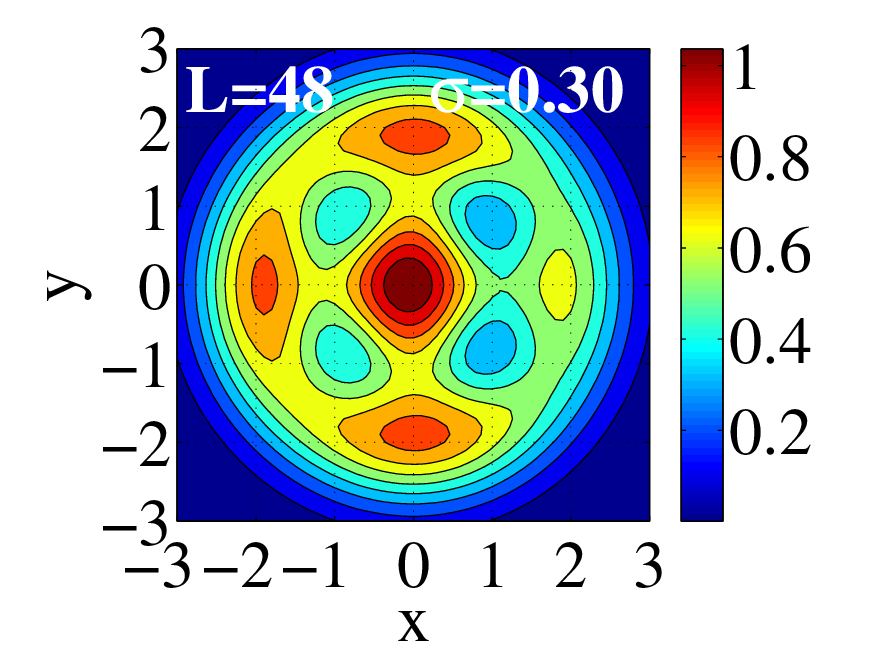}\label{fig:sp3_L48}
        \hspace{-5.0mm}
        \includegraphics[width=0.3\linewidth]{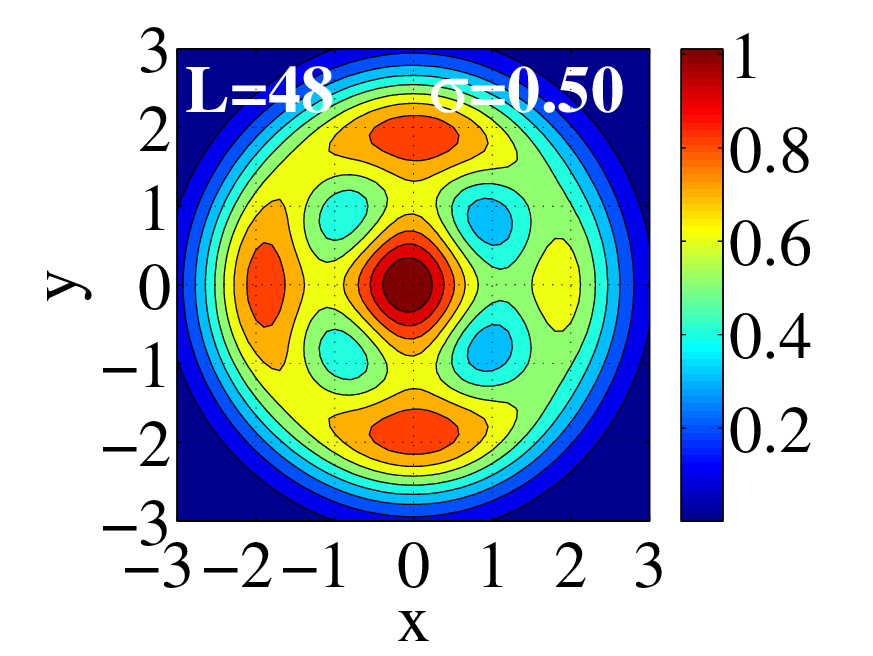}\label{fig:sp5_L48}
        \hspace{-5.0mm}
       \includegraphics[width=0.3\linewidth]{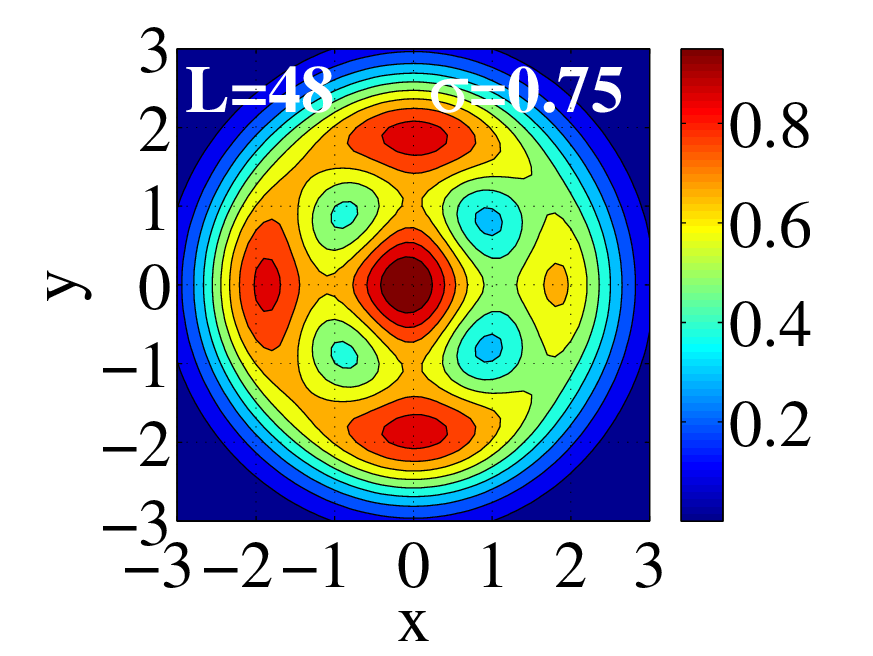}\label{fig:sp75_L48}}
  \subfigure[]{ \hspace{-5.0mm}
\begin{tikzpicture}[spy using outlines={magnification=10.0, circle, size=1.50cm, black, connect spies}]\node (n1) at (0,0)
     {\includegraphics[width=0.35\linewidth]{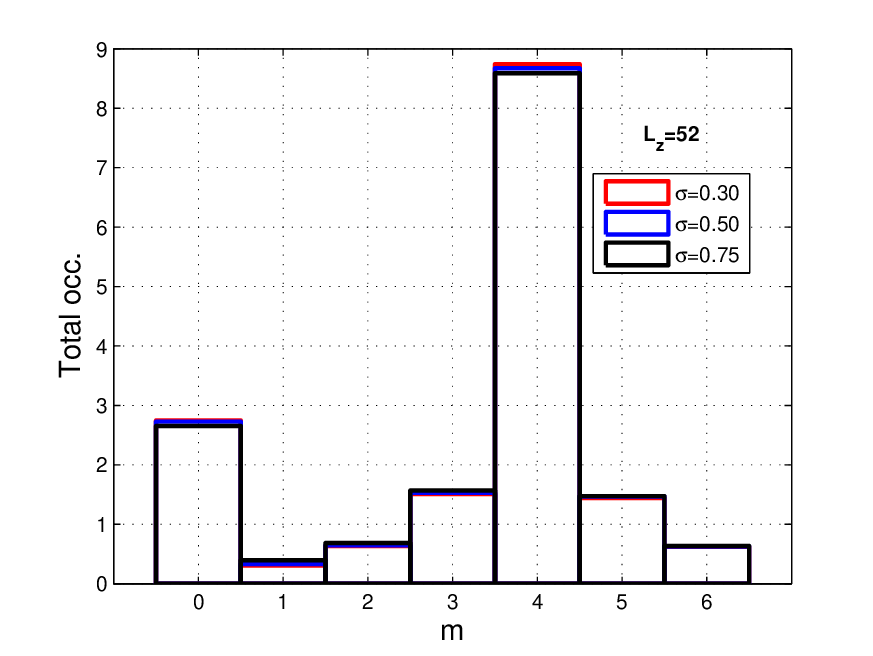}};
	\spy on (0.50,1.560) in node at (-1.0,0.50);
	\end{tikzpicture}
		\hspace{-6.50mm}
		\includegraphics[width=0.3\linewidth]{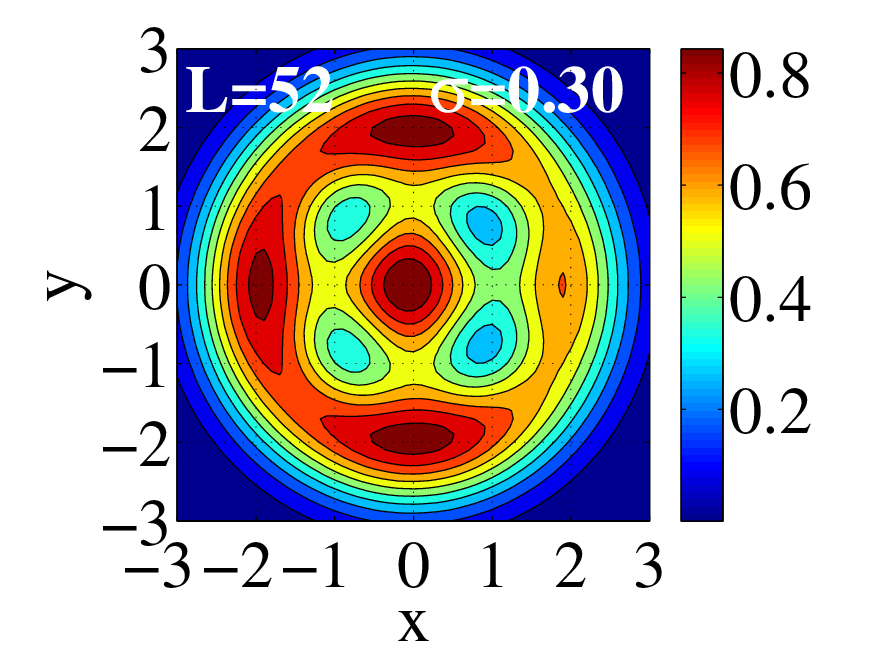}\label{fig:sp3_L52}
		\hspace{-5.0mm}
		\includegraphics[width=0.3\linewidth]{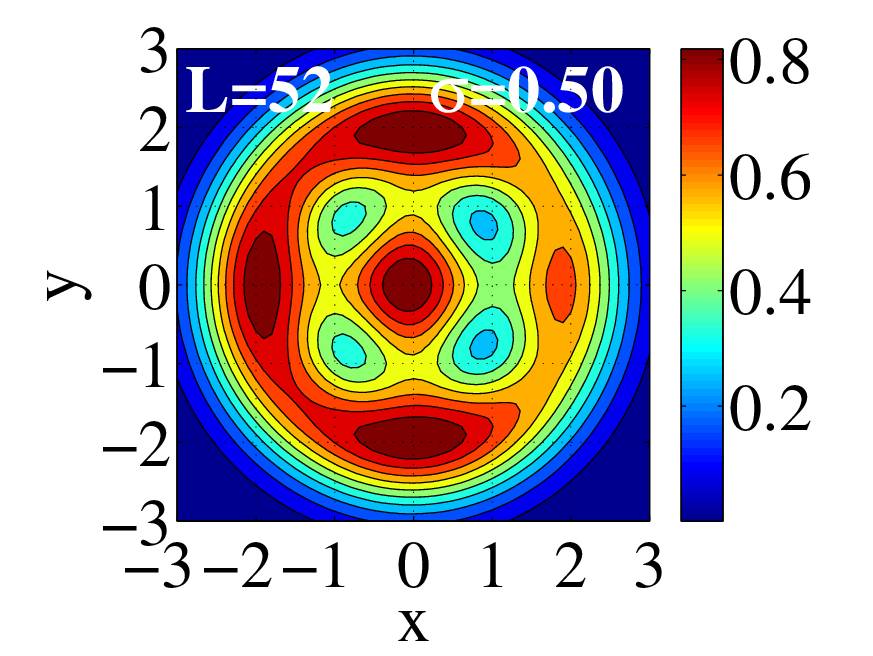}\label{fig:sp5_L52}
		\hspace{-5.0mm}
		\includegraphics[width=0.3\linewidth]{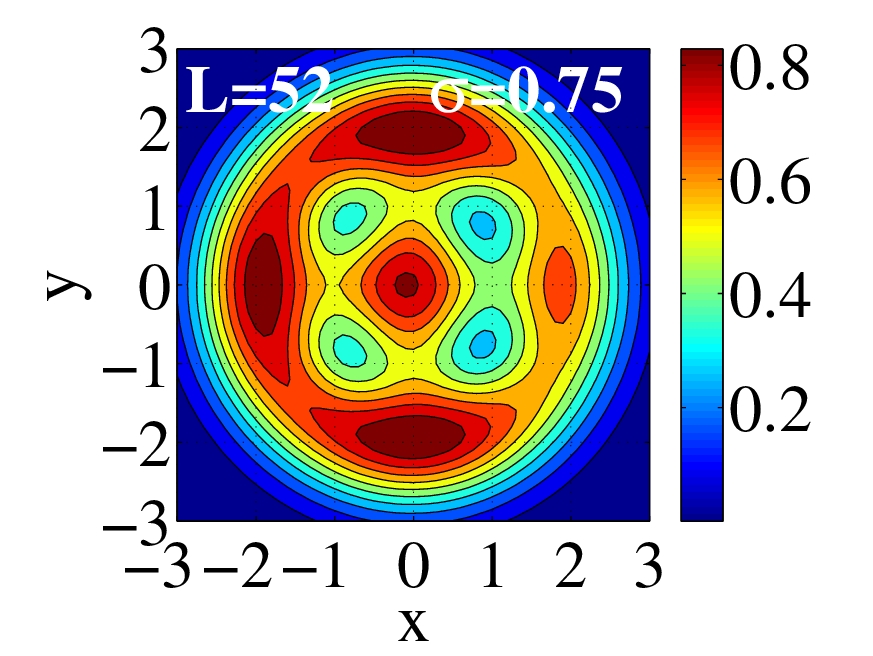}\label{fig:sp75_L52}}
\caption{\label{fig:fvs}(Color online) The occupations of one-particle angular momentum state of OPRDM {\it vs} one-particle angular momentum $m$ and the corresponding CPD plots depicting four-vortex state for $N=16$ bosons with interaction strength $U_{0}=0.2171$ and interaction range $\sigma=0.30,0.50,0.75$. The reference point has been chosen at $\mathbf{r}_{0}=(1.5,0)$. (a)-(b) for the meta-stable four-vortex $L_{z}=48,52$ states, with increase in interaction range $\sigma$, the macroscopic occupation of maximally occupied one-particle angular momentum state $m=4$ of OPRDM decreases whereas, in the CPD plots, the values of probability density decreases at the central peak and increases at the off-center peaks due to redistribution with increase in interaction range $\sigma$.}
\end{figure*}
\begin{figure*}
\subfigure[]{\hspace{-5.0mm}
{\includegraphics[width=0.35\linewidth]{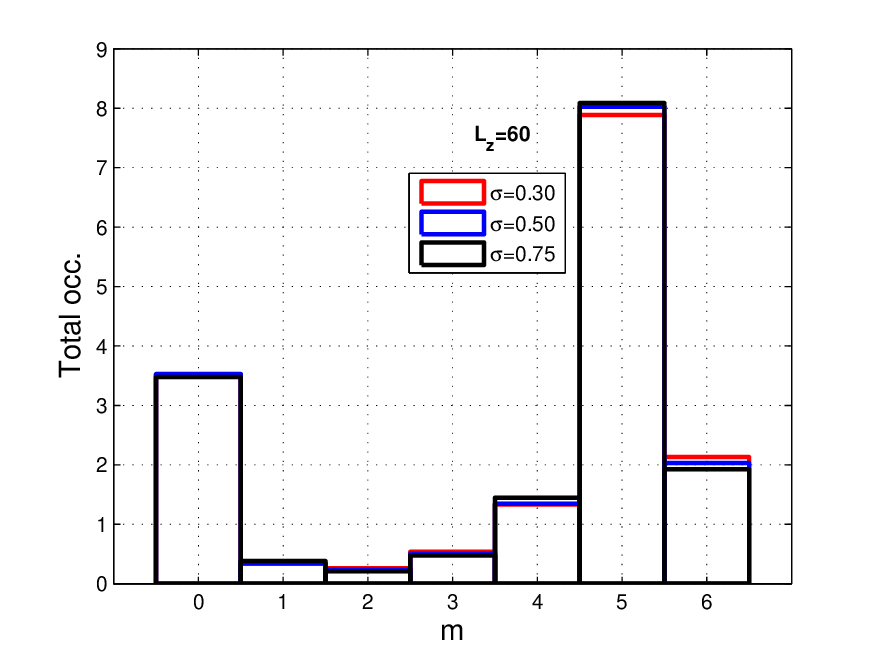}
\hspace{-6.0mm}
\includegraphics[width=0.3\linewidth]{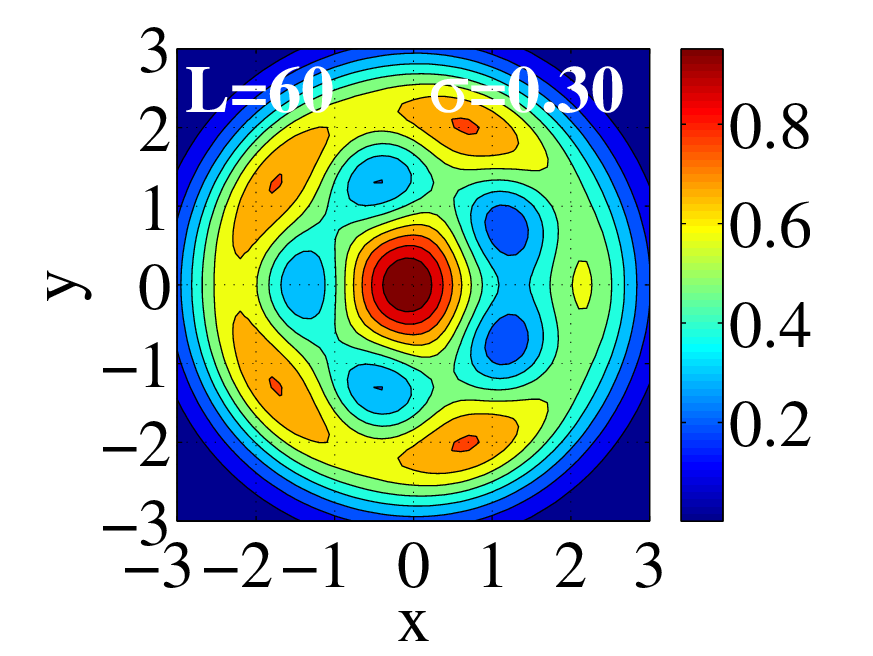}\label{fig:sp3_L60}
\hspace{-5.0mm}
\includegraphics[width=0.3\linewidth]{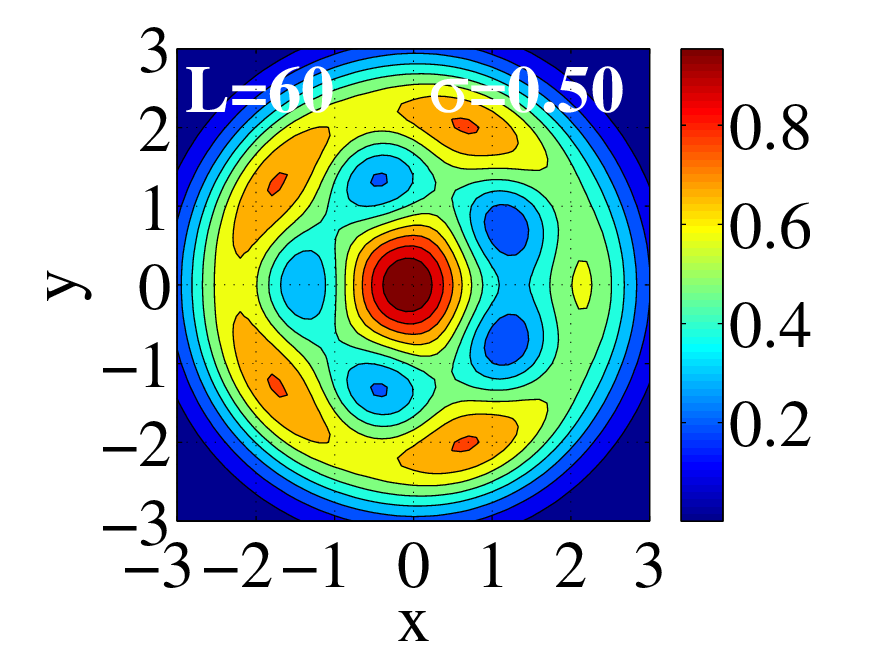}\label{fig:sp5_L60}
\hspace{-5.0mm}
\includegraphics[width=0.3\linewidth]{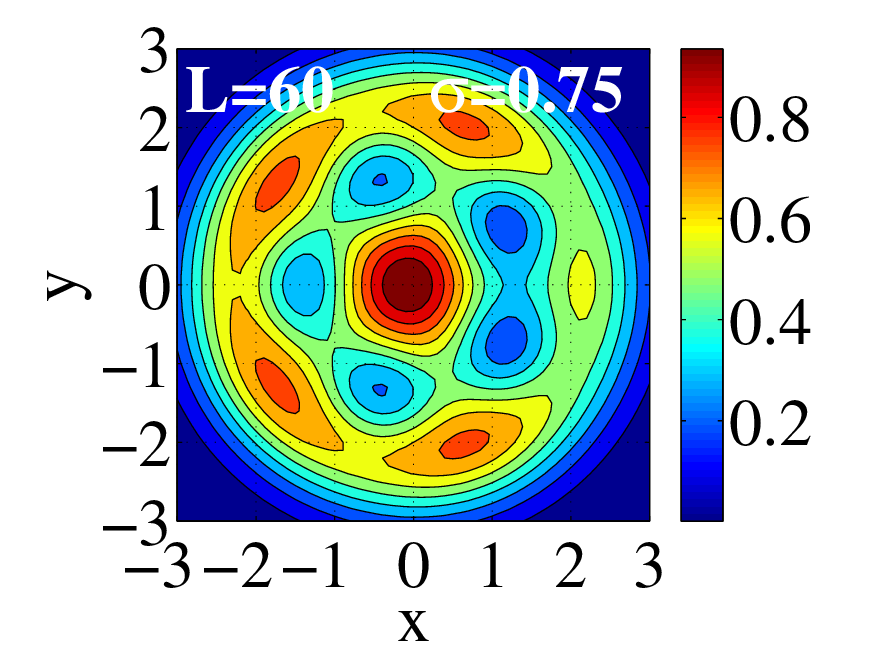}\label{fig:sp75_L60}}}
\caption{\label{fig:fivs}(Color online) The macroscopic occupations of one-particle angular momentum state of OPRDM {\it vs} one-particle angular momentum $m$ and the corresponding CPD plots depicting five-vortex state for $N=16$ bosons with interaction strength ${U_{0}}=0.2171$ and interaction range $\sigma=0.30,0.50,0.75$. The reference point has been chosen at $\mathbf{r}_{0}=(1.5,0)$. For the stable five-vortex $L_{z}=60$ state, with increase in interaction range $\sigma$, there is small increase in the macroscopic occupation of maximally occupied one-particle angular momentum state $m=5$ of OPRDM whereas, in the CPD plots, the values of probability density decreases at the central peak and increases at the off-center peaks due to redistribution with increase in interaction range $\sigma$.}	
\end{figure*}
\subsection{Multi-vortex states \label{many_vortex_st}}
\indent With increase in angular momentum $L_{z}$, vortices nucleate and arrange themselves in regular patterns~\cite{Madison_00} with respect to the reference line. For the angular momentum regime $2N<L_{z}\leq 4N$ with $N=16$, a $3,4,5$-fold symmetric vortex states are observed as shown in Figs.~\ref{fig:tvs}-\ref{fig:fivs}.
\\
\indent For the angular momentum regime $35\leq L_{z}\leq 39$, the macroscopic occupation of one-particle angular momentum state $m=3$ of OPRDM increases with increase in angular momentum 
and becomes maximum for $L_{z}=39$, whereas the occupation of $m=0$ state for the corresponding angular momentum states decreases, as shown in Fig.~\ref{fig:tvs}. 
For the meta-stable $L_{z}=35$ state, the $m=3$ state is maximally occupied and the occupation of $m=0$ and $m=2$ states are of the same order. With increase in interaction range $\sigma$, the occupation of $m=0,3$ states decreases whereas that of $m=1,2$ states increases, as shown in Fig.~\ref{fig:occL35s}. Further, it is also observed from the corresponding CPD plots in Fig.~\ref{fig:occL35s}, that the probability density decreases at the central peak whereas it increases at the off-center peaks. However, for the stable angular momentum states $L_{z}=36,39$, the one-particle angular momentum $m=3$ state of OPRDM becomes macroscopically occupied and its occupation increases with increase in interaction range $\sigma$, as shown in Figs.~\ref{fig:sp3_L36}-\ref{fig:sp3_L39}. The second maximally occupied state is the $m=0$ state occupying the cores of the vortices which contributes zero to the total energy of the system. 
Both $L_{z}=36$ and $L_{z}=39$ angular momentum states, form triangular-vortex states with the probability density at the central-peak higher than the probability density at the off-center peaks due to our system being non-uniform, as shown in the corresponding CPD plots in Fig.~\ref{fig:sp3_L36}-\ref{fig:sp3_L39}. 
Had the system been uniform (no confining potential), the probability densities at all the three peaks would have been equal. For a uniform system with translational invariance, this triangular-vortex state will become the building block for the regular Abrikosov lattice~\cite{Abrikosov_04}.
\\
\indent For angular momentum $L_{z}=48,52$, four-vortex states are observed, as shown in Fig.~\ref{fig:fvs}. The macroscopic occupation of one-particle angular momentum $m=4$ state increases whereas that of $m=0$ state decreases with increase in the angular momentum regime $48 \leq L_{z} \leq 52$. Further, for $L_{z}=48,52$ states, there is decrease in the macroscopic occupation of $m=4$ state with increase in interaction range $\sigma$ as shown in Figs.~\ref{fig:sp3_L48}-\ref{fig:sp3_L52}.
For angular momentum $L_{z}=60$, a five-fold symmetric vortex state is observed, as shown in Fig.~\ref{fig:fivs}. There is increase in macroscopic occupation of $m=5$ state with increase in the interaction range $\sigma$. In the CPD, corresponding to $L_{z}=48,52,60$ states, the probability density decreases at the central peak and increases at the off-center peaks with increase in interaction range $\sigma$ due to redistribution of particles. An increase(decrease) in probability density at the central-peak is accompanied by a decrease(increase) in probability density at the off-center peaks as shown in Table~\ref{tab:ccpd}.
Due to harmonic confining potential in 2D, the probability density at the central-peak will be higher than the probability density at the off-center peaks, placed symmetrically {\it w.r.t} the trap center, and will all be equal at off-center peaks due to circular symmetry of the trap as shown in Fig.~\ref{fig:L60sp3x0y0}.
By choosing a reference point, the circular symmetry is spontaneously broken. The off-center peaks with equal densities are now (after spontaneous symmetry breaking) placed symmetrically about the reference line as shown in CPD plots in Fig.~\ref{fig:sp3_L60}.
\\
\indent From Table~\ref{tab:ccpd}, we observe that there is either no central-peak or one central-peak whereas off-center peaks increase with increase in angular momentum $L_{z}$. The number of off-center peaks is proportional to the value $m$ of the maximally occupied one-particle angular momentum state of OPRDM. The $m=0$ state is the second maximally occupied state  excluding $L_{z}=17$ state$^{*}$
which is highly unstable in the vicinity of the stable $L_{z}=N=16$ single-vortex state.
\\
\begin{table*}[]
\caption{Summary of CPD plots in Fig.~\ref{fig:b4_vortx} to Fig \ref{fig:fivs} for $N=16$ bosons and angular momentum states $0\leq L_{z}\leq 60$. For a given angular momentum state, we list the number of central peak, the number of off-center peaks and the maximally occupied as well as the second maximally occupied one-particle angular momentum $m$ states. 
We, further list variation in the probability density at the central-peaks and the occupation of the maximally occupied one-particle angular momentum $m-$state with increase in interaction range $\sigma$.
Column $(6)$ and column $(9)$ corresponding to $\sigma=0.30$ with $\uparrow$-arrow are for reference.
An increase (decrease) in the probability density at the central-peak (column $6$ through $8$) or the occupation of maximally occupied $m$-state (column $9$ through $11$) with respect to the preceding column is denoted by an
$\uparrow$-arrow ($\downarrow$-arrow).}
\begin{tabular}{|c|c|c|c|c|c|c|c|c|c|c|}\hline\hline
\multirow{2}{*}{}&\multirow{2}{*}{Number}&\multirow{2}{*}{Number}&\multirow{2}{*}{Maximally}&\multirow{2}{*}{Second} &\multicolumn{3}{p{4.0cm}|}{Variation in the probability density at central-peak for various values of }&\multicolumn{3}{p{4.0cm}|}{Variation in the occupation of maximally occupied $m$-state for various values of} \\
$L_{z}$& of & of & occupied &Maximally & & &  & &  &  \\
&central- & off-center& $m$-state &occupied &$\sigma=0.30$ &$\sigma=0.50$ &$\sigma=0.75$ &$\sigma=0.30$ &$\sigma=0.50$ &$\sigma=0.75$\\
&peak     & peaks  &     &  $m$-state & & & & & &\\
\hline   
0&1&0&- &-&$\uparrow$ & $\uparrow$ & $\uparrow$ &- &- &- \\ \hline
11&1&0&-&-&$\uparrow$ & $\uparrow$ & $\uparrow$ &- &- &- \\ \hline
12&1&0&-&-&$\uparrow$ & $\uparrow$ & $\uparrow$ &- &- &- \\ \hline
13&1&0&-&-&$\uparrow$ & $\uparrow$ & $\uparrow$ &- &- &- \\ \hline
16&0&1&1 &0 &$\uparrow$ & $\uparrow$ & $\uparrow$ & $\uparrow$ & $\uparrow$ & $\uparrow$\\ \hline
17&0&1&1 &$2^{*}$ &$\uparrow$ & $\uparrow$ & $\uparrow$ & $\uparrow$ &$\uparrow$ & $\uparrow$\\ \hline
26&0&2&2 &0 &-&-&-&$\uparrow$ & $\downarrow$ & $\uparrow$ \\ \hline
28&0&2&2 &0 &-&-&-&$\uparrow$ & $\uparrow$ & $\uparrow$ \\ \hline
30&0&2&2 &0 &-&-&-&$\uparrow$ & $\uparrow$ & $\uparrow$ \\ \hline
32&0&2&2 &0 &-&-&-&$\uparrow$ & $\uparrow$ & $\uparrow$ \\ \hline
35&1&3&3 &0 &$\uparrow$ & $\downarrow$ & $\downarrow$  & $\uparrow$ & $\downarrow$ &$\downarrow$ \\ \hline
36& 1 & 3&3 &0 &$\uparrow$ & $\downarrow$ & $\downarrow$  & $\uparrow$ & $\uparrow$ &$\uparrow$ \\ \hline
39&1&3 &3 &0 &$\uparrow$ & $\downarrow$ & $\downarrow$ & $\uparrow$ & $\uparrow$ & $\uparrow$ \\ \hline
48&1&4 &4 &0 & $\uparrow$ & $\downarrow$ & $\downarrow$ & $\uparrow$ & $\downarrow$ & $\downarrow$ \\ \hline
52&1&4 &4 &0 &$\uparrow$ & $\downarrow$ & $\downarrow$ & $\uparrow$ & $\downarrow$ & $\downarrow$ \\ \hline
60&1&5 &5 &0 &$\uparrow$ & $\downarrow$ & $\downarrow$ & $\uparrow$ & $\uparrow$ & $\uparrow$ \\ \hline\hline
\end{tabular}
\label{tab:ccpd}
\end{table*}
\subsection{Effects of the interaction range $\sigma$, the number of particles $N$ and the interaction strength $U_{0}$ on the condensate fraction}
\indent We observe that for both the non-rotating and the single-vortex states, the largest condensate fraction $\lambda_{1}\left(N,L_{z},\sigma\right)$ increases with increase in interaction range $\sigma$ as shown in Table~\ref{tab:tlim}-\ref{tab:tlim2} for two values of interaction strength $U_{0}$. 
This may be attributed to the quantum mechanical phase coherence of the many-body state spreading over more number of particles with increase in interaction range $\sigma$.\\
\indent We, further, study the effects of the number of particles $N$ on the largest condensate fraction $\lambda_{1}$ for which the results are summarized in Table~\ref{tab:tlim}-\ref{tab:tlim2}. For the non-rotating $L_{z}=0$ state, the density-profile $n_{0}({\bf r})\sim e^{- r_{\perp}^{2}/a_{\perp}^{2}}$ is Gaussian and most of the particles are distributed around the centre of the trap $(0,0)$. In the mean-field approximation, the effective particle-particle interaction given by ${U_{0}}(N-1)n_{0}({\bf r})$ increases with increase in number of particles $N$, leading to the depletion of the condensate fraction $\lambda_{1}$.
On the other hand, for the single-vortex state, particles are spread out due to centrifugal force with the density profile given by $n_{1}({\bf r})\sim \left( r_{\perp}/a_{\perp}\right)^{2} e^{-r_{\perp}^{2}/a_{\perp}^{2}}$, which vanishes at the centre of the trap (0,0). Thus, the effective particle-particle interaction for the first-vortex state given by ${U_{0}}(N-1)n_{1}({\bf r})$, in the mean-field approximation, decreases due to the vanishing of the density at the center of the trap, leading to increase in the condensate fraction $\lambda_{1}$. Further, we observe that with increase in the interaction strength $U_{0}$ for a given number of particles $N$ and interaction range  $\sigma$, the largest condensate fraction $\lambda_{1}\left(N,L_{z},\sigma\right)$ decreases for both the non-rotating $L_{z}=0$ and the single-vortex $L_{z}=N$ states as shown in Table~\ref{tab:tlim}-\ref{tab:tlim2} 
for two different values of interaction strength $U_{0}$.
\\
\indent Further, for the two-vortex state $L_{z}=2N=32$,
there are significant rearrangement in the active Fock-space
with increase in $\sigma$ for the interaction range regime $0.6\leq \sigma \leq 0.75$,
as seen from the one-particle angular momentum states in OPRDM as shown in Table~\ref{tab:sigma_sp}. 
Also, in Table \ref{tab:occL32s}, for $L_{z}=2N=32$ state and various values of $\sigma$, we list only those one-particle angular momentum states $\left(n_{\mu},m^{\mu}\right), \mu=1,\cdots, 5 $ of OPRDM 
for which there are violent rearrangement in the active Fock-space.
The occupations corresponding to states $(n_{\mu},m^{\mu})$  of OPRDM decrease across the row and that of state $(n_{1},m^{1})$ increase along the column in the Table~\ref{tab:occL32s}. This rearrangement in states $(n_{\mu},m^{\mu})$ may lead to the violent rearrangement in the active Fock-space and may be a signature of quantum phase transition induced by the interaction range $\sigma$.\\
\indent The complete results presented in this work are summarised in tabular form in the Table~\ref{Tab:n16_sp3}. 
\section{\label{sec:summary} SUMMARY AND CONCLUSION}
\indent In summary, we present exact diagonalization study on the ground state of a system of $N=10,12,16,20,24$ spinless bosons, interacting via Gaussian potential, in the angular momentum regime $0\leq L_{z} \leq 4N$ for values of interaction range $\sigma=0.30,0.50,0.75$ and interaction strength $U_{0}=0.2171$. 
It is found~\cite{Ahsan_Kumar,Lewenstein09} that the critical angular velocity $\{\Omega_{c_{i}}\left(\sigma\right), i=1,2,3,\cdots \}$ decreases with increase in interaction strength $U_{0}$ and the number of particles $N$. In contrast, we find through our present study that an increase in interaction range $\sigma$ leads to an increase in critical angular velocity $\{ \Omega_{c_{i}}(\sigma)\}$. We further observe that the span of plateaus on the $L_{z} \ {  \it vs } \ \Omega$ stability graph, decreases with increase in $\sigma$.
It is found that some of the angular momentum states which are meta-stable state at small values of $\sigma$, evolve into stable states as $\sigma$ is increased beyond a critical value.
Further, with increase in $\sigma$, there is an increase in the value of the largest eigenvalue $\lambda_{1}\left(\sigma\right)$ of OPRDM corresponding to the largest condensate fraction.
It is found that with increase in $N$, the largest condensate fraction $\lambda_{1}$ decreases for the non-rotating $L_{z}=0$ state whereas it increases for the first vortex $L_{z}=N$ state.
For the two-vortex $L_{z}=2N=32$ state, with increase in interaction strength $U_{0}$ beyond a certain value in the strongly interacting regime, the maximally occupied one-particle angular momentum state of OPRDM changes from $m$ to $m+1$.
As a signature of quantum phase transition, a significant rearrangement in the active Fock-space is observed as interaction range $\sigma$ is increased.
A decrease in von Neumann entropy $S_{1}\left(N,L_{z},\sigma\right)$ is observed with increase in $\sigma$.
We observe crossings in von Neumann entropy $S_{1}\left(N,L_{z},\sigma\right)$ with variations in $\sigma$ among different stable and meta-stable angular momentum states.
With increasing angular momentum $L_{z}$, more number of quantized vortices appear as the system goes from one stable state to another, forming regular patterns of $2,3,4,5$-fold symmetric vortex states in the angular momentum regime $0\leq L_{z} \leq 4N$. 
The single-vortex state is significantly more stable than the multi-vortex states. Among the multi-vortex states, the three-vortex state is more stable in comparison to the two-vortex state.
In the CPD, for $L_{z}=0$ state, the probability density 
at the central peak increases with increase in $\sigma$.
For the first-vortex state, the density-peak appears as a crescent due to the off-center choice of the reference point,
the peak-density increases with increase in $\sigma$ resulting in a sharper and narrower crescent.
Further, for the two-vortex state, it is observed that the cores of the two-vortex state merge with increase in $\sigma$.
However, for the vortex states comprising of more than two vortices, we do not observe merging of vortex cores.
For the three-vortex $L_{z}=36,39$ states, the four-vortex $L_{z}=48,52$ states and five-vortex $L_{z}=60$ state, the probability density decreases at the central peak and increases at the off-center peaks, with increase in $\sigma$. The macroscopic occupation of the maximally occupied one-particle angular momentum $m$-state increases for angular momentum states $L_{z}=0,16,32,36,39,60$ and decreases for $L_{z}=26,35$ states with increase in $\sigma$.
By tuning the interaction range $\sigma$, we can create the Bose-Einstein condensate for the desired systems.\\
\indent Further work is in progress to study the effects of three-body Gaussian interaction, in addition to two-body Gaussian interaction, on the ground state of Bose-Einstein condensate. It is anticipated that even a small three-body interaction will significantly alter the ground state of the many-body system~\cite{Beliaev}.
\begin{table*}[t]
	\vspace{-2mm}
\caption{The three largest eigenvalues $\lambda_{1}>\lambda_{2}>\lambda_{3}$ of OPRDM in Eq.~(\ref{eq:rdm}) for 
the non-rotating state $L_{z}=0$ and the single-vortex state $L_{z}=N$ for $N=10,12,16$ particles with interaction strength ${U_{0}}=0.2171$ and interaction range $\sigma=0.30,0.50, 0.75$.}
		\begin{tabular}{|c|c|ccc |c|ccc|}
			\hline\hline
			\multirow{2}{*}{$N$}	&
			{$\lambda_{\mu}$}&\multicolumn{3}{c|}{$L_{z}=0$} & 
			{$\lambda_{\mu}$}&\multicolumn{3}{c|}{$L_{z}=N$} \\
			& $\mu=1,2,3.$	&$\sigma=0.30$  &$\sigma=0.50$  &$\sigma=0.75$ &$\mu=1,2,3.$  &$\sigma=0.30$ &$\sigma=0.50$ &$\sigma=0.75$ \\ \hline
			&	$\lambda_{1}$   &0.99240 &0.99518  &0.99790 &$\lambda_{1}$ &0.81775 &0.82176 &0.82451\\ 
     		$10$&	$\lambda_{2}$  &0.00295 &0.00200  &0.00094 &$\lambda_{2}$ &0.09399 &0.09210 &0.09080\\ 
			&	$\lambda_{3}$   &0.00295 &0.00200  &0.00094 &$\lambda_{3}$ &0.07767 &0.07749 &0.07777\\ \hline
			&	$\lambda_{1}$   &0.99130 &0.99436  &0.99747 &$\lambda_{1}$ &0.84247 &0.84676 &0.84988\\ 
			$12$&	$\lambda_{2}$  &0.00329 &0.00230  &0.00112 &$\lambda_{2}$ &0.08039 &0.07850 &0.07716\\ 
			&	$\lambda_{3}$   &0.00329 &0.00230  &0.00112 &$\lambda_{3}$ &0.06796 &0.06753 &0.06756\\ \hline
			&	$\lambda_{1}$   &0.98949 &0.99292  &0.99664 &$\lambda_{1}$ &0.87484 &0.87949 &0.88316\\ 
			$16$&	$\lambda_{2}$   &0.00378 &0.00278  &0.00146 &$\lambda_{2}$ &0.06268 &0.06083 &0.05944\\ 
			&	$\lambda_{3}$   &0.00378 &0.00278  &0.00146 &$\lambda_{3}$ &0.05460 &0.05387 &0.05355\\ \hline\hline
		\end{tabular}
	\label{tab:tlim}
\end{table*}
\begin{table*}[!htb]
	\vspace{-2mm}
\caption{The three largest eigenvalues $\lambda_{1}>\lambda_{2}>\lambda_{3}$ in Eq.~(\ref{eq:rdm}) for the ground state of the non-rotating state $L_{z}=0$ and the single-vortex state $L_{z}=N$ for $N=10,12,16$ particles with the lesser interaction strength ${U_{0}}=0.02171$ and the interaction range $\sigma=0.30,0.50, 0.75$.}
		\begin{tabular}{|c|c|ccc |c|ccc|}
			\hline\hline
			\multirow{2}{*}{$N$}	&
			{$\lambda_{\mu}$}&\multicolumn{3}{c|}{$L_{z}=0$} & 
			{$\lambda_{\mu}$}&\multicolumn{3}{c|}{$L_{z}=N$} \\
			&$\mu=1,2,3.$	&$\sigma=0.30$  &$\sigma=.50$  &$\sigma=0.75$ &$\mu=1,2,3.$  &$\sigma=0.30$ &$\sigma=0.50$ &$\sigma=0.75$ \\ \hline
			&	$\lambda_{1}$   &0.99990 &0.99994  &0.99998 &$\lambda_{1}$ &0.82505 &0.82531 &0.82541\\
			$10$&	$\lambda_{2}$  &0.00005  &0.00003  &0.00001 &$\lambda_{2}$ &0.09063 &0.09047 &0.09038\\ 
			&	$\lambda_{3}$   &0.00005  &0.00003  &0.00001 &$\lambda_{3}$ &0.07829 &0.07831 &0.07838\\ \hline
			&	$\lambda_{1}$   &0.99987  &0.99993  &0.99997 &$\lambda_{1}$ &0.85078 &0.85105 &0.85119\\ 
		    $12$&	$\lambda_{2}$  &0.00005  &0.00003  &0.00001 &$\lambda_{2}$ &0.07690 &0.07673 &0.07663\\ 
			&	$\lambda_{3}$   &0.00005  &0.00003  &0.00001 &$\lambda_{3}$ &0.06789 &0.06789 &0.06793\\ \hline
			&	$\lambda_{1}$   &0.99983  &0.99991  &0.99996 &$\lambda_{1}$ &0.88472 &0.88500 &0.88517\\ 
			$16$&	$\lambda_{2}$  &0.00007  &0.00004  &0.00002 &$\lambda_{2}$ &0.05899 &0.05883 &0.05873\\ 
			&	$\lambda_{3}$   &0.00007  &0.00004  &0.00002 &$\lambda_{3}$ &0.05358 &0.05355 &0.05357\\ \hline\hline
		\end{tabular}
	\label{tab:tlim2}
\end{table*}
\begin{table*}[t]
\caption{\label{tab:sigma_sp} For $N=16$ bosons, the largest five eigenvalues $\lambda_{1}>\lambda_{2}>\lambda_{3}>\lambda_{4}>\lambda_{5}$ of OPRDM and the corresponding one-particle basis states $\left(n_{\mu}, m^{\mu}\right), \  \mu=1,\cdots,5$ in Eq.~(\ref{eq:rdm}) for the angular momentum $L_{z}=2N=32$ state with interaction strength $\mbox{g}_{2}=0.2171$ and interaction range $\leq \sigma \leq 1$. The largest eigenvalue $\lambda_{1}$ increases with increase in interaction range $\sigma$ $i.e$ $\lambda_{1}\left(\sigma^{\prime}\right)> \lambda_{1}\left(\sigma\right)$ for $\sigma^{\prime}>\sigma$.}
	\medskip
	\begin{tabular}{|ccc ccc ccc cc|}
		\hline\hline
 $\sigma$&$n_{1},m^{1}$&$\lambda_{1}$&$n_{2},m^{2}$&$\lambda_{2}$&$n_{3},m^{3}$&
		$\lambda_{ 3}$&$n_{4},m^{4}$&$\lambda_{ 4}$&$n_{5},m^{5}$&$\lambda_{ 5}$ \\
		\hline
		0.00&2,2&0.42220 &0,0&0.19470 &4,4&0.14026 &1,1&0.10217 &3,3&0.10028 \\
		0.10&2,2&0.42318 &0,0&0.19457 &4,4&0.14096 &1,1&0.10199 &3,3&0.09933 \\
		0.20&2,2&0.42675 &0,0&0.19365 &4,4&0.14231 &1,1&0.10153 &3,3&0.09703 \\
		0.30&2,2&0.43495 &0,0&0.19058 &4,4&0.14261 &1,1&0.10069 &3,3&0.09468 \\
		0.40&2,2&0.45124 &0,0&0.18335 &4,4&0.13974 &1,1&0.09924 &3,3&0.09351 \\
		0.50&2,2&0.48124 &0,0&0.16906 &4,4&0.13125 &1,1&0.09698 &3,3&0.09385 \\
		0.60&2,2&0.53230 &0,0&0.14395 &4,4&0.11418 &3,3&0.09455 &1,1&0.09435 \\
		0.70&2,2&0.60341 &0,0&0.10799 &3,3&0.09401 &1,1&0.09303 &4,4&0.08811 \\
		0.75&2,2&0.63923 &3,3&0.09346 &1,1&0.09329 &0,0&0.08937 &4,4&0.07415 \\ 
		0.80&2,2&0.67015 &1,1&0.09402 &3,3&0.09303 &0,0&0.07299 &4,4&0.06163 \\
		0.90&2,2&0.71317 &1,1&0.09588 &3,3&0.09272 &0,0&0.04952 &4,4&0.04342 \\ 
		1.00&2,2&0.73738 &1,1&0.09742 &3,3&0.09284 &0,0&0.03589 &4,4&0.03271 \\
		\hline \hline
	\end{tabular}
\end{table*}
\section*{Acknowledgement}
\indent We thank Mohd Imran and Mohd Talib for useful comments on the manuscript.
\appendix
\begin{table}[h]
\caption{For $N=16$ bosons, the one-particle angular momentum states $\left(n_{\mu},m^{\mu}\right), \mu=1,\cdots,5$ of OPRDM with the corresponding occupancies $\left(N\lambda_{\mu}\right), \mu=1,\cdots,5$ for angular momentum $L_{z}=2N=32$ state and interaction strength $U_{0}=0.2171$. There is significant rearrangement in the active Fock-space with increase in $\sigma$.
\label{tab:occL32s}}
    \centering
    \begin{tabular}{|c|c|c|c|c|c|}
    \hline\hline
\multirow{2}{*}{}&$n_{1},m^{1}$&$n_{2},m^{2}$&$n_{3},m_{3}$&      $n_{4},m^{4}$&$n_{5},m^{5}$ \\ 
    &$N\lambda_{1}$&$N\lambda_{2}$&$N\lambda_{3}$& $N\lambda_{4}$&$N\lambda_{5}$ \\ \hline
\multirow{2}{*}{$\sigma=0.30$} 
    &(2,2)&(0,0)&(4,4)&(1,1)&(3,3)  \\ 
    &6.9592&3.0493&2.2818&1.6109&1.5481\\ \hline
\multirow{2}{*}{$\sigma=0.60$}&(2,2)&(0,0)&(4,4)&(3,3)&(1,1)\\
    &8.5168&2.3032&1.8268&1.5127&1.5095\\ \hline
\multirow{2}{*}{$\sigma=0.75$}&(2,2)&(3,3)&(1,1)&(0,0)&(4,4)\\
    &10.2277&1.4954&1.4926&1.4299&1.1864\\ \hline
\multirow{2}{*}{$\sigma=0.80$}&(2,2)&(1,1)&(3,3)&(0,0)&(4,4)\\
    &10.7223&1.5044&1.4884&1.1677&0.9862\\ \hline\hline
    \end{tabular}
\end{table}
\section{One-particle density in beyond lowest Landau level approximation\label{app:intfr}} 
\indent The variationally obtained many-particle wavefunction for a given number of particles $N$ in a given subspace of angular momentum $L_{z}$ is given by 
\begin{eqnarray}
&&\vert\Psi_{N,L_{z}}\rangle=\sum_{\nu}^{\prime}C_{\nu}\vert\nu;N,L_{z}\rangle\nonumber\\
&=&\sum_{\underbrace{\nu_{1},\nu_{2},\cdots,\nu_{M}}_{\sum_{\alpha}\nu_{\alpha=N}; \sum_{\alpha}\nu_{\alpha m_{\alpha}=L_{z}}}}C_{\nu_{1},\nu_{2},\cdots,\nu_{M}}\vert \nu_{1},\nu_{2},\cdots,\nu_{M} \rangle
\nonumber \\
\mbox{where}
&&\vert \nu_{1},\nu_{2},\cdots,\nu_{M} \rangle =\frac{(a^{\dagger}_{1})^{\nu_{1}}(a^{\dagger}_{2})^{\nu_{2}}\cdots (a^{\dagger}_{M})^{\nu_{M}}}{\sqrt{\nu_{1}!\nu_{2}!\cdots \nu_{M}!}}\vert 0 \rangle.\nonumber 
\label{eq:PsiNL}
\end{eqnarray}
The one-particle density in the first quantized form is given by:
\begin{eqnarray}
{n}(\bf{r})&=&\sum_{i=1}^{N}\delta(\bf{r}-\bf{r}_{i}) \nonumber
\end{eqnarray}
where $N$ is the number of particles.
 In the second quantized form the number-density operator becomes
\begin{eqnarray}
\hat{n}(\bf{r})&=&\int d{\bf r}^{\prime} \hat{\Psi}^{\dagger}({\bf r}^{\prime})\delta({\bf{r}}-{\bf{r}}^{{\prime}})\hat{\Psi}({\bf r}^{\prime})\nonumber\\
&=&\hat{\Psi}^{\dagger}({\bf r})\hat{\Psi}({\bf r})
=\sum^{M}_{{\bf n}^{\prime},{\bf n}=1}u^{*}_{\bf{n}^{\prime}}({\bf r})u^{}_{\bf{n}}({\bf r})a^{\dagger}_{\bf{n}^{\prime}}a^{}_{\bf{n}}\nonumber
\end{eqnarray}
where $M$ is the number of one-particle basis states with composite quantum number ${\bf n}\equiv (n_{r},m)$ in 2D. The expectation value of the number-density operator over the many-particle state $\vert \Psi_{N,L_{z}} \rangle$ is given by:
\begin{eqnarray}
&&n({\bf r})=\langle \Psi_{N,L_{z}}\vert \hat{n}({\bf r})\vert \Psi_{N,L_{z}} \rangle \nonumber\\
&=& \sum_{{\nu}^{\prime}}^{\prime} \sum_{{\nu}}^{\prime} \sum_{{\bf n}^{\prime}{\bf n}} u^{*}_{\bf{n}^{\prime}}({\bf r}) u_{\bf{n}}({\bf r}) C^{*}_{\nu^{\prime}}C_{\nu}\nonumber\\
&& \langle{\nu_{1}^{\prime} \nu_{2}^{\prime} \cdots \nu_{M}^{\prime}} \vert{a^{\dagger}_{\bf{n}^{\prime}}a_{\bf{n}}} \vert{ \nu_{1} \nu_{2} \cdots \nu_{M}}\rangle \nonumber\\
\end{eqnarray}
\begin{eqnarray}
&=&\sum_{\bf n} \vert u_{\bf n}({\bf r}) \vert ^{2} \sum_{\nu_{1}\cdots \nu_{\bf n}\cdots\nu_{M}}\nu_{{\bf n}} \vert C^{*}_{\nu_{1}\cdots\nu_{\bf n}\cdots\nu_{M}} \vert^{2}\nonumber\\
&+& \sum_{{\bf n}^{\prime}\neq{\bf n}} u^{*}_{\bf{n}^{\prime}}({\bf r}) u^{}_{\bf{n}}({\bf r})\nonumber \\
&&\sum_{\nu_{1}\cdots \nu_{\bf n}\cdots \nu_{\bf n^{\prime}}\cdots\nu_{M}} C^{*}_{{\nu_{1}}\cdots {(\nu_{\bf n}-1)}\cdots {(\nu_{\bf n^{\prime}}+1)} \cdots {\nu_{M}}}\nonumber \\
&& C_{{\nu_{1}}\cdots {\nu_{\bf n}} \cdots {\nu_{\bf n^{\prime}}} \cdots {\nu_{M}}} \sqrt{{\nu_{\bf n}}({\nu_{\bf n'}+1})}. \nonumber
\end{eqnarray}
The above equation takes the following form~\cite{Ahsan_Kumar}
\begin{eqnarray}
n({\bf r})&=& \sum_{n'_{r},{n_{r}}} \ {\sum_{m}}
\rho_{{n'_{r},m};\ {n_{r},m}}
u^{*}_{{n'_{r},m}}(\textbf{r})u_{n_{r},m}(\textbf{r})\nonumber\\
~~~~~~~~~~\label{eq:nrrp}
\end{eqnarray}
which after diagonalization becomes
\begin{eqnarray}
n({\bf r})=\sum_{\mu} (N \lambda_{\mu})\ \vert
\chi_{\mu,m^{\mu}}\left({\bf r}\right)\vert^{2}~~~~~~~~
\label{eq:densit}
\end{eqnarray}
where $\chi_{{\bf{\mu}},m^{\mu}}({\bf r})$ is as defined in Eq.~(\ref{eq:chi_mu}). It is clear that the density operator will not exhibit any interference pattern due to $\vert
\chi_{\mu,m^{\mu}}\left({\bf r}\right)\vert^{2}$ being real.\\
\indent In the lowest Landau level approximation  ${\bf n}\equiv(0,m)$ with $n_{r}=0$, the one-particle basis state is labelled by only the angular momentum quantum number $m$ and Eq.~(\ref{eq:densit}) in the beyond lowest Landau level approximation reduces to Eq.~(9) in reference~\cite{lewenstein06} in the lowest Landau level approximation.
\section{Conditional probability distribution (CPD) in beyond lowest Landau level approximation \label{app:cpdCyld}}
\begin{figure*}[!htb]
\subfigure[\label{fig:L60sp3x0y0} ${\bf r}_{0}=(0,0)$]
{\includegraphics[width=0.32\linewidth]{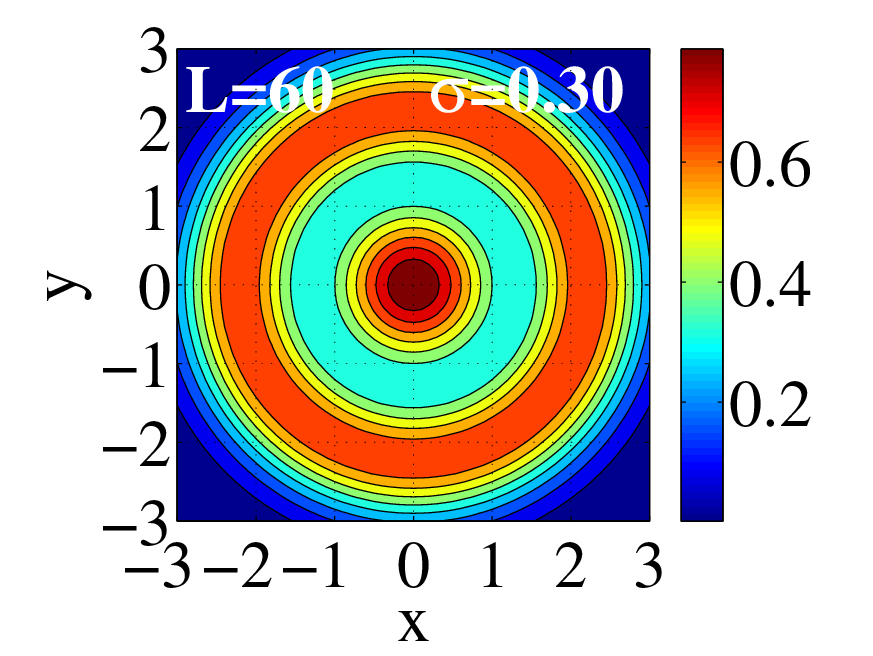}}
\subfigure[\label{fig:x0y0} ${\bf r}_{0}=(0,0)$]
{\includegraphics[width=0.32\linewidth]{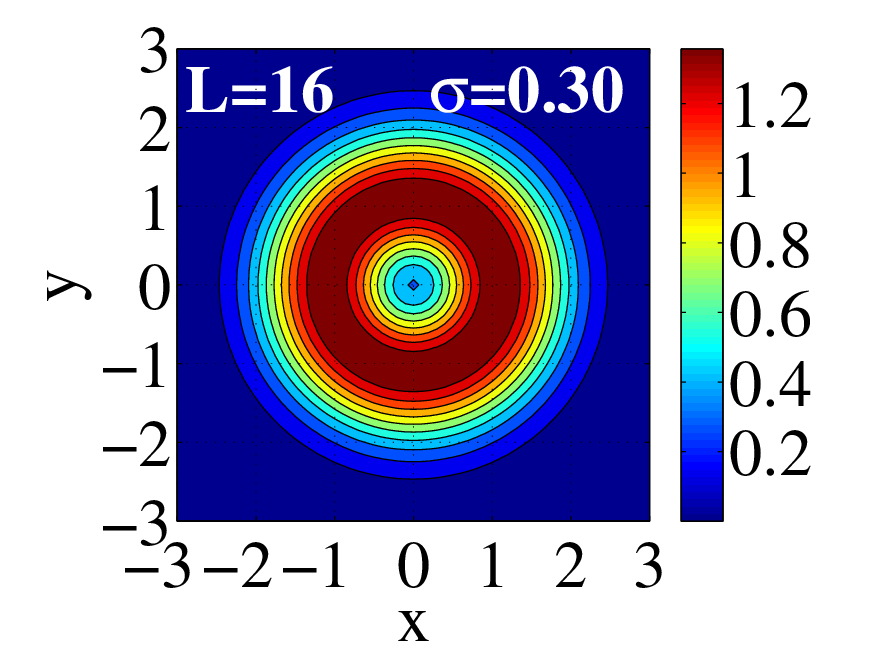}\label{fig:00c}}
\subfigure[\label{fig:x0p5y0} ${\bf r}_{0}=(0.5,0)$]
{\includegraphics[width=0.32\linewidth]{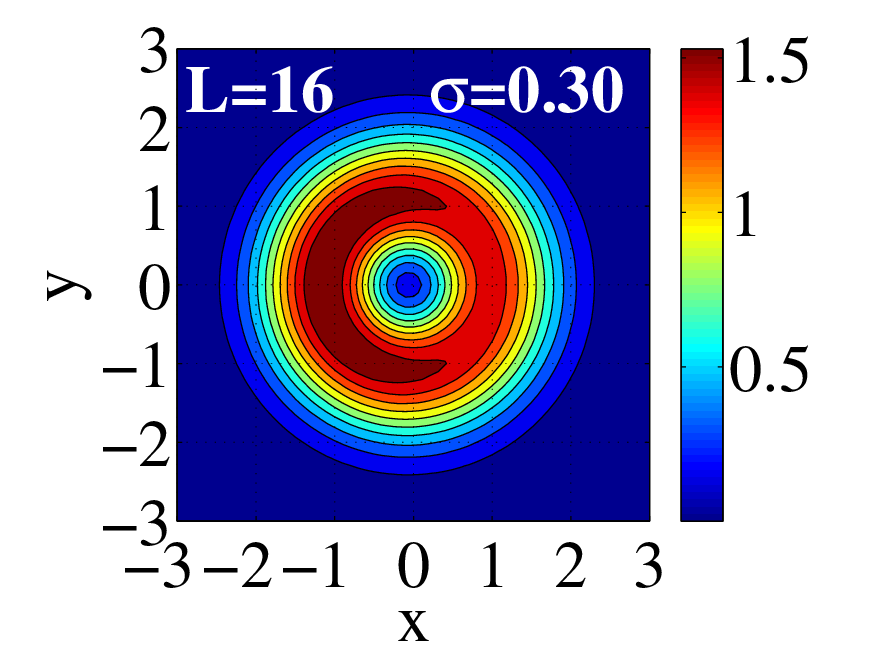}}
\caption{\label{fig:cpd_ent_1st_vor}(Color online) CPD plots depicting (a) a central-peak for $L_{z}=60$ and (b, c) off-center-peak for $L_{z}=N=16$ bosons for the interaction strength ${U_{0}}=0.2171$ and the interaction range $\sigma=0.30$. The CPD exhibits cylindrical symmetry when reference point ${\bf r}_{0}$ coincides with trap centre, whereas, the cylindrical symmetry is broken as the reference point is moved from the trap centre.}
\end{figure*}

The conditional probability distribution (CPD) is defined as the probability of finding a boson at ${\bf r}$ given that the other is at ${\bf r}_{0}$
and is given by: 
\begin{eqnarray}
{\cal{P}}({\bf r},{\bf r}_{0})=\sum^{N}_{i < j}\delta({\bf r}-{\bf r}_{i})\delta({\bf r}_{0}-{\bf r}_{j}) \nonumber
\end{eqnarray}
where $N$ is the number of particles. In the second quantized form, this becomes:
\begin{eqnarray}
\hat{\cal{P}}({\bf r},{\bf r}_{0})&=&\frac{1}{2}\left[\int d{\bf r}^{\prime} d{\bf r}^{''} \right]\Psi^{\dagger}({\bf r}^{\prime})
\Psi^{\dagger}({\bf r}^{''})\nonumber\\
&\times&(\delta({\bf r}-{\bf r}^{\prime})\delta({\bf r}_{0}-{\bf r}^{''})+ \delta({\bf r}-{\bf r}^{''})\delta({\bf r}_{0}-{\bf r}^{'}) \nonumber\\	
&\times& \Psi^{}({\bf r}^{''})\Psi^{}({\bf r}^{\prime})
\nonumber\\
&=& \Psi^{\dagger}({\bf r}) \Psi^{\dagger}({\bf r}_{0}) \Psi({\bf r}_{0}) \Psi({\bf r}) 
\nonumber\\
&=& \sum_{{\bf n'},{\bf n}}\sum_{{\bf m'},{\bf m}} u^{*}_{\bf n'}\left({\bf r}\right)
u_{\bf n}\left({\bf r}\right) u^{*}_{\bf m'}\left({\bf r}_{0}\right) u_{\bf m}\left({\bf r}_{0}\right) \nonumber\\	
&&{a^{\dagger}_{\bf n'} a^{\dagger}_{\bf m'} a_{\bf m} a_{\bf n}}.  \nonumber
\end{eqnarray}
The expectation value of CPD operator over the many-particle state $\vert \Psi_{N,L_{z}}\rangle$ defined in Eq.~(\ref{eq:PsiNL}) above is given by:
\begin{eqnarray}
&&{\cal P}({\bf r},{\bf r_{0}})=\langle \Psi_{N,L_{z}}\vert \hat{\cal{P}}({\bf r},{\bf r_{0}})\vert \Psi_{N,L_{z}} \rangle \nonumber\\
&=& \langle \Psi_{N,L_{z}}\vert \sum_{{\bf m'},{\bf m}} \sum_{{\bf n'},{\bf n}} u^{*}_{\bf n'}\left({\bf r}\right)
u_{\bf n}\left({\bf r}\right) u^{*}_{\bf m'}\left({\bf r}_{0}\right) u_{\bf m}\left({\bf r}_{0}\right) \nonumber\\&& {a^{\dagger}_{\bf n'} a^{\dagger}_{\bf m'} a_{\bf m} a_{\bf n}} \vert \Psi_{N,L_{z}} \rangle  \nonumber\\
&=& \sum_{\nu^{\prime},\nu}^{\prime}C^{*}_{\nu^{\prime}}C^{}_{\nu^{}}\sum_{{\bf m'},{\bf m}} \sum_{{\bf n'},{\bf n}} u^{*}_{\bf n'}\left(\bf r\right)
u_{\bf n}\left({\bf r}\right) u^{*}_{\bf m'}\left({\bf r_{0}}\right)u_{\bf m}\left({\bf r}_{0}\right)  \nonumber\\
&& \langle \nu^{\prime}\vert {a^{\dagger}_{\bf n'} a^{\dagger}_{\bf m'} a_{\bf m} a_{\bf n}} \vert \nu \rangle \nonumber
\end{eqnarray}
\begin{eqnarray}
&&
=\sum_{{\bf m'},{\bf m}} \sum_{{\bf n'},{\bf n}} u^{*}_{\bf n'}\left({\bf r}\right)
u_{\bf n}\left({\bf r}\right) u^{*}_{\bf m'}\left({\bf r}_{0}\right) u_{\bf m}\left({\bf r}_{0}\right)\nonumber\\
&&\sum^{\prime}_{{\nu}_{\bf n},{\nu}_{\bf n'},  {\nu}_{\bf m},{\nu}_{\bf m'}}
C^{*}_{\nu_{\bf 1}\cdots(\nu_{\bf n}-1)\cdots(\nu_{\bf n'}+1)
\cdots(\nu_{\bf m}-1)
\cdots(\nu_{\bf m'}+1) \cdots \nu_{\bf M}}
\nonumber\\
&& C_{\nu_{\bf 1}\cdots(\nu_{\bf n})\cdots(\nu_{\bf n'})\cdots(\nu_{\bf m})
\cdots(\nu_{\bf m'})\cdots \nu_{\bf M}}
\nonumber\\
&&\sqrt{\nu_{\bf n}\nu_{\bf m}(\nu_{\bf n'}+1)(\nu_{\bf m'}+1)}.
\label{eq:cpd_2quant}
\end{eqnarray}
Equation (\ref{eq:cpd_2quant}) above is the conditional probability distribution(CPD) in the beyond lowest Landau level approximation which with ${\bf m}^{\prime}\equiv(0,m'),\ {\bf m}\equiv(0,m), \ {\bf n}^{\prime}\equiv(0,m''), \ {\bf n}\equiv(0,m''') $ in the lowest Landau level approximation reduces to 
\begin{eqnarray}
&=&\sum_{{ m'},{ m}} \sum_{{m''},{m'''}} u^{*}_{m''}\left({\bf r}\right)
u_{m'''}\left({\bf r}\right) u^{*}_{ m'}\left({\bf r}_{0}\right) u_{m}\left({\bf r}_{0}\right)
\nonumber\\
&&\sum^{\prime}_{{\nu}_{m'''},{\nu}_{m''},{\nu}_{ m},{\nu}_{ m'}}\nonumber\\
&&C^{*}_{\nu_{1}\cdots(\nu_{m'''}-1)\cdots(\nu_{m''}+1)\cdots(\nu_{m}-1)
	\cdots(\nu_{m'}+1) \cdots \nu_{M}}\nonumber\\
&& C_{\nu_{1}\cdots(\nu_{m'''})\cdots(\nu_{m''})\cdots(\nu_{m})
\cdots(\nu_{m'})\cdots \nu_{M}}\nonumber\\
&&\sqrt{\nu_{m'''}\nu_{m}(\nu_{m''}+1)(\nu_{m'}+1)}. \nonumber\\
\label{eq:cpd_lll}
\end{eqnarray} 
\indent In order to understand the role of reference point ${\bf r}_{0}$
in Eq.~(\ref{eq:cpd_lll}), consider the function in the Lowest Landau Level approximation
\begin{eqnarray}
&&u^{*}_{m''}\left({\bf r}\right)
u_{m'''}\left({\bf r}\right) u^{*}_{ m'}\left({\bf r}_{0}\right) u_{m}\left({\bf r}_{0}\right)\nonumber\\
&=&\frac{1}{\pi^{2}} \frac{r^{m''}}{\sqrt{m''!}} \frac{r^{m'''}}{\sqrt{m'''!}}\frac{r^{m'}_{0}}{\sqrt{m'!}}\frac{r^{m}_{0}}{\sqrt{m!}} \nonumber\\
&&e^{i(m-m')\theta_{0}} \ e^{i(m'''-m'')\theta}. 
\label{eq:unrmmm}
\nonumber
\end{eqnarray}
It follows from the conservation of angular momentum that $m'''-m''=-(m-m')$ whereby the angular dependence in the above equation 
becomes
\begin{eqnarray}
&&u^{*}_{m''}\left({\bf r}\right)
u_{(m''-(m-m'))}\left({\bf r}\right) u^{*}_{ m'}\left({\bf r}_{0}\right) u_{m}\left({\bf r}_{0}\right)\nonumber\\
&=&\frac{1}{\pi^{2}} \frac{r^{m''}}{\sqrt{m''!}} \frac{r^{(m''-(m-m'))}}{\sqrt{(m''-(m-m'))!}}\frac{r^{m'}_{0}}{\sqrt{m'!}}\frac{r^{m}_{0}}{\sqrt{m!}}\nonumber\\
&& e^{i(m-m')(\theta-\theta_{0}).} 
\label{eq:unrmm}
\end{eqnarray}
Taking our reference point ${\bf r}_{0}$ to coincide with the origin(trap centre) {\it i.e.} ${\bf r}_{0}=(0,\theta_{0})$, 
the function in (\ref{eq:unrmm}) above is non-zero only when $m=m'=0$ and the conditional probability distribution in Eq.~(\ref{eq:cpd_lll}), with $n^{\mu}_{r}=0$ in Eq.~(\ref{eq:chi_mu}), becomes  
\begin{eqnarray}
&&{\cal P}({\bf r},{\bf 0})= \vert u_{0}\left({0}\right)\vert^{2}\nu_{0}\nonumber\\
&& \sum_{m} \underbrace{ \vert u_{m}\left({\bf r}\right)\vert^{2}  \sum^{\prime}_{  {\nu}_{m}}\nu_{ m}
\vert C_{\nu_{0}\cdots \nu_{m} \cdots \nu_{M-1}}\vert^{2}}_{(N\lambda_{m}) \ \vert \chi_{m}({\bf r})\vert^{2}} ~~~~~~~~
\label{eq:cpdlll}
\end{eqnarray}
{\it i.e.} the CPD exhibits cylindrical symmetry~\cite{lewenstein06} as shown in Fig.~\ref{fig:x0y0}. 
\\
\indent For the reference point ${\bf r}_{0}=(r_{0},\theta_{0})$ with ${r}_{0}\neq 0$ {\it i.e.} reference point not coinciding with the center of the trap, it follows from Eq.~(\ref{eq:unrmm}) that now $m, m^{\prime}$ can take non-zero, unequal values, leading to breaking of cylindrical symmetry of CPD as shown in Fig.~\ref{fig:x0p5y0} for ${\bf r}_{0}=(0.5,0)$.
\section{The perturbation term in the Hamiltonian breaking $x$-$y$ symmetry in beyond lowest Landau level approximation \label{App:anisot}}
\indent The stirring potential with  
anisotropy in $x$-$y$ plane 
after scaling by the unit of energy $\hbar\omega_{\perp}$, in the first quantized form is given by:
\begin{eqnarray}
V_{p}&=&\frac{1}{\hbar\omega_{\perp}}\frac{A}{2}M\omega_{\perp}^{2}\left(x^2-y^2\right)= \frac{A}{2} \left(\frac{r_{\perp }}{a_{\perp }}\right)^{2}\left(\cos{2}\phi_{}\right)\nonumber\\
\label{eq:vp}
\end{eqnarray}
where $A$ is the dimensionless small parameter of the perturbation.
The equation (\ref{eq:vp}) in the second quantized form becomes
\begin{eqnarray}
&&\hat{V}_{p}=\frac{A}{2} \int_{}^{} r_{\perp}dr_{\perp}d\phi \ \Psi^{\dagger}({r_{\perp},\phi}) \left(\frac{r_{\perp}}{a_{\perp}}\right)^{2} \cos(2\phi) \Psi({r_{\perp},\phi})\nonumber\\
&=&\frac{A}{2}\sum_{\bf n'}\sum_{\bf n} a^{\dagger}_{{\bf n}^{\prime}} a_{\bf n} \int r_{\perp}dr_{\perp}d\phi u^{*}_{{\bf n}^{\prime}}({r_{\perp},\phi}) \left(\frac{r_{\perp}}{a_{\perp}}\right)^{2} \cos(2\phi) u_{\bf n}({r_{\perp},\phi}) \nonumber\\
\label{eq:vp1}
\end{eqnarray}
with composite quantum number written explicitly as ${\bf n}\equiv(n_{r},m)$. The Eq.~(\ref{eq:vp1}) becomes   
\begin{eqnarray}
\hat{V}_{p}&=& \frac{A}{2}\sum_{n'_{r},n_{r}}^{0,1,2\cdots}\sum_{m',m=-\infty}^{\infty}  a^{\dagger}_{n^{\prime}_{r},m^{\prime}} a_{n_{r},m}\nonumber\\
&& \sqrt{\frac{1}{\pi^{2}}\frac{n'_{r}!n_{r}!}{(n'_{r}+|m'|)!(n_{r}+|m|)!}} \nonumber\\
&& \int_{\Lambda=0}^{\infty} \int_{\phi=0}^{2\pi} \frac{1}{2} d\Lambda^{2}\ d\phi L_{n'_{r}}^{|m'|}\left(\Lambda^{2} \right)L_{n_{r}}^{|m|}\left(\Lambda^{2} \right) \nonumber\\ &&\Lambda^{|m'|+|m|} e^{-\Lambda^{2}}\ e^{i(m-m')\phi}  \Lambda^{2} \left(\frac{e^{2i\phi}+e^{-2i\phi}}{2}\right)
\nonumber
\end{eqnarray}
where $\Lambda=r_{\perp}/a_{\perp}$ is the new radial coordinate.
We first integrate over $\phi$ which gives $\delta$-function
\begin{eqnarray}
&&\hat{V}_{p}= \frac{A}{8\pi} \sum_{n_{r},n'_{r}}^{0,1,2\cdots}\sum_{m,m'=-\infty}^{\infty} 2\pi\ [ \delta_{|m|,|m'|-2} \nonumber\\
&&a^{\dagger}_{n^{\prime}_{r},m^{\prime}} a_{n_{r},m} \sqrt{\frac{n_{r}!n'_{r}!}{(n_{r}+|m|)!(n'_{r}+|m'|)!}}\nonumber\\
&& \int_{\Lambda=0}^{\infty}d\Lambda^{2}   
\Lambda^{|m|+|m'|+2} e^{-\Lambda^{2}} L_{n_{r}}^{|m|}\left(\Lambda^{2} \right)L_{n'_{r}}^{|m'|}\left(\Lambda^{2} \right)\nonumber\\
&+& 
\delta_{|m|,|m'|+2}\ 
a^{\dagger}_{n^{\prime}_{r},m^{\prime}} a_{n_{r},m}\sqrt{\frac{n'_{r}!n_{r}!}{(n'_{r}+|m'|)!(n_{r}+|m|)!}}\nonumber\\
&&\int_{\Lambda=0}^{\infty} d\Lambda^{2}  
\Lambda^{|m|+|m'|+2}\ e^{-\Lambda^{2}} 
L_{n'_{r}}^{|m'|}\left(\Lambda^{2} \right) L_{n_{r}}^{|m|}\left(\Lambda^{2} \right)] \nonumber
\end{eqnarray}
and carrying out the summation over $m$ using the $\delta$-functions, we obtain 
\begin{eqnarray}
&&\hat{V}_{p}=\frac{A}{4} \sum_{n_{r},n'_{r}}^{0,1,2\cdots}\sum_{m'=-\infty}^{\infty}[a^{\dagger}_{n^{\prime}_{r},m^{\prime}} a_{n_{r},m'-2} \nonumber\\
&&  \sqrt{\frac{n_{r}!n'_{r}!}{(n_{r}+|m'|-2)!(n'_{r}+|m'|)!}}\nonumber\\
&& \int_{\Lambda=0}^{\infty} e^{-\Lambda^{2}} \Lambda^{2|m'|} L_{n_{r}}^{|m'|-2}\left(\Lambda^{2} \right)L_{n'_{r}}^{|m'|}\left(\Lambda^{2} \right)d\Lambda^{2} \nonumber\\
&+&
a^{\dagger}_{n^{\prime}_{r},m^{\prime}} a_{n_{r},m'+2} 
\nonumber\\
&&
\sqrt{\frac{n_{r}!n'_{r}!}{(n_{r}+|m'|+2)!(n'_{r}+|m'|)!}}\nonumber\\
&&\int_{\Lambda=0}^{\infty} e^{-\Lambda^{2}} \Lambda^{2(|m'|+2)} L_{n_{r}}^{|m'|+2}\left(\Lambda^{2} \right)\nonumber\\
&&
L_{n'_{r}}^{|m'|}\left(\Lambda^{2} \right) d\Lambda^{2}]. \nonumber
\end{eqnarray}
Expanding the associated Laguerre polynomial as 
$L_{q}^{k}\left(x\right)=\sum_{p=0}^{q}\left(-1\right)^{p}\frac{(r+k)!}{(q-p)!(k+p)!p!} x^{p}$ and carrying out the integration, we obtain
\begin{eqnarray}
&&\hat{V}_{p}=\frac{A}{4}\sum_{n^{\prime}_{r},n_{r}}\sum_{m'=-\infty}^{\infty}  [ \ a^{\dagger}_{n^{\prime}_{r},m^{\prime}} a_{n_{r},m^{\prime}-2}\nonumber\\
&&\sqrt{\frac{n'_{r}!n_{r}!}{(n'_{r}+|m'|)!(n_{r}+|m'|-2)!}} \nonumber\\
&& \sum^{n_{r}}_{p=0}(-1)^{p} \frac{(|m'|+n_{r}-2)!}{(n_{r}-p)!(|m'|+p-2)!p!} \nonumber\\
&& \sum^{n'_{r}}_{q=0}(-1)^{q} \frac{(n'_{r}+|m'|)!}{(n'_{r}-q)!(|m'|+q)!q!} \ (|m'|+p+q)! \nonumber\\
&+& 
a^{\dagger}_{n^{\prime}_{r},m^{\prime}} a_{n_{r},m^{\prime}+2} 
\sqrt{\frac{n_{r}!n'_{r}!}{(n_{r}+|m'|+2)!(n'_{r}+|m'|)!}} \nonumber\\
&&\sum^{n_{r}}_{p=0}(-1)^{p} \frac{(n_{r}+|m'|+2)!}{(n_{r}-p)!(|m'|+p+2)!p!} \nonumber\\
&& \sum^{n'_{r}}_{q=0}(-1)^{q} \frac{(|m'|+n'_{r})!}{(n'_{r}-q)!(|m'|+q)!q!} \ (|m'|+p+q+2)! \ ] \nonumber\\
\label{eq:anisbll}
\end{eqnarray}
The Eq.~(\ref{eq:anisbll}) for the perturbation $\hat{V}_{p}$  is the beyond lowest Landau level approximation. The perturbation, in the lowest Landau level approximation, can be obtained from Eq.~(\ref{eq:anisbll}) by setting $n'_{r}=n_{r}=0$ and $m=0,1,2,\cdots$ 
\begin{eqnarray}
\hat{V}_{p}&=&\frac{A}{4}\sum_{m=0,1,2,\cdots} (\ \sqrt{m(m-1)}\ a^{\dagger}_{m} a_{m-2}\nonumber\\
&+& \sqrt{(m+1)|((m+2)}\ a^{\dagger}_{m} a_{m+2}\ )~~~~~~~
\end{eqnarray}
which is the same as equation (17) in reference~\cite{lewenstein06}
\begin{table*}[t]
	\caption{
		\label{Tab:n16_sp3} 
		The many-body ground state energy $E_{0}^{lab}$ in the laboratory frame for $N=16$ spinless bosons interacting via Gaussian potential
		with interaction strength ${U_{0}}=0.2171$ and interaction range $\sigma=0.30,0.50,0.75$. The three largest eigenvalues $\lambda_{1} > \lambda_{ 2} > \lambda_{ 2}$ of the OPRDM and the corresponding one-particle quantum numbers $(n_{1},m^{1})$, $(n_{2},m^{2}), (n_{3},m^{3})$ in the total angular momentum regime $0\leq L_{z} \leq 4N$ are listed.}
	\footnotesize\setlength{\tabcolsep}{1.0pt}
	\centering
	\begin{longtable}{|c |cccc |cccc |cccc|}\hline
		\rowcolor{gray}
		&\multicolumn{4}{c}{$\sigma=0.30$}&\multicolumn{4}{|c}{$\sigma=0.50$}&\multicolumn{4}{|c|}{$\sigma=0.75$}
		\\ \hline
		\multirow{2}{*}{$L_{z}$}	&\multirow{2}{*}{$E_{0}^{lab}$} &$\lambda_{ 1}$ &$\lambda_{2}$ &$\lambda_{3}$ &\multirow{2}{*}{$E_{0}^{lab}$} &$\lambda_{1}$ &$\lambda_{2}$ &$\lambda_{3}$ &\multirow{2}{*}{$E_{0}^{lab}$} &$\lambda_{1}$ &$\lambda_{2}$ &$\lambda_{3}$\\
		&  &$(n_{1},m^{1})$&$(n_{2},m^{2})$ &$(n_{3},m^{3})$ & &$(n_{1},m^{1})$ &$(n_{2},m^{2})$ &$(n_{3},m^{3})$ & &$(n_{1},m^{1})$ &$(n_{2},m^{2})$ & $(n_{3},m^{3})$ \\ \hline
		\multirow{2}{*}{0}	&\multirow{2}{*}{46.86428} &0.98949 &0.00378 &0.00378 &\multirow{2}{*}{46.35902} &0.99292 &0.00278 &0.00278 &\multirow{2}{*}{45.36785} &0.99664& 0.00146& 0.00146\\
		&  &(0,0)&(1,-1)&(1,1)& &(0,0)&(1,-1)&(1,1) & &(0,0) &(1,1) &(1,-1) \\ \hline
		\multirow{2}{*}{1}	&\multirow{2}{*}{47.87008} &0.92247&0.06723& 0.00643&\multirow{2}{*}{47.36087} &0.92678&0.06622&0.00485&\multirow{2}{*}{46.36791} &0.93204&0.06462&0.00263\\
		&  &(0,0) &(1,1) &(1,-1)& &(0,0) &(1,1) &(1,-1)& &(0,0) &(1,1) &(1,-1)\\ \hline
		\multirow{2}{*}{2}	&\multirow{2}{*}{48.63374} &0.91733& 0.05309& 0.02279&\multirow{2}{*}{48.08484} &0.92363&0.05542&0.01677 &\multirow{2}{*}{47.08297} &0.92901&0.05695&0.01227\\
		&  &(0,0) &(2,2) &(1,1) & &(0,0) &(2,2) &(1,1) & &(0,0) &(2,2) &(1,1)\\ \hline
			\multirow{2}{*}{3}	&\multirow{2}{*}{49.32981} &0.91477& 0.05037& 0.01778&\multirow{2}{*}{48.81060} &0.91811&0.05051&0.01665 &\multirow{2}{*}{47.87304} &0.92162&0.05066&0.01533\\
			&  &(0,0) &(3,3) &(1,1) & &(0,0) &(3,3) &(1,1) & &(0,0) &(3,3) &(1,1)\\ \hline
			\multirow{2}{*}{4}	&\multirow{2}{*}{50.25820} &0.83190&0.09654&0.03397 &\multirow{2}{*}{49.73297} &0.83691&0.08557&0.05098&\multirow{2}{*}{48.77732} &0.84923&0.08871&0.05363\\
			&  &(0,0) &(1,1) &(2,2) & &(0,0) &(1,1) &(2,2) & &(0,0) &(2,2) &(1,1)\\ \hline
			\multirow{2}{*}{5}	&\multirow{2}{*}{51.02044} &0.81554&0.07699&0.06578&\multirow{2}{*}{50.48376} &0.82582&0.06967&0.06365&\multirow{2}{*}{49.56409} &0.83417&0.07245&0.05350\\
			&  &(0,0) &(1,1) &(2,2) & &(0,0) &(2,2) &(1,1) & &(0,0) &(2,2) &(1,1)\\ \hline
			\multirow{2}{*}{6}	&\multirow{2}{*}{51.76595} &0.80862&0.07690&0.07077&\multirow{2}{*}{51.25036} &0.81479&0.07857&0.06500&\multirow{2}{*}{50.38120} &0.81743&0.07690&0.06302\\
			&  &(0,0) &(3,3) &(2,2) & &(0,0) &(3,3) &(1,1) & &(0,0) &(3,3) &(1,1)\\ \hline
				\multirow{2}{*}{7}	&\multirow{2}{*}{52.59151} &0.70849&0.17101&0.07419&\multirow{2}{*}{52.08038} &0.71652&0.15698&0.08601 &\multirow{2}{*}{51.22436} &0.72779&0.13523&0.10344\\
	&  &(0,0) &(1,1) &(2,2) & &(0,0) &(3,3) &(1,1) & &(0,0)&(1,1)&(2,2)\\ \hline
	\multirow{2}{*}{8}	&\multirow{2}{*}{53.34973} &0.66720&0.19594&0.08829&\multirow{2}{*}{52.84751} &0.67944&0.17796&0.09439&\multirow{2}{*}{52.03192} &0.68952&0.16253&0.10159\\
	&  & (0,0) &(1,1) &(2,2) & &(0,0) &(1,1) &(2,2)& &(0,0) &(1,1) &(2,2)\\ \hline
	\multirow{2}{*}{9}	&\multirow{2}{*}{54.10657} &0.61417&0.24315&0.09162 &\multirow{2}{*}{53.62127} &0.62186&0.23161&0.09790&\multirow{2}{*}{52.84731} &0.62580&0.22300&0.10954\\
	&  &(0,0) &(1,1) &(2,2) & &(0,0) &(1,1) &(2,2)& &(0,0) &(1,1) &(2,2)\\ \hline
				\multirow{2}{*}{10}	&\multirow{2}{*}{54.86196} &0.54235&0.31429&0.10388&\multirow{2}{*}{54.39106} &0.54967&0.30222&0.11176&\multirow{2}{*}{53.65663} &0.55727&0.28850&0.12212\\
			&  &(0,0) &(1,1) &(2,2) & &(0,0) &(1,1) &(2,2)& & (0,0) &(1,1) &(2,2)\\ \hline
			\multirow{2}{*}{11}	&\multirow{2}{*}{55.60626} &0.47917&0.37575&0.10950&\multirow{2}{*}{55.15183} &0.48501&0.36641&0.11592&\multirow{2}{*}{54.46197} &0.48999&0.35737&0.12412\\
			&  &(0,0) &(1,1) &(2,2) & &(0,0) &(1,1) &(2,2)& &(0,0) &(1,1) &(2,2)\\ \hline
			\multirow{2}{*}{12}	&\multirow{2}{*}{56.34683} &0.44883&0.40950&0.11146&\multirow{2}{*}{55.90975} &0.44355&0.41282&0.11674&\multirow{2}{*}{55.26553} &0.43835&0.41559&0.12329\\
			&  &(0,0) &(1,1) &(2,2) & &(1,1) &(0,0) &(2,2)& &(1,1) &(0,0) &(2,2)\\ \hline
			\multirow{2}{*}{13}	&\multirow{2}{*}{57.08395} &0.53212&0.33439&0.10924&\multirow{2}{*}{56.66452} &0.53037&0.33575&0.11280&\multirow{2}{*}{56.06727} &0.52917&0.33649&0.11684\\
			&  &(1,1) &(0,0) &(2,2) & &(1,1) &(0,0) &(2,2) & &(1,1) &(0,0) &(2,2)\\ \hline
			\multirow{2}{*}{14}	&\multirow{2}{*}{57.81826} &0.62887&0.25247&0.10044&\multirow{2}{*}{57.41706} &0.63071&0.33575&0.10200&\multirow{2}{*}{56.86801} &0.63353&0.25064&0.10350\\
			&  &(1,1) &(0,0) &(2,2) & &(1,1) &(0,0) &(2,2)& &(1,1) &(0,0) &(2,2)\\ \hline
			\multirow{2}{*}{15}	&\multirow{2}{*}{58.54967} &0.74481&0.16083&0.08223&\multirow{2}{*}{58.16768} &0.74901&0.15910&0.08204&\multirow{2}{*}{57.66821} &0.75436&0.15669&0.08146\\
			&  &(1,1) &(0,0) &(2,2)& &(1,1) &(0,0) &(2,2)& &(1,1) &(0,0) &(2,2)\\ \hline
			\multirow{2}{*}{16}	&\multirow{2}{*}{59.25225} &0.87484&0.06268&0.05460&\multirow{2}{*}{58.90377} &0.87949&0.06083&0.05387&\multirow{2}{*}{58.46294} &0.88316&0.05944&0.05355 \\
			&  &(1,1) &(0,0) &(2,2)& &(1,1) &(0,0) &(2,2)& &(1,1) &(0,0) &(2,2)\\ \hline
			\multirow{2}{*}{17}	&\multirow{2}{*}{60.25409} &0.72000&0.14957&0.11366&\multirow{2}{*}{59.90512} &0.72520&0.14895&0.11138&\multirow{2}{*}{59.46389} &0.72910&0.14918&0.10962\\
			&  &(1,1) &(2,2) &(0,0) & &(1,1) &(2,2) &(0,0)& &(1,1) &(2,2) &(0,0)\\ \hline
			\multirow{2}{*}{18}	&\multirow{2}{*}{61.23338} &0.62690&0.19541&0.14181&\multirow{2}{*}{60.87598} &0.65539&0.16395&0.13487 &\multirow{2}{*}{60.41157} &0.68344&0.13243&0.12793\\
			&  &(1,1) &(2,2) &(0,0)& &(1,1) &(2,2) &(0,0)& &(1,1) &(2,2) &(0,0)\\ \hline
		\multirow{2}{*}{19}	&\multirow{2}{*}{62.05475} &0.73381&0.10344&0.09326&\multirow{2}{*}{61.70337} &0.74832&0.09687&0.08710&\multirow{2}{*}{61.26707} &0.76392&0.09034&0.08000\\
		&  &(1,1) &(2,2) &(0,0)& &(1,1) &(2,2) &(0,0) & &(1,1) &(2,2) &(0,0)\\ \hline
		\multirow{2}{*}{20}	&\multirow{2}{*}{63.00573} &0.54544&0.22910&0.15422&\multirow{2}{*}{62.66299} &0.56506&0.21911&0.14556&\multirow{2}{*}{62.23797} &0.58188&0.21401&0.13675\\
		&  &(1,1) &(2,2) &(0,0)& &(1,1) &(2,2) &(0,0)& &(1,1) &(2,2) &(0,0)\\ \hline
		\multirow{2}{*}{21}	&\multirow{2}{*}{63.93098} &0.37760&0.33382&0.21138&\multirow{2}{*}{63.59491} &0.40591&0.31222&0.20091&\multirow{2}{*}{63.16968} &0.45490&0.27007&0.18285\\
		&  &(1,1) &(2,2) &(0,0)& &(1,1) &(2,2) &(0,0)& &(1,1) &(2,2) &(0,0)\\ \hline
		\multirow{2}{*}{22}	&\multirow{2}{*}{64.79399} &0.36431&0.30116&0.23872&\multirow{2}{*}{64.46708} &0.41047&0.29314&0.19153&\multirow{2}{*}{64.05903} &0.51903&0.22401&0.14461\\
		&  &(2,2) &(1,1) &(0,0) & &(1,1) &(2,2) &(0,0)& &(1,1) &(2,2) &(0,0)\\ \hline
		\multirow{2}{*}{23}	&\multirow{2}{*}{65.70191} &0.40268&0.27154&0.22220&\multirow{2}{*}{65.38521} &0.38009&0.30566&0.20860&\multirow{2}{*}{64.99627} &0.36228&0.34470&0.18875\\
		&  &(2,2) &(1,1) &(0,0)& &(2,2) &(1,1) &(0,0)& &(2,2) &(1,1) &(0,0)\\ \hline
		\multirow{2}{*}{24}	&\multirow{2}{*}{66.54940} &0.53143&0.28750&0.08470&\multirow{2}{*}{66.25488} &0.52244&0.27979&0.27979&\multirow{2}{*}{65.89361} &0.51307&0.26090&0.13135\\
		&  &(2,2) &(1,1) &(0,0)& &(2,2) &(0,0) &(1,1) & &(2,2) &(0,0) &(1,1)\\ \hline
		\multirow{2}{*}{25}	&\multirow{2}{*}{67.49024} &0.46150&0.21617&0.18994&\multirow{2}{*}{67.19100} &0.43787&0.21557&0.20891&\multirow{2}{*}{66.82809} &0.40627&0.26644&0.18659\\
		&  &(2,2) &(1,1) &(0,0) & &(2,2) &(1,1) &(0,0)& &(2,2) &(1,1) &(0,0)\\ \hline
		\end{longtable}
\end{table*}
\begin{table*}[t]
\footnotesize\setlength{\tabcolsep}{1.10pt}
\begin{longtable}{|c |cccc |cccc |cccc|}\hline
		\multirow{2}{*}{26}	&\multirow{2}{*}{68.31986} &0.58021&0.24529&0.07453&\multirow{2}{*}{68.04368} &0.57653&0.24110&0.07248&\multirow{2}{*}{67.71570} &0.58305&0.22595&0.08725\\
		&  &(2,2) &(0,0) &(4,4) & &(2,2) &(0,0) &(4,4) & &(2,2) &(0,0) &(4,4)\\ \hline
		\multirow{2}{*}{27}	&\multirow{2}{*}{69.31222} &0.50609&0.19009&0.14434&\multirow{2}{*}{69.02539} &0.49911&0.18611&0.15505&\multirow{2}{*}{68.68383} &0.49899&0.18282&0.16587\\
		&  &(2,2) &(0,0) &(1,1) & &(2,2) &(0,0) &(1,1)& &(2,2) &(1,1) &(0,0)\\ \hline
		\multirow{2}{*}{28}	&\multirow{2}{*}{70.13310} &0.61063&0.20011&0.08265&\multirow{2}{*}{69.86716} &0.61129&0.19537&0.07853&\multirow{2}{*}{69.55952} &0.62821&0.17685&0.07645\\
		&  &(2,2) &(0,0) &(4,4) & &(2,2) &(0,0) &(4,4)& &(2,2) &(0,0) &(1,1) \\ \hline
		\multirow{2}{*}{29}	&\multirow{2}{*}{71.16949} &0.47328&0.19354&0.17233&\multirow{2}{*}{70.88586} &0.47026&0.18210&0.17080&\multirow{2}{*}{70.55437} &0.50003&0.15967&0.14859\\
		&  &(2,2) &(0,0) &(3,3) & &(2,2) &(0,0) &(3,3) & &(2,2) &(1,1) &(3,3)  \\ \hline
		\multirow{2}{*}{30}	&\multirow{2}{*}{71.99933} &0.61755&0.20508&0.10182&\multirow{2}{*}{71.73464} &0.63313&0.15321&0.09048&\multirow{2}{*}{71.43247} &0.67711&0.12229&0.07103\\
		&  &(2,2) &(0,0) &(4,4) & &(2,2) &(0,0) &(4,4) & &(2,2) &(0,0) &(1,1)\\ \hline
			\multirow{2}{*}{31}	&\multirow{2}{*}{73.03732} &0.42735&0.20508&0.19018&\multirow{2}{*}{72.76240} &0.44022&0.20842&0.17746&\multirow{2}{*}{72.44891} &0.48912&0.20769&0.14204\\
	&  &(2,2) &(3,3) &(0,0) & &(2,2) &(3,3) &(0,0)& &(2,2) &(3,3) &(0,0)\\ \hline
	\multirow{2}{*}{32}	&\multirow{2}{*}{73.83171} &0.43495&0.19058&0.14261&\multirow{2}{*}{73.59262} &0.48124&0.16906&0.13125&\multirow{2}{*}{73.31758} &0.63923&0.09346&0.09329\\
	&  &(2,2) &(0,0) &(4,4)& &(2,2) &(0,0) &(4,4)& &(2,2) &(3,3) &(1,1)\\ \hline
	\multirow{2}{*}{33}	&\multirow{2}{*}{74.77374} &0.26996&0.24653&0.22551&\multirow{2}{*}{74.54991} &0.30196&0.21557&0.19393&\multirow{2}{*}{74.29530} &0.35610&0.24362&0.19006\\
	&  &(3,3) &(0,0) &(2,2) & &(2,2) &(0,0) &(3,3) & &(2,2) &(3,3) &(0,0)\\ \hline
		\multirow{2}{*}{34}	&\multirow{2}{*}{75.70208} &0.26981&0.23826&0.23264&\multirow{2}{*}{75.48376} &0.25519&0.25365&0.22252&\multirow{2}{*}{75.24622} &0.29534&0.24975&0.19127\\
		&  &(3,3) &(2,2) &(0,0)& &(3,3) &(2,2) &(0,0)& &(2,2) &(3,3) &(0,0)\\ \hline
		\multirow{2}{*}{35}	&\multirow{2}{*}{76.62143} &0.32727&0.24476&0.18294&\multirow{2}{*}{76.41198} &0.31683&0.23683&0.19349&\multirow{2}{*}{76.18597} &0.25519&0.25365&0.22252\\
		&  &(3,3) &(0,0) &(2,2)& &(3,3) &(0,0) &(2,2)& &(3,3) &(2,2) &(0,0)\\ \hline
		\multirow{2}{*}{36}	&\multirow{2}{*}{77.55398} &0.48498&0.24302&0.10486&\multirow{2}{*}{77.34967} &0.48347&0.23920&0.18098&\multirow{2}{*}{77.12801} &0.50812&0.23021&0.10629\\
		&  &(3,3) &(0,0) &(2,2)& &(3,3) &(0,0) &(2,2)& &(3,3) &(0,0) &(2,2)\\ \hline
		\multirow{2}{*}{37}	&\multirow{2}{*}{78.52752} &0.28141&0.21992&0.20067&\multirow{2}{*}{78.32315} &0.27538&0.21411&0.20491&\multirow{2}{*}{78.10569} &0.31124&0.20665&0.19600\\
		&  &(3,3) &(0,0) &(2,2)& &(3,3) &(0,0) &(2,2)& &(3,3) &(2,2) &(0,0)  \\ \hline
		\multirow{2}{*}{38}	&\multirow{2}{*}{79.45758} &0.36911&0.21714&0.15734&\multirow{2}{*}{79.26053} &0.36638&0.21275&0.16164&\multirow{2}{*}{79.05078} &0.38258&0.19200&0.17805\\
		&  &(3,3) &(0,0) &(2,2)& &(3,3) &(0,0) &(2,2)& &(3,3) &(0,0) &(2,2)\\ \hline
		\multirow{2}{*}{39}	&\multirow{2}{*}{80.39032} &0.47656&0.20374&0.12276&\multirow{2}{*}{80.19670} &0.47883&0.20164&0.12226&\multirow{2}{*}{79.98856} &0.53684&0.18769&0.09989\\
		&  &(3,3) &(0,0) &(4,4) & &(3,3) &(0,0) &(4,4)& &(3,3) &(0,0) &(4,4)\\ \hline
		\multirow{2}{*}{40}	&\multirow{2}{*}{81.37679} &0.24484&0.22016&0.19725&\multirow{2}{*}{81.18208} &0.23981&0.22654&0.18993&\multirow{2}{*}{80.98327} &0.35689&0.18428&0.14831\\
		&  &(0,0) &(4,4) &(3,3) & &(0,0) &(4,4) &(3,3)& &(3,3) &(0,0) &(2,2)\\ \hline
		\multirow{2}{*}{41}	&\multirow{2}{*}{82.34432} &0.32571&0.19444&0.15778&\multirow{2}{*}{82.14866} &0.32007&0.19086&0.16184&\multirow{2}{*}{81.94333} &0.36876&0.17156&0.16218\\
		&  &(3,3) &(0,0) &(4,4) & &(3,3) &(0,0) &(4,4)& &(3,3) &(0,0) &(2,2)\\ \hline
		\multirow{2}{*}{42}	&\multirow{2}{*}{83.28308} &0.35277&0.19483&0.16750&\multirow{2}{*}{83.09047} &0.38254&0.18530&0.16036&\multirow{2}{*}{82.88402} &0.53418&0.14530&0.11736\\
		&  &(3,3) &(0,0) &(4,4)& &(3,3) &(0,0) &(4,4)& &(3,3) &(0,0) &(4,4)\\ \hline
		\multirow{2}{*}{43}	&\multirow{2}{*}{84.23940} &0.33911&0.24764&0.16590&\multirow{2}{*}{84.05134} &0.33797&0.24222&0.17261&\multirow{2}{*}{83.86494} &0.29744&0.23747&0.20865\\
		&  &(4,4) &(0,0) &(3,3) & &(4,4) &(0,0) &(3,3)& &(4,4) &(3,3) &(0,0)\\ \hline
		\multirow{2}{*}{44}	&\multirow{2}{*}{85.17724} &0.44093&0.26281&0.10992&\multirow{2}{*}{84.99470} &0.43838&0.25920&0.10867&\multirow{2}{*}{84.82073} &0.39876&0.23720&0.14083\\
		&  &(4,4) &(0,0) &(3,3)& &(4,4) &(0,0) &(5,5) & &(4,4) &(0,0) &(3,3) \\ \hline
		\multirow{2}{*}{45}	&\multirow{2}{*}{86.19175} &0.25767&0.24947&0.23945&\multirow{2}{*}{86.00803} &0.25919&0.24227&0.23546&\multirow{2}{*}{85.81515} &0.49713&0.16417&0.10927\\
		&  &(4,4) &(0,0) &(5,5) & &(4,4) &(0,0) &(5,5)& &(3,3) &(4,4) &(0,0)\\ \hline
		\multirow{2}{*}{46}	&\multirow{2}{*}{87.16767} &0.28202&0.19966&0.19827&\multirow{2}{*}{86.97846} &0.27755&0.19614&0.19235&\multirow{2}{*}{86.78861} &0.25276&0.21999&0.18615\\
		&  &(4,4) &(0,0) &(5,5)& &(4,4) &(5,5) &(0,0)& &(4,4) &(3,3) &(5,5)\\ \hline
		\multirow{2}{*}{47}	&\multirow{2}{*}{88.09247} &0.37420&0.21216&0.16681&\multirow{2}{*}{87.91174} &0.37577&0.20809&0.16238&\multirow{2}{*}{87.73690} &0.36723&0.19508&0.17510\\
		&  &(4,4) &(0,0) &(5,5)& &(4,4) &(0,0) &(5,5)& &(4,4) &(0,0) &(3,3) \\ \hline
		\multirow{2}{*}{48}	&\multirow{2}{*}{88.89032} &0.48443&0.22052&0.10773&\multirow{2}{*}{88.73774} &0.47930&0.21845&0.11075 &\multirow{2}{*}{88.60174} &0.46760&0.21079&0.11811\\
		&  &(4,4) &(0,0) &(3,3) & &(4,4) &(0,0) &(3,3)& &(4,4) &(0,0) &(3,3)\\ \hline
		\multirow{2}{*}{49}	&\multirow{2}{*}{89.92046} &0.32123&0.21793&0.17497&\multirow{2}{*}{89.76270} &0.32100&0.21662&0.17762&\multirow{2}{*}{89.62116} &0.32397&0.21245&0.17784\\
		&  &(4,4) &(0,0) &(5,5)& &(4,4) &(0,0) &(5,5)& &(4,4) &(0,0) &(5,5)\\ \hline
		\multirow{2}{*}{50}	&\multirow{2}{*}{90.90461} &0.21761&0.20342&0.20182&\multirow{2}{*}{90.74682} &0.23463&0.20335&0.19545&\multirow{2}{*}{90.59611} &0.37857&0.15893&0.15433\\
		&  &(4,4) &(0,0) &(3,3)& &(4,4) &(3,3) &(0,0)& &(4,4) &(3,3) &(0,0)\\ \hline
		\multirow{2}{*}{51}	&\multirow{2}{*}{91.86018} &0.28196&0.21103&0.18267&\multirow{2}{*}{91.70336} &0.29217&0.21128&0.17811&\multirow{2}{*}{91.55611} &0.32979&0.21110&0.16150\\
		&  &(4,4) &(3,3) &(0,0) & &(4,4) &(3,3) &(0,0)& &(4,4) &(3,3) &(0,0)\\ \hline
		\multirow{2}{*}{52}	&\multirow{2}{*}{92.79163} &0.54632&0.17165&0.09413&\multirow{2}{*}{92.64065} &0.54190&0.17028&0.09559&\multirow{2}{*}{92.50030} &0.53680&0.16591&0.09784\\
		&  &(4,4) &(0,0) &(3,3)& &(4,4) &(0,0) &(3,3)& &(4,4) &(0,0) &(3,3)\\ \hline
		\multirow{2}{*}{53}	&\multirow{2}{*}{93.81503} &0.29268&0.19396&0.17741&\multirow{2}{*}{93.65995} &0.30013&0.19670&0.17472&\multirow{2}{*}{93.51604} &0.32485&0.19984&0.16602\\
		&  &(4,4) &(5,5) &(0,0)& &(4,4) &(5,5) &(0,0)& &(4,4) &(5,5) &(0,0)\\ \hline
		\multirow{2}{*}{54}	&\multirow{2}{*}{94.79663} &0.23017&0.21764&0.19550&\multirow{2}{*}{94.64215} &0.22788&0.21393&0.21335&\multirow{2}{*}{94.50029} &0.29504&0.19093&0.18550\\
		&  &(5,5) &(0,0) &(4,4) & &(5,5) &(0,0) &(3,3)& &(4,4)& (0,0)&(5,5) \\ \hline
\end{longtable}
\end{table*}
\begin{table*}[t]
\footnotesize\setlength{\tabcolsep}{1.10pt}
\begin{longtable}{|c |cccc |cccc |cccc|}\hline
			\multirow{2}{*}{55}	&\multirow{2}{*}{95.76346} &0.40131&0.24870&0.13557&\multirow{2}{*}{95.61163} &0.41169&0.2505&0.12700&\multirow{2}{*}{95.47868} &0.39847&0.23822&0.11948\\
	&  &(5,5) &(0,0) &(6,6) & &(5,5) &(0,0) &(6,6)& &(5,5) &(0,0) &(4,4)\\ \hline
		\multirow{2}{*}{56}	&\multirow{2}{*}{96.74138} &0.36293&0.22038&0.11831&\multirow{2}{*}{96.59324} &0.32819&0.19334&0.12312&\multirow{2}{*}{96.44362} &0.53301&0.11707&0.11096\\
		&  &(6,6) &(1,1) &(0,0) & &(6,6) &(1,1) &(0,0)& &(4,4) &(0,0) &(5,5)\\ \hline
		\multirow{2}{*}{57}	&\multirow{2}{*}{97.74751} &0.31343&0.17286&0.16518&\multirow{2}{*}{97.59379} &0.27007&0.22202&0.17065&\multirow{2}{*}{97.45045} &0.34580&0.17726&0.14997\\
		&  &(6,6) &(5,5) &(0,0) & &(6,6) &(5,5) &(0,0)& &(5,5) &(0,0) &(6,6)\\ \hline
		\multirow{2}{*}{58}	&\multirow{2}{*}{98.73114} &0.26638&0.22223&0.17347&\multirow{2}{*}{98.57570} &0.25901&0.22335&0.17368&\multirow{2}{*}{98.43219} &0.22502&0.21346&0.20474\\
		&  &(6,6) &(5,5) &(0,0)& &(6,6) &(5,5) &(0,0)& &(6,6) &(4,4) &(5,5)\\ \hline
		\multirow{2}{*}{59}	&\multirow{2}{*}{99.68387} &0.33968&0.20227&0.19941&\multirow{2}{*}{99.53192} &0.34524&0.20254&0.19324&\multirow{2}{*}{99.39629} &0.34703&0.19718&0.18262\\
		&  &(5,5) &(0,0) &(6,6) & &(5,5) &(0,0) &(6,6)& &(5,5) &(0,0) &(6,6)\\ \hline
		\multirow{2}{*}{60}	&\multirow{2}{*}{100.63525}&0.49277&0.21973&0.13320&\multirow{2}{*}{100.48698} &0.50136&0.22034&0.12698&\multirow{2}{*}{100.35792} &0.50544&0.21727&0.12034\\
		& &(5,5) &(0,0) &(6,6)& &(5,5) &(0,0) &(6,6)& &(5,5) &(0,0) &(6,6)\\ \hline
		\multirow{2}{*}{61}	&\multirow{2}{*}{101.62990} &0.45397&0.22843&0.09139&\multirow{2}{*}{101.48789} &0.42762&0.20063&0.12101&\multirow{2}{*}{101.36445} &0.36603&0.18254&0.14930\\
		&  &(6,6) &(1,1) &(5,5) & &(6,6) &(1,1) &(5,5)& &(6,6) &(5,5) &(1,1)\\ \hline
		\multirow{2}{*}{62}	&\multirow{2}{*}{102.64995} &0.41259&0.14677&0.10603&\multirow{2}{*}{102.50076} &0.40369&0.15460&0.14726&\multirow{2}{*}{102.36492} &0.28829&0.28503&0.13531\\
		&  &(6,6) &(0,0) &(5,5)& &(6,6) &(5,5) &(0,0)& &(6,6) &(5,5) &(0,0)\\ \hline
		\multirow{2}{*}{63}	&\multirow{2}{*}{103.66118} &0.41207&0.14829&0.13460&\multirow{2}{*}{103.50724} &0.39121&0.16719&0.12877&\multirow{2}{*}{103.36137} &0.32674&0.22540&0.11743\\
		&  &(6,6) &(5,5) &(0,0) & &(6,6) &(5,5) &(0,0)& &(6,6) &(5,5) &(0,0)\\ \hline
		\multirow{2}{*}{64}	&\multirow{2}{*}{104.61332} &0.34839&0.28352&0.12550&\multirow{2}{*}{104.46445} &0.31898&0.24512&0.14863&\multirow{2}{*}{104.3154} &0.25069&0.19113&0.17554\\
		&  &(1,1) &(7,7) &(4,4) & &(1,1) &(7,7) &(4,4)& &(1,1) &(4,4) &(5,5)\\ \hline
	\end{longtable}
\end{table*}

\end{document}